\documentclass[
 reprint,
 superscriptaddress,
 preprintnumbers,
 amsmath
 amssymb,
 prx,
 floatfix,
]{revtex4-2}

\DeclareMathAlphabet{\mathsfit}{T1}{\sfdefault}{\mddefault}{\sldefault}
\SetMathAlphabet{\mathsfit}{bold}{T1}{\sfdefault}{\bfdefault}{\sldefault}

\usepackage{graphicx}
\graphicspath{{fig/}}   
\usepackage{dcolumn}
\usepackage{bm}
\usepackage{makecell}
\usepackage{braket} 
\usepackage{siunitx}
\usepackage{color}
\usepackage{array}
\usepackage{xr}
\usepackage{upgreek}
\usepackage{hyperref}
\usepackage{hyperref}
\hypersetup{
    colorlinks = true,
    linkcolor = black,
    citecolor = black
}
\usepackage[caption=false]{subfig}
\usepackage{multirow} 
\usepackage{amsmath}

\setcitestyle{super}

\begin{document} 

\title{Non-degenerate noise-resilient superconducting qubit}

\def\RLEaffil{Research Laboratory of Electronics, Massachusetts Institute of Technology, Cambridge, MA 02139, USA}
\def\Physaffil{Department of Physics, Massachusetts Institute of Technology, Cambridge, MA 02139, USA}
\def\EECSaffil{Department of Electrical Engineering and Computer Science, Massachusetts Institute of Technology, Cambridge, MA 02139, USA}

\author{Max~Hays}
\email{maxhays@mit.edu}
\affiliation{\RLEaffil}

\author{Junghyun~Kim}
\affiliation{\RLEaffil}
\affiliation{\EECSaffil}

\author{William~D.~Oliver}
\affiliation{\RLEaffil}
\affiliation{\EECSaffil}
\affiliation{\Physaffil}

\date{\today}

\begin{abstract}
\noindent We propose a superconducting qubit based on engineering the first and second harmonics of the Josephson energy and phase relation $E_{J1}\cos \varphi$ and $E_{J2}\cos 2\varphi$. By constructing a circuit such that $E_{J2}$ is negative and $|E_{J1}| \ll |E_{J2}|$, we create a periodic potential with two non-degenerate minima. The qubit, which we dub ``harmonium'', is formed from the lowest-energy states of each minimum. Bit-flip protection of the qubit arises due to the localization of each qubit state to their respective minima, while phase-flip protection can be understood by considering the circuit within the Born-Oppenheimer approximation. We demonstrate with time-domain simulations that single- and two-qubit gates can be performed in approximately one hundred nanoseconds. 
Finally, we compute the qubit coherence times using numerical diagonalization of the complete circuit in conjunction with state-of-the-art noise models. 
We estimate out-of-manifold heating times on the order of milliseconds, which can be treated as erasure errors using conventional dispersive readout. 
We estimate pure-dephasing times on the order of many tens of milliseconds, and bit-flip times on the order of seconds. 

\end{abstract}

\maketitle


\section{I\MakeLowercase{ntroduction}}

Superconducting circuits are a leading platform in the effort to develop fault-tolerant quantum computers\cite{krantz2019quantum,devoret2013superconducting,blais2021circuit}.
While the field has made tremendous progress since the first manipulation of a superconducting quantum bit twenty-five years ago, the two most widely-used superconducting qubits today can be considered optimized versions of the original charge and flux qubits\cite{nakamura1999coherent, mooij1999josephson}. 
The transmon, a capacitively-shunted charge qubit, was the fundamental building block underpinning the first demonstrations of quantum supremacy and implementations of lattice-based error correction\cite{koch2007charge,arute2019quantum,krinner2022realizing,googleEC}. Meanwhile, the fluxonium, a flux qubit with large shunt inductance, was used to demonstrate the highest coherence and highest fidelity two-qubit gates in superconducting circuits to date\cite{manucharyan2009fluxonium,somoroff2023millisecond,ding2023high,zhang2024tunable,lin202424}. 

There has been a significant effort to develop superconducting qubits with intrinsic noise protection relative to charge and flux qubits\cite{gyenis2021moving,kitaevCurrent,brooks2013protected,kalashnikov2020bifluxon,gyenis2021experimental}. 
Typically, these proposed qubits are imagined to offer bit-flip protection by hosting logical wavefunctions with disjoint support, meaning the wavefunctions occupy separate, non-overlapping regions of the circuit phase space. 
Phase-flip protection, on the other hand, would be achieved by ensuring that the qubit frequency has a minimal dependence on Hamiltonian parameters such as voltage/flux biases or circuit element energies. 

\begin{figure}
\includegraphics[width=\columnwidth]{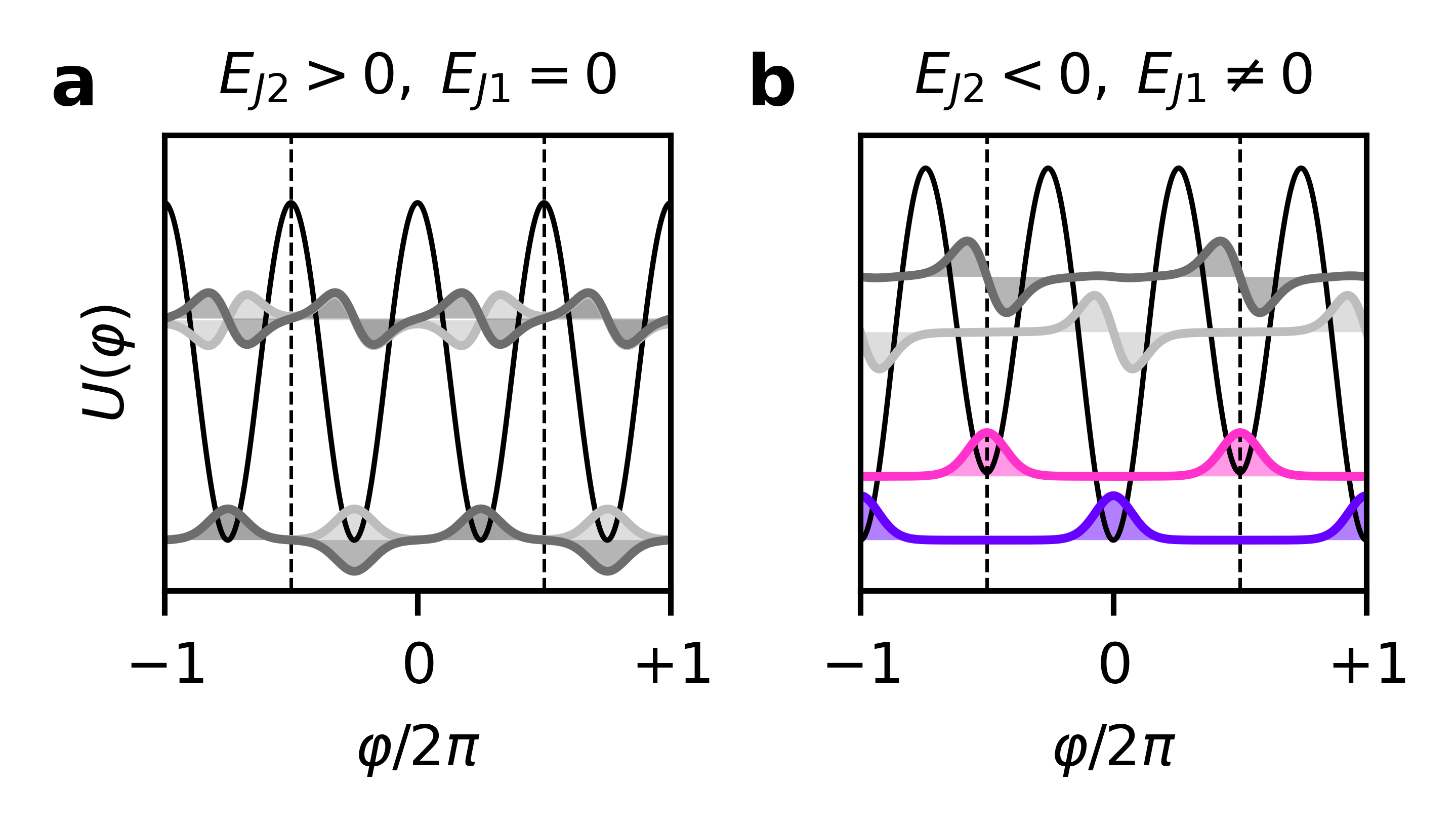}
\caption{\textbf{Breaking the degeneracy of a $\pi$-periodic Josephson element.} (a) Potential (black) and wavefunctions (gray) of a $\pi$-periodic element with a energy and phase relation of $E_{J2} \cos 2\varphi$, with $E_{J2} > 0$. The phase is defined from $-\pi$ to $\pi$ (black dashed lines); the x-axis is only expanded for visual clarity. The splitting between the lowest two eigenstates is exponentially suppressed in $E_{J2}/E_C$, with $E_C$ the charging energy, rendering a qubit formed from these two states extremely sensitive to flux noise. (b) This degeneracy is broken by flipping the sign of $E_{J2}$ and adding a term $E_{J1}\cos\varphi$ to the potential. The lowest two eigenstates (purple and pink wavefunctions) are localized to the wells at even/odd multiples of $\pi$, suppressing bit flips due to both charge and phase noise. } 
\label{fig:fig1}
\end{figure}

An approach that has received much attention is the implementation of $\pi$-periodic Josephson elements [Fig. ~\ref{fig:fig1}(a)]~\cite{ioffe2002possible, douccot2005protected, gladchenko2009superconducting, bell2014protected, smith2020superconducting, larsen2020parity, schrade2022protected}. 
Because the superconducting phase is fundamentally $2\pi$-periodic, a $\pi$-periodic element hosts two degenerate minima. 
The low-energy states of these two minima can be used as a qubit, which is resilient against certain noise channels.  
Consider a $\pi$-periodic element with a $\cos2\varphi$ energy and phase relationship. 
The noise protection of such an element can be understood by noting that it only allows the tunneling of \textit{pairs} of Cooper pairs, as can be seen by writing the operator $\cos2\hat{\varphi}$ in the charge basis, where $n$ counts the number of Cooper pairs to have tunnelled through the element: 

\begin{equation*} \label{cos_2phi}
\cos 2\hat{\varphi} = \frac{1}{2}\sum_n |n\rangle \langle n+2| + \mathrm{h.c.}
\end{equation*}

\noindent Because no odd-order Josephson term is present in the Hamiltonian, the parity of the number of Cooper pairs to have tunneled through the element is a good quantum number, and in particular the two quasi-degenerate qubit states $|g\rangle,|e\rangle$ are of opposite parity. 
As such, fluctuations that couple to the charge operator $\hat{n}$ are unable to induce bit-flips because $\langle e |\hat{n}|g \rangle = 0$. 
However, phase-coupled noise can in general induce bit flips because $\langle e |e^{i\hat{\varphi}}|g \rangle \neq 0$. 
This is a severe problem for $\pi$-periodic elements, since the qubit is formed from two quasi-degenerate states~[Fig.\ref{fig:fig1}(a)] and phase-coupled flux noise is known to follow a $1/f$ divergence at low frequencies.
In other words, while charge noise only couples to the qubit longitudinally, flux noise can couple transversely. 
It has been proposed to deal with these problems by using arrays of $\pi$-periodic elements, but these approaches either require unrealistically symmetric fabrication or \textit{in-situ} tuning of Josephson energies, which will itself add another non-protected noise channel~\cite{gladchenko2009superconducting, bell2014protected, schrade2022protected}. 

Here we propose to intentionally break the degeneracy of a $\pi$-periodic element $E_{J2} \cos2\hat{\varphi}$ by adding a term $E_{J1}\cos\hat{\varphi}$ [Fig.~\ref{fig:fig1}(b)], thus forming the potential $U = E_{J1} \cos\hat{\varphi} + E_{J2} \cos2\hat{\varphi}$.
If $|E_{J1}| \ll |E_{J2}|$ and $E_{J2}$ is negative, then this non-degenerate qubit will have disjoint support in $\hat{\varphi}$, resulting in phase $\langle e |\hat{\varphi}|g \rangle$ and charge $\langle e |\hat{n}|g \rangle =-i\langle e |\frac{\partial}{\partial\hat{\varphi}}|g \rangle$ matrix elements that are  exponentially suppressed in $E_{J2}/E_{C}$, where $E_{C}$ is the charging energy of the circuit. 
In this work, we show how to construct such a bit-flip-protected superconducting circuit, and in addition how this circuit can simultaneously be made resilient against phase flips. 
While the qubit frequency on the order of MHz is much smaller than typical superconducting qubit operation frequencies, we stress the weakly-broken degeneracy of the \textit{potential} as the salient feature of our circuit. 

The remainder of this paper is organized as follows. In section \ref{intuition}, we present a sketch of the ideas driving the circuit design; we hope this will give intuition for the operating principles of the qubit. 
This includes qualitative arguments for how our proposed circuit realizes the effective potential Fig.~\ref{fig:fig1}(b), as well as a Born-Oppenheimer treatment of a reduced circuit model. 
We also discuss the significant practical advantages of our proposed qubit over the $0-\pi$ qubit, which has a low-energy structure similar to that of Fig.~\ref{fig:fig1}(b). 
In section \ref{quantitative}, we present a complete analysis using numerically-exact diagonalization. 
We quantitatively estimate bit-flip and phase-flip rates using contemporary noise models. 
We discuss the implementation of single- and two-qubit gates, and how out-of-manifold heating events can be treated as erasure errors using conventional dispersive readout~\cite{stace2009thresholds,shim2016semiconductor,campbell2020universal,chou2023demonstrating,ma2023high, levine2024demonstrating, koottandavida2024erasure}. 

\begin{figure*}
\includegraphics[width=\textwidth]{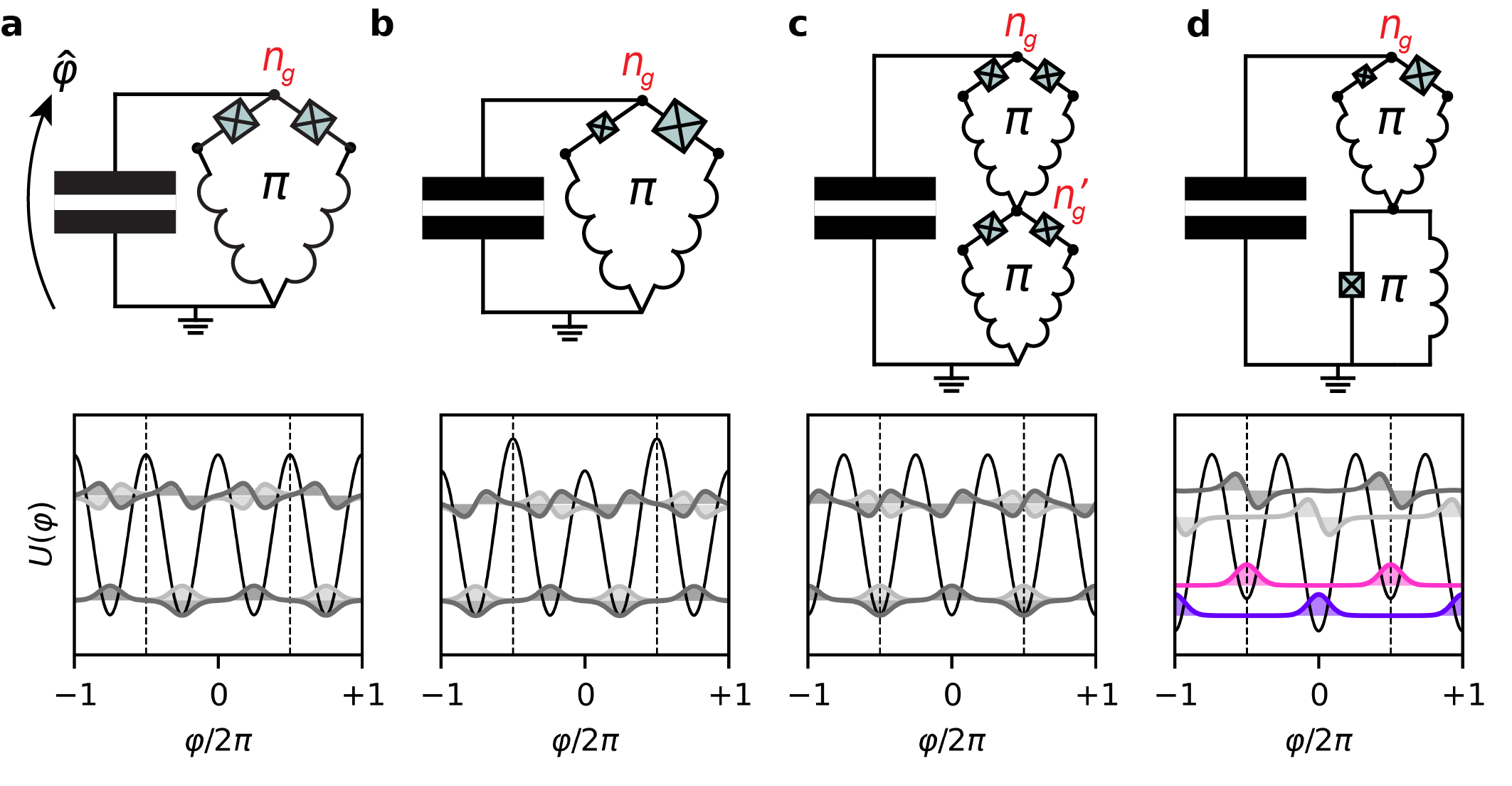}
\caption{\textbf{Constructing the Harmonium Circuit.} (a) The kite circuit. At external flux $\varphi_\mathrm{e} = \pi$, the effective potential (lower panel) for the phase $\varphi$ across the element is $\pi$-periodic with minima at $\pm \pi/2$ (positive $E_{J2}$). As the potential is periodic, the circuit is sensitive to the gate charge $n_g$. (b) Asymmetry of the JJs results in an additional $\cos\hat{\varphi}$ component of the potential, leading to an alternation of the barrier heights in the potential. (c) Connecting two kites in series results in an effective negative $E_{J2}$ potential for the entire phase $\hat{\varphi}$ across the device (note the shift of the potential by $\pi/2$). The addition of a second kite also adds a second noisy gate charge $n_g'$. (d) Now, the addition of a $\cos\varphi$ component results in alternation in the potential minima instead of the potential heights. The second kite can be replaced with a fluxonium with appropriate $E_J,E_L$ values, thus rendering the circuit energy levels insensitive to the noisy gate charge $n_g'$. 
}

\label{fig:fig2}
\end{figure*}

\section{Q\MakeLowercase{ualitative description of the qubit}} \label{intuition}

\subsection{Engineering the qubit potential}

The central goal of this work is to engineer a superconducting circuit such that the low-energy degrees of freedom move in a potential qualitatively similar to the one depicted in Fig.~\ref{fig:fig1}(b). 
In this section, we sketch for the reader how our proposed circuit accomplishes this, at the expense of rigor. 
We stress that a rigorous circuit analysis can be found in section \ref{quantitative}.

There have been many proposals and experiments geared towards the realization of $\pi$-periodic elements~\cite{ioffe2002possible, douccot2005protected, gladchenko2009superconducting, bell2014protected, smith2020superconducting, larsen2020parity, schrade2022protected}. 
In this work, we will primarily focus on the kite~\cite{smith2020superconducting} [Fig.~\ref{fig:fig2}(a)]. 
However, we note that many of the arguments presented throughout this manuscript could also apply to $\pi$-periodic elements fabricated from other weak links, such as superconductor-semiconductor heterostructures~\cite{larsen2020parity, schrade2022protected}. 

The operating principle of the kite can be understood as follows. 
Two identical branches of a JJ in series with an inductor are wired in parallel, and an external flux of half a flux quantum $\varphi_\mathrm{e} = \pi$ is threaded between them~[Fig.~\ref{fig:fig2}(a)] (throughout this work we will express external flux in units of reduced flux quanta $\Phi_0/2\pi$).  
Classically, there are then two degenerate states of the loop corresponding to zero and one flux quanta.
In the case of zero flux quanta, the flux $\varphi_\mathrm{e} = \pi$ is split evenly between the two identical branches, such that average value of the phase drop $\varphi$ across the element is $\varphi_\mathrm{e}/2 = +\pi/2$. 
In the single flux quantum case, the average value is $(\varphi_\mathrm{e}-1)/2 = -\pi/2$; the potential for the phase drop $\hat{\varphi}$ across the kite thus has two minima at $\pm \pi/2$ corresponding to the cases of zero and one flux quanta. 
Furthermore, one can show that the potential is symmetric around each of the minima, rendering the potential $\pi$-periodic. 
Because the minima are at $\pm \pi/2$ instead of $0, \pi$, the effective $E_{J2}$ of the kite is \textit{positive} [Fig.~\ref{fig:fig2}(a), lower panel]; the importance of the distinction between positive and negative $E_{J2}$ will become clear below. 
As far as we are aware, the same is true for all proposed and realized $\pi$-periodic elements (excluding the $0-\pi$ qubit~\cite{brooks2013protected}, which we comment on in detail later).

In an attempt to break the degeneracy of the wells, one can introduce an asymmetry between the two JJs of the kite [Fig.~\ref{fig:fig2}(b)]. 
This results in a non-zero $E_{J1}\cos \hat{\varphi}$ term due to incomplete destructive interference; this is the same reason there is a non-zero $E_{J1}$ term at $\pi$ flux in an asymmetric DC squid. 
However, the additional $E_{J1} \cos \hat{\varphi}$ term does not break the degeneracy between wells as we desire, but instead results in alternating peak heights of the potential [Fig.~\ref{fig:fig2}(b), lower panel].  
The central issue is that the wells of the positive $E_{J2}\cos 2\hat{\varphi}$ potential are located at $\pm \pi/2$ as in Fig.~\ref{fig:fig1}(a) instead of $0, \pi$ as in Fig.~\ref{fig:fig1}(b). 

We must therefore design a circuit that has effective negative $E_{J2}$ instead of positive. 
The key observation underlying our proposed qubit circuit is that two serially-connected \textit{positive} $E_{J2}$ $\pi$-periodic elements together function as a \textit{negative} $E_{J2}$ $\pi$-periodic element [Fig.~\ref{fig:fig2}(c)]. 
To understand this qualitatively, first recall that each element individually has minima at $\pm \pi/2$. 
When we consider the total phase drop across the two elements $\hat{\varphi}$, these minima can either add together $\pi/2 + \pi/2 = \pi$ or subtract  $\pi/2 - \pi/2 = 0$; from the point of view of the $\hat{\varphi}$ variable, the two elements thus look like a single element with \textit{negative} $E_{J2}$. 
With an effective negative $E_{J2}$, the addition of a non-zero $E_{J1} \cos \hat{\varphi}$ term will break the degeneracy as desired. 

Unfortunately, the addition of the second phase-periodic kite also results in an additional noisy gate charge $n_g'$~[Fig.~\ref{fig:fig2}(c)]. 
Luckily, the job of the second kite is only to provide minima at $\pm \pi/2$, which can also be accomplished by a circuit without gate-charge sensitivity: a fluxonium with $E_J \approx \pi E_L/2$  (though as we comment on later, this relationship between $E_J$ and $E_L$ is not at all strict) [Fig.~\ref{fig:fig2}(d)]~\cite{manucharyan2009fluxonium}. 
Making this swap and adding JJ asymmetry to the first kite to introduce the requisite $E_{J1}$ term, we arrive at our final qubit circuit and the desired potential [Fig.~\ref{fig:fig2}(d), lower panel]. 
In the next section, we present a more thorough description of the circuit Fig.~\ref{fig:fig2}(d) using a reduced circuit model. 

\begin{figure}
\includegraphics[width=\columnwidth]{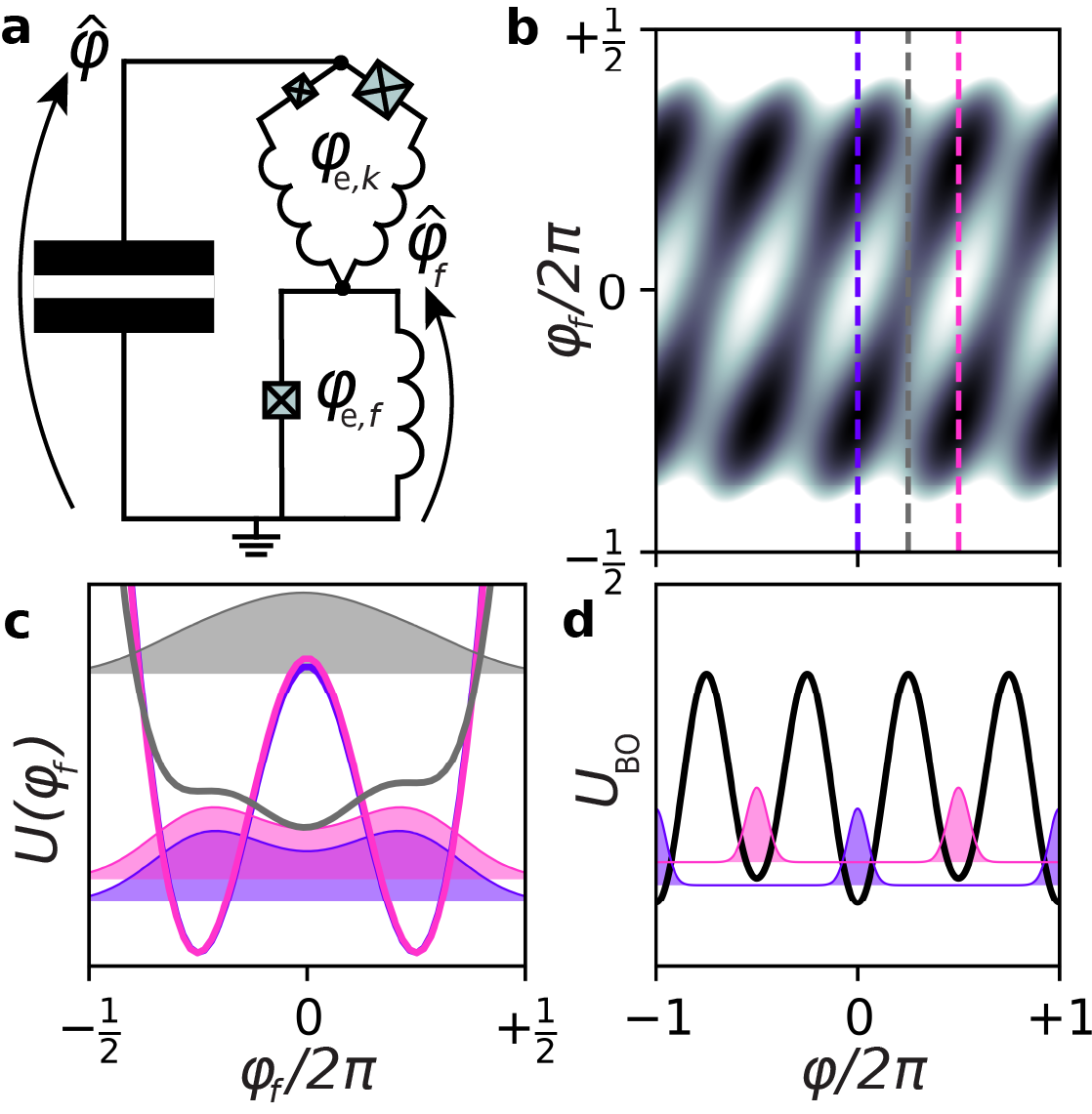}
\caption{\textbf{Born-Oppenheimer Treatment of the Harmonium Circuit} (a) Reduced model of the harmonium circuit in terms of the phase $\hat{\varphi}$ across the entire device and the phase $\hat{\varphi}_f$ across the fluxonium. Unless otherwise specified, we set $\varphi_{\mathrm{e},f} = \varphi_{\mathrm{e},k} = \pi$. (b) 2D potential of the circuit. Darker corresponds to lower potential energy. (c) Three cuts of the 2D potential at $\varphi = 0, \pi/2, \pi$ (dashed lines in panel (b)) and associated ground state wavefunctions of $H_\mathrm{fast}$. The $E_{J1,k}$ of the kite introduces a slight asymmetry in the tunnel barrier between $\varphi = 0, \pi$, breaking the degeneracy. (d) BO potential and associated qubit states of $H_\mathrm{slow}$. 
}. 
\label{fig:fig3}
\end{figure}

\subsection{Born-Oppenheimer treatment of the harmonium circuit}

Our discussion in the previous section focused on the effective potential for $\hat{\varphi}$, completely neglecting the internal nodes (degrees of freedom) of the kite and fluxonium. 
While in section \ref{quantitative} we present a full circuit treatment of all degrees of freedom, to further develop intuition, we will first consider a reduced circuit model that only treats the fluxonium degree of freedom and neglects those internal to the kite [Fig.~\ref{fig:fig3}(a)].
This reduced model thus treats the kite and fluxonium on different levels, when in fact they should be placed on equal footing. 
As such, one should be careful not to draw quantitative conclusions from this model; we present it only for intuition. 
The two imperfect assumptions this model makes are: 

\begin{enumerate}
    \item It ignores the internal degrees of freedom of the kite, instead treating it as an element with potential $E_{J2,k}\cos \hat{\varphi}_k + E_{J1,k}\cos{\hat{\varphi}_k}$. Here $\hat{\varphi}_k$ is the phase drop across the kite, and the $E_{J1,k}$ term is due to the JJ asymmetry~\cite{smith2020superconducting}. 
    \item It assumes that there is no capacitance across the kite, only that there is a small charging energy $E_C$ associated with $\hat{\varphi}$, and a large charging energy $E_{C,f}$ associated with the fluxonium phase $\hat{\varphi}_f$.
\end{enumerate}

\noindent 
With these assumptions and substituting $\hat{\varphi}_k = \hat{\varphi} - \hat{\varphi}_f$, the Hamiltonian of the circuit can be written as 

\begin{align*}
H = & + 4E_C(\hat{n} - n_g)^2 + 4 E_{C,f} \hat{n}_f^2 + \frac{E_L}{2}\hat{\varphi}_f^2 + E_J \cos \hat{\varphi}_f \\& + E_{J2,k} \cos 2(\hat{\varphi} - \hat{\varphi}_f) - E_{J1,k} \cos (\hat{\varphi} - \hat{\varphi}_f) \\
\end{align*}

\noindent where $\hat{n}, \hat{n}_f$ are the Cooper pair number operators conjugate to $\hat{\varphi}, \hat{\varphi}_f$ and we have set the flux through the fluxonium and kite to $\pi$. 
The second line shows how the kite couples the phase $\hat{\varphi}$ to the fluxonium phase $\hat{\varphi}_f$. 
Note that the sign of $E_{J2,k}$ is positive. 
The full potential of this circuit is depicted in Fig.~\ref{fig:fig3}(b). 

We are interested in the effective dynamics of $\hat{\varphi}$. 
To this end, it is helpful to apply a technique originally developed for understanding molecular dynamics known as the Born-Oppenheimer (BO) approximation. 
This method consists of separating the system degrees of freedom into ``fast'' and ``slow'' variables; in the context of molecular dynamics, the electron positions are the fast variables and the nuclei positions are the slow variables. 
In superconducting circuits, the BO approximation can be used to eliminate higher-energy degrees of freedom, allowing one to focus on the effective dynamics of the low-energy degrees of freedom~\cite{divincenzo2006decoherence, rymarz2023consistent, egusquiza2024consistent}. 
This is often useful because one is usually interested in the lowest-energy states of a circuit; for a qubit, usually the lowest two energy states. 

For the circuit Fig.~\ref{fig:fig3}(a), we operate in the regime $E_{C} \ll E_{C,f}$, rendering $\hat{\varphi}_f$ the fast variable and $\hat{\varphi}$ the slow. 
The qualitative idea behind the BO approximation is that, due to this separation of energy scales, as $\hat{\varphi}$ varies, the $\hat{\varphi}_f$ variable can quickly find its ground state. 
This idea is operationalized by defining the fast part of the Hamiltonian, where $\varphi$ has been demoted to a scalar variable: 

\begin{align*}
H_\mathrm{fast}(\varphi) = & +4 E_{C,f} \hat{n}_f^2 + \frac{E_L}{2}\hat{\varphi}_f^2 + E_J \cos \hat{\varphi}_f \\& + E_{J2,k} \cos 2(\varphi - \hat{\varphi}_f) + E_{J1,k} \cos (\varphi - \hat{\varphi}_f)
\end{align*}

\noindent Here, $\varphi$ functions a bit like an external flux on the modified fluxonium potential terms provided by the kite (second line). 
The next step in the BO approximation is to find the ground state energy $\epsilon_{g,\mathrm{fast}}(\varphi)$ of $H_\mathrm{fast}(\varphi)$ as a function of the scalar $\varphi$. 
The $\varphi$-dependent ground state energy then forms an effective potential for the $\hat{\varphi}$ operator $\epsilon_{g,\mathrm{fast}}(\varphi) = U_\mathrm{BO}(\varphi)$, which determines the dynamics of $\hat{\varphi}$ via the effective Hamiltonian 

\begin{equation*}
H_\mathrm{slow} = 4E_C(\hat{n} - n_g)^2 + U_\mathrm{BO}(\hat{\varphi})
\end{equation*}

\noindent where $\hat{\varphi}$ has once again been promoted to an operator. 

We now aim to show that the effective potential $U_\mathrm{BO}(\varphi)$ is qualitatively similar to our desired potential Fig.~\ref{fig:fig1}(b). 
The two important qualities are the effective negative $E_{J2}$ and the effective $E_{J1}$ term introduced by the junction asymmetry of the kite $-E_{J1,k} \cos (\hat{\varphi} - \hat{\varphi}_f)$. 
First, we focus on the effective negative $E_{J2}$. 
Consider three line cuts of the 2D potential at $\varphi = 0, \pi/2, \pi$ [Fig.~\ref{fig:fig3}(b)]; these cuts function as the potential of the fast degree of freedom $\hat{\varphi}_f$ for the corresponding value of $\varphi$ [Fig.~\ref{fig:fig3}(c)]. 
At $\varphi = 0, \pi$, the potential of $\hat{\varphi}_f$ has a double-well form, with ground states de-localized between the two wells (pink, purple curves). 
At $\varphi = \pi/2$, however, the barrier is eliminated, resulting in a ground state of $\hat{\varphi}_f$ with higher energy (gray curves). 
Qualitatively, this is the structure required for an effective negative $E_{J2}$: energy minima at $\epsilon_{g,\mathrm{fast}}(\varphi = 0), \epsilon_{g,\mathrm{fast}}(\varphi = \pi)$ and a maximum at $\epsilon_{g,\mathrm{fast}}(\varphi = \pi/2)$. 

Now we turn to the effective $E_{J1}$ term required to break the degeneracy between the minima of $\varphi$ at $0, \pi$. 
Here, this is achieved via $E_{J1,k}$, which is due to an asymmetry between the JJs of the kite.  
This term results in a slight increase in the height of the tunnel barrier for $\varphi = \pi$ compared to $\varphi = 0$ [Fig.~\ref{fig:fig3}(c)]. 
This weakens the hybridization between the two wells for $\varphi = \pi$, resulting in a higher-energy ground state as compared to the solution of $H_\mathrm{fast}$ for $\varphi = 0$. 
The degeneracy between $\varphi = 0$ and $\varphi = \pi$ is thus broken, the qualitative structure required for non-zero $E_{J1}$. 
In Fig.~\ref{fig:fig3}(d), we plot $U_\mathrm{BO}(\varphi) = \epsilon_{g,\mathrm{fast}}(\varphi)$
to demonstrate that the above arguments for $\varphi = 0, \pi/2, \pi$ generalize to all $\varphi$ to create an effective potential for the slow degree of freedom $\hat{\varphi}$ that has the desired structure [Fig. ~\ref{fig:fig1}(b)]. 
The ground and excited state wavefunctions of $H_\mathrm{slow}$ are localized to their respective wells, rendering the circuit exponentially insensitive to bit-flips due to phase- and charge-coupled noise. 

\subsection{Dephasing protection} \label{dephasing_qualitative}

Our proposed circuit realizes an important condition of qubit states with disjoint support (bit-flip protection). 
However, this circuit is only useful if it can simultaneously suppress pure dephasing noise. 
In this section we discuss dephasing from charge, flux and critical current noise, but we will expand this list in further sections of the paper. 

Charge noise dephasing arises due to voltage fluctuations in the qubit environment that cause the gate charge $n_g$ to diffuse continuously. 
The original triumph of the transmon was to exponentially suppress the qubit sensitivity to the gate charge, so much so that $n_g$ is often forgotten from the transmon Hamiltonian~\cite{koch2007charge}. 
One way to understand this reduced sensitivity is that $n_g$-dependence arises due to tunneling events where $\hat{\varphi}$ winds by $2\pi$; with reduced tunneling, there is therefore reduced charge sensitivity. 
In the transmon, this is achieved by a large ratio of $E_J/E_C$; 
we can use the same trick for harmonium. 
By reducing $E_C$ with respect to the amplitude of $U_\mathrm{BO}$, we exponentially suppress the tunneling between wells of $U_\mathrm{BO}$ and therefore exponentially suppress charge noise. 
Note that this is the same condition required for disjoint support; the qubit states should be well-localized to their respective wells. 

Flux noise dephasing arises due to fluctuating magnetic fields in the qubit environment. 
With proper shielding, global flux noise (i.e. noise common to multiple loops of a circuit) can be mitigated, leaving mainly local flux noise generated by defect spins (though one must be careful to minimize metallization shared between loops)~\cite{gustavsson2011noise, kou2017fluxonium, braumuller2020characterizing}. 
To understand harmonium's insensitivity to independent noisy fluxes through the two loops of the device, we first consider the case where the kite has no junction asymmetry $E_{J1,k} = 0$. 
Deep in the BO regime $E_C \rightarrow 0$, the flux dispersion of the ground and excited state energies of $H_\mathrm{slow}$ are completely determined by $H_\mathrm{fast}(\varphi = 0)$ and $H_\mathrm{fast}(\varphi = \pi)$, respectively. 
In particular, the qubit frequency is determined by $\omega_q = \epsilon_{\mathrm{fast}}(\varphi = \pi) -\epsilon_{\mathrm{fast}}(\varphi = 0)$.
Re-introducing the two external fluxes $\varphi_{\mathrm{e},f}, \varphi_{\mathrm{e},k}$ to the Hamiltonian and computing $H_\mathrm{fast}$ at $\varphi = 0, \pi$, we have~\cite{smith2020superconducting}

\begin{align*}
H_\mathrm{fast}(\varphi = 0, \pi) = & +4 E_{C,f} \hat{n}_f^2 + \frac{E_L}{2}\hat{\varphi}_f^2 \\ &- E_J \cos (\varphi_{\mathrm{e},f} - \hat{\varphi}_f) \\& + E_{J2,k} \cos 2\hat{\varphi}_f  \mp E_{k,\Phi} (\pi - \varphi_{\mathrm{e},k})\sin \hat{\varphi}_f
\end{align*}

\noindent where the energy scale of the $\varphi_{\mathrm{e},k}$-dependent term $E_{k,\Phi}$ is mainly determined by the inductive energy of the kite inductors; note that this is only the leading order term describing the true kite flux dependence, within the same spirit of ignoring the internal degrees of freedom of the kite.

Harmonium's insensitivity to flux noise is due to a symmetry between $H_\mathrm{fast}(\varphi = 0)$ and $H_\mathrm{fast}(\varphi = \pi)$. In this case of $E_{J1,k} = 0$, the only difference between $H_\mathrm{fast}(\varphi = 0)$ and $H_\mathrm{fast}(\varphi = \pi)$ is the sign on the $\varphi_{\mathrm{e},k}$-dependent term.
Consider the dependence of $\omega_q$ on $\varphi_{\mathrm{e},f}$ for the nominal kite flux bias $\varphi_{\mathrm{e},k} = \pi$. 
In this case, the $\varphi_{\mathrm{e},k}$-dependent term is zero, and $H_\mathrm{fast}(\varphi = 0) = H_\mathrm{fast}(\varphi = \pi)$ for any value of $\varphi_{\mathrm{e},f}$. Therefore, $\omega_q = \epsilon_{\mathrm{fast}}(\varphi = \pi) -\epsilon_{\mathrm{fast}}(\varphi = 0) = 0$ for any value of $\varphi_{\mathrm{e},f}$. 
As such, the qubit frequency is insensitive to local fluctuations in $\varphi_{\mathrm{e},f}$ at all orders within this rough model. 
On the other hand, for noise in $\varphi_{\mathrm{e},k}$ with $\varphi_{\mathrm{e},f} = \pi$, the argument is slightly more subtle because the symmetry between $\varphi = 0, \pi$ requires that we also send $\hat{\varphi}_f \rightarrow -\hat{\varphi}_f$ for $H_\mathrm{fast}(\varphi = \pi)$. 
Still, this symmetry guarantees $\omega_q = 0$ for any value of $\varphi_{\mathrm{e},k}$ within this rough model. 
We also note that the fact that $H_\mathrm{fast}(\varphi = 0) = H_\mathrm{fast}(\varphi = \pi)$ at nominal bias ensures that critical current noise of the fluxonium JJ (fluctuation of $E_{Jf}$) does not dephase the qubit. 
Critical current noise of the kite JJs, however, can affect the qubit; indeed, it is the kite JJ asymmetry that we use to break the degeneracy. 

All of the above arguments for immunity to flux noise and fluxonium JJ critical current noise were based on $E_{J1,k} = 0$. 
However, we previously showed that we needed $E_{J1,k} \neq 0$ to realize bit-flip protection via the broken degeneracy. 
These requirements are thus at odds with each other. 
This trade-off could in principle be broken by realizing the effective $E_{J1}$ with a separate JJ shunting both the kite and fluxonium, but this would introduce another flux loop. 
Besides, current fabrication techniques limit the minimum average JJ asymmetry to about 
$1\%$; we show quantitatively in section \ref{quantitative}, this level of asymmetry is sufficient to maintain bit-flip protection. 
As such, for the requisite $E_{J1,k} \neq 0$, harmonium will have a weak second-order flux noise sensitivity; first-order noise is suppressed because the circuit is operated at the double sweet-spot $\varphi_{\mathrm{e},f} = \varphi_{\mathrm{e},k} = \pi$. 

\subsection{Fighting heating events} \label{heating}

While our goal is to encode a qubit into the lowest two energy levels of the circuit, there are of course higher energy levels as well. 
As such, not only can the environment cause transitions between the computational states (bit flips), it can also cause transitions from the computational states to the higher energy states. 
We refer to such transitions as heating events. 
Ideally, the transition frequencies $\omega$ to these higher excited states are large compared to temperature $T$ such that heating events are exponentially suppressed $e^{-\hbar \omega /k_B T}$. 
Suppose we have a circuit described by a Hamiltonian $H$. 
To reduce the rate of heating events by a power $\alpha$ but maintain the circuit matrix element structure and therefore bit-flip protection, we can scale the Hamiltonian $H \rightarrow \alpha H$ such that $\omega \rightarrow \alpha \omega$; note that this only linearly increases dephasing rates.  

Unfortunately, fabrication places a limit on how far this strategy can be pushed. 
To scale the Hamiltonian by $\alpha$, the capacitance matrix for the circuit must be scaled by the inverse $C\rightarrow C/\alpha$. 
While the harmonium circuit does minimize parasitic capacitances to ground compared to e.g. the $0-\pi$ (see next section), the minimal capacitance is fundamentally set by island length scales $C_i \sim \epsilon L$ where $\epsilon$ is the effective permittivity. 
Moreover, for modern fabrication techniques there is a much harder limit in the form of intrinsic JJ capacitance. 
A JJ Josephson energy $E_J$ and capacitance $C_J$ are related via the critical current density $j_c$ and specific capacitance $C_s$ as $\frac{E_J}{C_J} = \frac{\Phi_0}{2\pi}\frac{j_c}{C_s}$. 
While $C_s$ varies somewhat with $j_c$ due to changing oxidation parameters, the dependence is generally fairly weak~\cite{van1994one}. 
As such, to minimize $C_J$ for a given $E_J$ one should use the maximum critical current density possible; put another way, maximizing $j_c$ allows one to fabricate a given $E_J$ with a minimal area JJ thus minimizing the capacitance. 
We note that not only will minimizing the JJ capacitance help with heating, but it will also enforce the energy hierarchy necessary for the BO approximation. 

For most of the calculations we present in the later sections of the paper, we will take a value at the upper limit of the typical range used in Al/AlOx JJs of $j_c = 4 \; \mathrm{\mu A/\mu m}^2$ and take $C_s = 50 \mathrm{fF/\mu m ^2}$ for the specific capacitance~\cite{randeria2024dephasing}. 
However, we note that critical current densities as high as $30 \mathrm{\mu A/\mu m}^2$ have been demonstrated for Al/AlOx JJs~\cite{moskalev2023optimization}. 
Additionally, higher critical current densities could also be achieved using niobium/aluminum/aluminum oxide trilayer JJs~\cite{anferov2024superconducting}.
Of course, higher critical densities are only advantageous if the corresponding JJ size does not shrink below the limit imposed by lithography, and only if the parasitic capacitance associated with the JJ leads does not become dominant. 
Higher $E_J/C$ ratios could also be achieved with non-parallel-plate structures such as semiconductor JJs, though the material quality of these structures is not yet competitive with Al/AlOx processes~\cite{de2015realization, larsen2015semiconductor,casparis2018superconducting}. 

While the strategy of scaling $H\rightarrow \alpha H$ allows one to find a reasonable balance between heating and dephasing events, within an error correction architecture heating events (leakage) can often be much more detrimental than computational space errors. 
Fortunately, if heating events can be robustly detected during error correction, they can be flagged as erasure errors~\cite{stace2009thresholds}. 
As we discuss in section \ref{readout}, such erasure detection can be achieved in harmonium using conventional dispersive readout. 
This erasure-based strategy for fighting heating events thus allows the circuit to be optimized against computational space errors only. 
A third strategy to combat heating events could be cooling of the higher-energy states via bath engineering, though we will not explore this avenue in detail here~\cite{paolo2019control, harrington2022engineered}. 

\subsection{Comparison to the 0-$\pi$ qubit}

The low-energy structure of harmonium is qualitatively similar to the $0-\pi$ qubit; indeed the ground and excited state wavefunctions are localized to $0$ and $\pi$ phase, respectively~\cite{brooks2013protected}. 
However, while it has proven experimentally challenging to push into the protected regime of the $0-\pi$ circuit, we argue that the same is not true for harmonium. 

Perhaps the most significant challenge faced by the $0-\pi$ qubit is the realization of the necessary capacitance hierarchy. 
Similar to harmonium, $0-\pi$ requires that one mode have a very small capacitance (a light, high-energy mode), and another very large (a heavy, low-energy mode). 
However, in the $0-\pi$ circuit this is challenging because the light mode shares nodes with the heavy mode, so in fabricating the large capacitors required by the heavy mode, one also creates parasitic capacitances of the light mode to ground~\cite{gyenis2021experimental}. 
In harmonium, the story is similar: we desire that all degrees of freedom of the circuit besides $\hat{\varphi}$ have minimal capacitance. 
However, unlike $0-\pi$, only $\hat{\varphi}$ participates in the node associated with the large capacitor. 
This makes it possible to realize a suitable capacitance hierarchy in a way that is currently challenging for the fabrication of the $0-\pi$. 
Among other outcomes, this results in a large ``hybridization gap'' for harmonium, i.e. a significant energy gap between the computational subspace and higher-energy states. 

There are several other challenges faced by $0-\pi$ that harmonium evades. 
One is the lack of low-frequency parasitic modes in harmonium. 
In the ideal $0-\pi$ regime one undesired mode of the circuit approaches zero frequency;  in the presence of inevitable circuit disorder, this will result in significant thermal shot-noise dephasing of the $0-\pi$~\cite{paolo2019control}. 
The second is qubit control; while high-fidelity gates for the $0-\pi$ have been difficult to realize~\cite{premkumar2023hamiltonian}, we demonstrate with time-domain simulations in section \ref{single_qubit} that this should not be the case for harmonium. 
 
\section{Q\MakeLowercase{uantitative analysis}} \label{quantitative}

To accurately estimate the performance of harmonium, it is necessary to perform a complete treatment of the circuit. 
Namely, we must include the internal degrees of freedom of the kite, and not simply assume that it is a perfect $\cos2\varphi$ element. 
This significantly complicates the analysis of the circuit, increasing the number of degrees of freedom from two to four; for a discussion of how we perform this analysis we direct the reader to Appendix \ref{complete_hamiltonian}. 
With this complete treatment in hand, we can estimate decoherence rates, circuit element parametrics, gate performance, and readout. 
We present these results in this section, beginning with an analysis of decoherence rates and their dependence on circuit parameters. 

\subsection{Optimizing against decoherence} \label{optimizing}

Decoherence of a qubit can be grouped into three categories: bit flips, dephasing, and out-of-manifold heating.
For superconducting qubits, many of the physical noise channels inducing these decoherence events are well-understood, and therefore the error rates can be estimated. 
In this section we will present the results of these calculations for optimized circuit parameters; details can be found in appendices \ref{state_transitions} and \ref{dephasing}. 

Our goal is to optimize the circuit parameters to minimize the decoherence rates. 
We optimize over the JJ energies of the kite and the fluxonium $E_{Jk}, E_{Jf}$, the superinductances $E_{Lk}, E_{Lf}$, and the large shunt capacitance $C_\mathrm{sh}$ [see appendix Fig.~\ref{fig:fig4} for details]. 
During optimization, the asymmetry between the kite JJs is fixed to $\Delta E_{Jk}/E_{Jk} = 1\%$; we discuss this choice below. 
The JJ capacitances $C_{Jk}, C_{Jf}$ are locked to the JJ energies via $\frac{E_J}{C_J} = \frac{\Phi_0}{2\pi}\frac{j_c}{C_s}$ with $j_c = 4 \mathrm{\mu A/\mu m}^2$ and $C_s = 50 \mathrm{fF/\mu m ^2}$ as motivated in section \ref{heating}. 
Optimization is performed using the LIPO algorithm~\cite{malherbe2017global,pyLIPO}. 
For all presented calculations, we have carefully checked the convergence of the decoherence rates with the discretization of the Hilbert space (see Appendix~\ref{complete_hamiltonian}).

\begin{figure}
\includegraphics[width=\columnwidth]{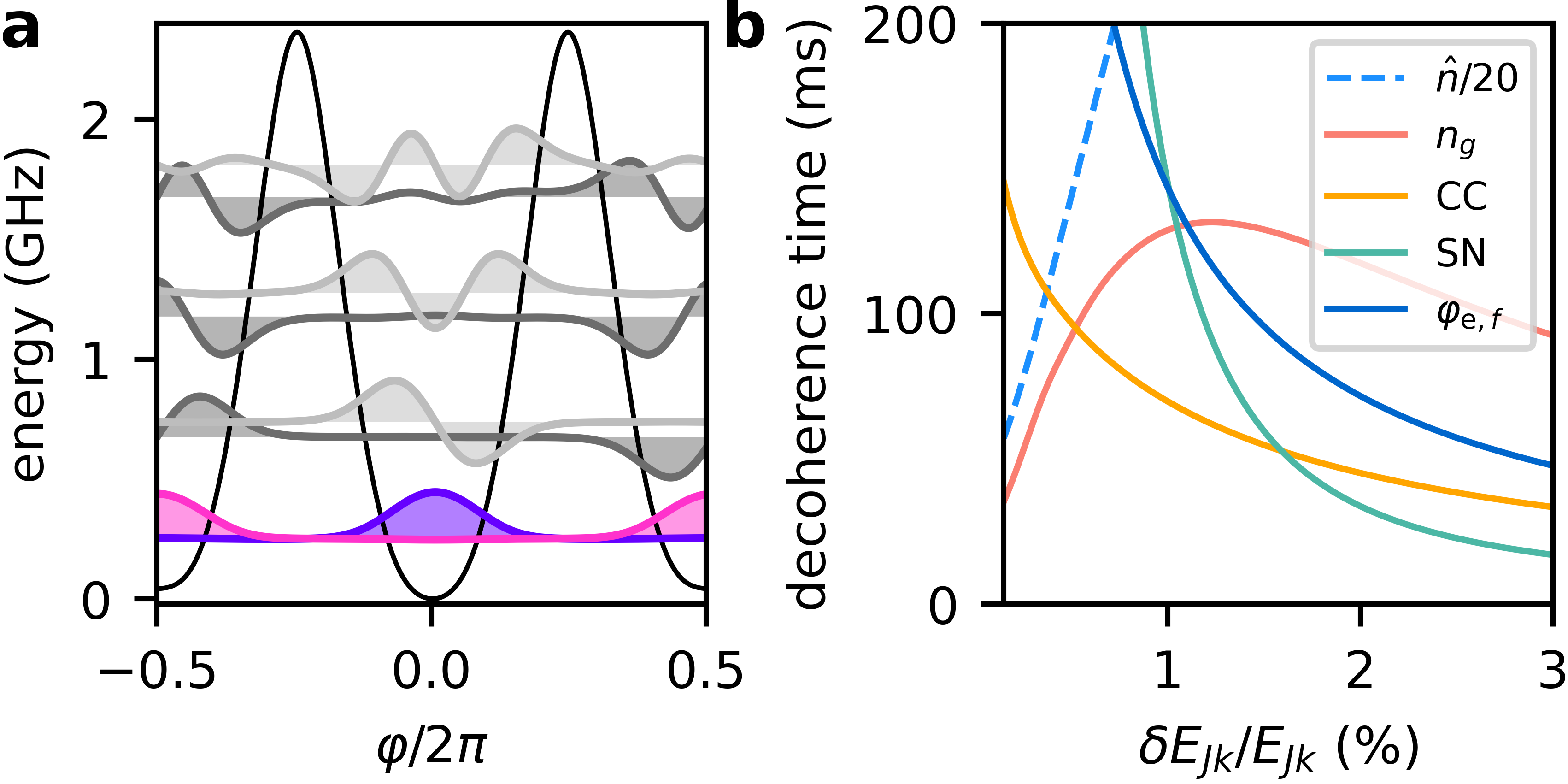}
\caption{\textbf{Results for optimized circuit parameters} (a) BO potential and qubit wavefunctions. 
For visual clarity, the value of $\delta E_{Jk}/E_{Jk}$ has been set to $3\%$ to increase the energy difference between the wells at $\varphi = 0, \pi$. 
(b) Limiting decoherence channels as a function of $\delta E_{Jk}/E_{Jk}$ (dephasing from $\varphi_{\mathrm{e}, k}$ is above 200 milliseconds for the entire range). Dashed blue line is the estimated bit-flip time from charge noise divided by 20, while solid lines indicate dephasing times for gate charge $n_g$, critical current CC, resonator shot noise SN, and fluxonium flux $\varphi_{\mathrm{e},f}$. 
} 
\label{fig:fig_erasure}
\end{figure}

The central objective function that we focus on is the maximum rate among all bit-flip and dephasing channels, but not heating events. 
The idea behind this objective is that, if the qubit is embedded within an erasure-based error correction architecture, heating events can be treated as erasure errors. 
Of course, this also relies on the successful detection of heating events as erasure errors without disturbing the computational subspace; we discuss this in detail in section \ref{readout}. 
Using this objective, we find optimal circuit parameters of $E_{Jk}/h = 35 $ GHz, $E_{Jf}/h = 50 $ GHz, $E_{Lk}/h = 11 $ GHz, $E_{Lf}/h = 14 $ GHz, and $C_\mathrm{sh} = 1500$ fF. 
We present the BO potential of $\hat{\varphi}$ for these optimized parameters in Fig. \ref{fig:fig_erasure}(a) (though we stress that we are not employing the BO approximation to calculate decoherence rates).
The effective potential has the desired characteristics of wells located at $\varphi = 0, \pi$ with the degeneracy very weakly broken by the kite JJ asymmetry $\Delta E_{Jk} /E_{Jk}$. 
This is despite the fact that $E_{Jf} \neq \frac{\pi}{2} E_{Lf}$ as we originally motivated for the fluxonium; we attribute this to the non-separability of the kite and fluxonium degrees of freedom. 

We present the estimated state-transition times and phase-flip times in Tables \ref{table:1} and \ref{table:2}, respectively. 
We expect that bit-flips of the qubit will be dominated by charge noise at the level of a few seconds. 
The qubit dephasing is predicted to be limited either by charge noise or critical current fluctuations, with the optimizer finding a balance between the two at the level of 80 milliseconds. 
However, we note that critical current noise is extremely poorly understood because it does not limit contemporary charge and flux qubits, and as such its noise amplitude has not been accurately measured and can only be bounded based on existing measurements~\cite{van2004decoherence, wang2019cavity}. 
If the critical current noise spectral density is in fact weaker than what we assume here, then the circuit parameters could be re-optimized to more heavily favor charge-noise dephasing protection. 
As both critical current noise and charge noise are expected to possess a $1/f$ spectral density, we note that dynamical decoupling could be used to mitigate their effect~\cite{bylander2011noise}. 
Out-of-manifold heating is predicted to be dominated by charge noise at the level of 2 milliseconds; we discuss how these can be treated as erasure errors in section \ref{readout}. 

An important requirement for a realistic superconducting qubit is insensitivity to variation in circuit parameters due to the imperfect targeting during fabrication.  
Among the circuit parameters, we find that the bit-flip and phase-flip rates are by far the most sensitive to the asymmetry between the kite JJs $\Delta E_{Jk}$ (heating rates are robust within $10\%$) [Fig. \ref{fig:fig_erasure}(b)]. 
As motivated in section \ref{dephasing_qualitative}, dephasing rates due to flux noise and critical current increase with asymmetry, while charge-noise dephasing and bit-flip rates decrease. 
In practice, modern fabrication techniques result in relative asymmetries  with a spread of a couple percent for nominally identical JJs. 
Here, we assume a normal distribution for $\Delta E_{Jk}/E_{Jk}$ with a standard deviation of $1\%$; the decoherence estimates in Tables \ref{table:1} and \ref{table:2} are averaged over this distribution. 
As the qubit frequency varies linearly from 200 kHz to 1.2 MHz for the range of asymmetries plotted in Fig. \ref{fig:fig_erasure}(b), it will also vary stochastically. 
While one might think that this lack of qubit frequency control could lead to unwanted frequency collisions between qubits in a larger device, we argue that this is not an issue because charge coupling matrix elements between adjacent qubits are on the order of $|\langle e| \hat{n} |g\rangle | \approx 2 \times 10^{-3}$. 
We note that a gate-tunable JJ could be used to control this asymmetry, but the efficacy of this strategy will depend on how much the noise amplitudes are increased by the addition of a more complex JJ, in particular charge and critical current noise~\cite{larsen2020parity, schrade2022protected}. 

While for the remainder of the paper we will focus on the parameter set discussed in the previous section, for some applications one may not want to rely on erasure detection.
As such, we briefly discuss the strategy presented in section \ref{heating} of scaling $H \rightarrow \alpha H$. 
For $\alpha \approx 3$, we find that the charge-induced heating times climbs from approximately $2$ to $20$ milliseconds, while the dephasing rates from charge noise and critical current drop to the same range. 
However, as previously discussed, for a physical device this would require increasing the junction $E_J/C_J$ ratio by the same factor of $\alpha \approx 3$.
While the corresponding critical current density $j_c \approx 12 \; \mu\mathrm{A}/\mu \mathrm{m}^2$ has been previously demonstrated~\cite{moskalev2023optimization}, this will also require small JJ areas on the order of 6000 nm$^2$ which may introduce yield issues and increased asymmetry. 
As a final comment, we note that while scaling $\alpha$ increases the dephasing rates, it also theoretically allows one to increase the speed of the gates discussed later in section \ref{control} by the same factor $\alpha$ while maintaining the same level of coherent error at the Hamiltonian level. 

\begin{table}[h!]
\begin{center}
\begin{tabular}{||c c c||} 
\hline
\multicolumn{3}{|c|}{ \rule{0pt}{3ex} \textbf{Estimated State Transition Times}} \\ [0.5ex]
 \hline
 \rule{0pt}{3ex} \textbf{Channel} & \textbf{Bit-Flip Time (s)} & \textbf{Heating Time (ms)} \\ [0.5ex] 
 \hline
 charge & 3 & 2  \\ 
 \hline
 flux & $\gg$ 1000 & 1000  \\
 \hline
 quasiparticles & 200 & 5000  \\
 \hline
\end{tabular}
\end{center}
\caption{Values are averaged over the distribution of $\Delta E_{Jk}/E_{Jk}$ (see text). Quasiparticles rates include contributions from the JJs and the superinductors. Heating times are averaged between the ground and excited states.}
\label{table:1}
\end{table}

\begin{table}[h!]
\begin{center}
\begin{tabular}{||c c||} 
\hline
\multicolumn{2}{|c|}{ \rule{0pt}{3ex} \textbf{Estimated Dephasing Times}} \\ [0.5ex]
 \hline
 \rule{0pt}{3ex} \textbf{Channel} & \textbf{Time} (ms) \\ [0.5ex] 
 \hline
 charge & 80  \\ 
 \hline
 flux & 170  \\
 \hline
 critical current & 80  \\
 \hline
   thermal resonator shot & 150 \\
 \hline
 Aharonov-Casher & 1000 \\
 \hline
\end{tabular}
\end{center}
\caption{Values are averaged over the distribution of $\Delta E_{Jk}/E_{Jk}$ (see text). These values can be interpreted as rough estimates of the echo times (see Appendix \ref{dephasing} for details).}
\label{table:2}
\end{table}

\subsection{C\MakeLowercase{ontrol and readout}} \label{control}

A challenge faced by noise-resilient qubits is that, in decoupling from their environment, they also decouple from control and readout apparatus. 
As such, it is important to consider how this problem can be circumvented or alleviated in harmonium. 
Here we present proof-of-principle approaches to single-qubit gates, two-qubit gates, and readout. 
However, we note that, by considering the history of charge and flux qubits, it is clear that perfecting gates and readout will require a much larger effort than what we present here. 

\subsubsection{Single-qubit gates} \label{single_qubit}

\begin{figure}
\includegraphics[width=\columnwidth]{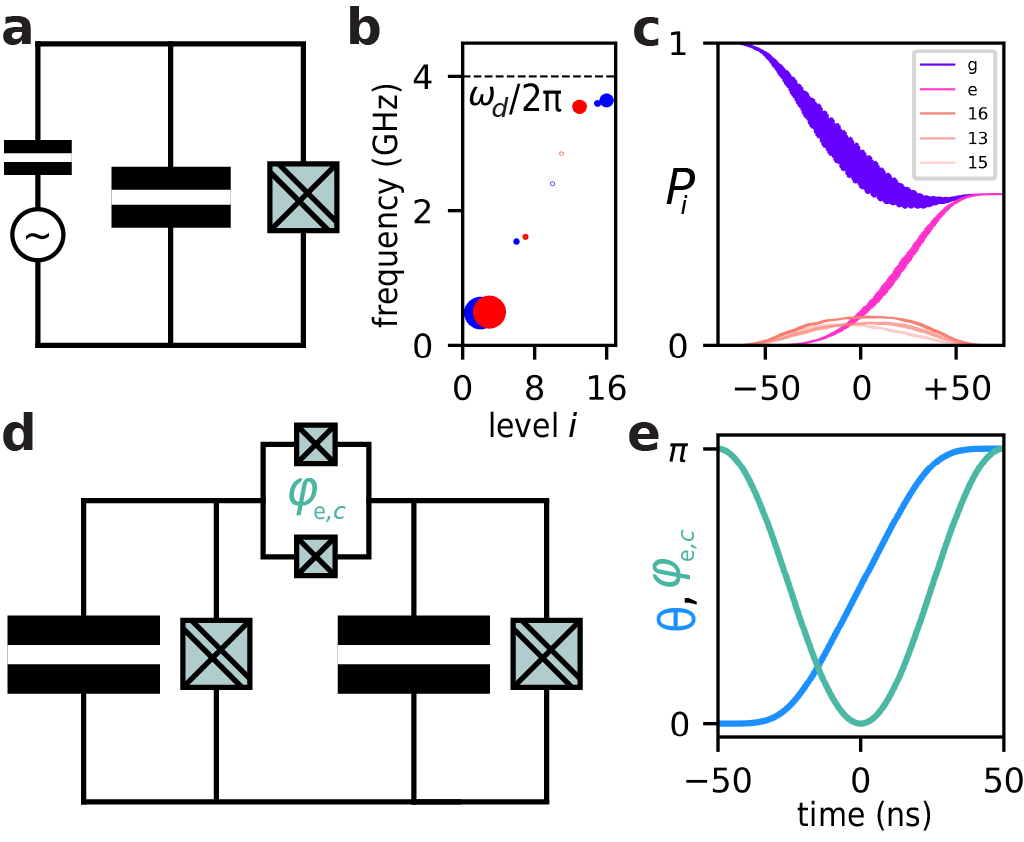}
\caption{\textbf{Qubit control} (a) A capacitively-coupled RF voltage drives transitions of the harmonium circuit. The series kite and fluxonium are indicated by the cross-hatched symbol. (b) Transition frequency from the $|g\rangle$ to $|i\rangle$, with marker determined by the Raman amplitude $\langle e |\hat{n}|i\rangle\langle i |\hat{n}|g\rangle$, which we have chosen to be real. Size is proportional to the magnitude of the Raman amplitude and positive/negative is indicated by red/blue. Presented simulations are with $\omega_d \approx 4$ GHz (dashed line) (c) Simulated action of a single-drive Raman $\pi/2$ pulse with a cosine envelope at $\omega_d \approx$ 4 GHz and $n_g = 0$. Levels 13, 15, and 16 have non-negligible population during the gate, but return to $P_i\approx 0$ by the end of the gate. (d) Two harmonium qubits coupled by a symmetric squid coupler. (e) Pulsing $\Phi_c$ away from $\pi$ with a cosine envelope causes a conditional phase $\theta$ between the qubits. For $\theta = \pi$, the pulse implements a CZ gate. 
} 
\label{fig:fig_gates}
\end{figure}

Of the critical control primitives, single-qubit gates are the most challenging for harmonium. 
Universal single-qubit control requires coherent population swapping between $|g\rangle$ and $|e\rangle$, and as demonstrated by the bit-flip rates presented in Table~\ref{table:1}, the matrix elements of all operators between the qubit states are much smaller than in charge or flux qubits. 
To circumvent this, we propose to use a single-tone Raman process through higher excited states of the circuit. 
Note that we are not employing a two-tone Raman process as is more typical for $\Lambda$ system control, which would require calibration of two separate drives. 

The single-tone Raman process can be understood as follows. 
Consider a higher excited state of the circuit $|i\rangle$. 
A single drive detuned by $\Delta$ from the $|g\rangle \leftrightarrow |i\rangle$ transition will activate a Raman process through the Hamiltonian terms

\begin{equation*}
    \omega_{eg} |e\rangle\langle e| + \Delta |i\rangle\langle i| + \frac{\Omega_{ig}}{2}|i\rangle\langle g| + \frac{\Omega_{ie}}{2}|i\rangle\langle e| + h.c.
\end{equation*}

\noindent where we have made a rotating-wave approximation and assume $\Omega_{ig}, \Omega_{ie}$ are real and positive without loss of generality.
Adiabatic elimination of $|i\rangle$ then yields the effective Hamiltonian in the computational subspace $\frac{\delta_\mathrm{eff}}{2}\sigma_z + \frac{\Omega_\mathrm{eff}}{2}\sigma_x$ where $\Omega_\mathrm{eff} = -\frac{\Omega_{ig} \Omega_{ie}}{2\Delta}$ and $\delta_\mathrm{eff} = \omega_{ge} + \frac{\Omega_{ig}^2}{4\Delta} - \frac{\Omega_{ie}^2}{4\Delta}$. 
Importantly, if the coupling of the drive field to the branches of the $\Lambda$ system is approximately (but not exactly) equal $\Omega_{ig} \approx \Omega_{ie}$, then for small qubit frequency the Stark shift terms can be used to reach $\delta_\mathrm{eff} = 0$ and therefore resonant Rabi oscillations. 

For a circuit with a rich level structure such as harmonium, a strong microwave tone will in general drive many such Raman processes in parallel. 
These parallel processes can destructively interfere with each other, cause unwanted stark shifts, and, in the worst case, result in leakage to non-computational states. 
Fortunately, so long as $\Omega_\mathrm{eff} \gtrsim \delta_\mathrm{eff}$, stark shift effects can be regularized for $\pi/2$ pulses~\cite{rower2024suppressing}. 
As such, here we will focus on the analysis of $\pi/2$ pulses in harmonium. 
Specifically, we imagine driving single-tone Raman gates via a capacitively-coupled drive [Fig.~\ref{fig:fig_gates}(a)]; the drive dominantly couples to the $\hat{n}$ operator. 
The dominant matrix elements are to the second and third excited states of the system [Fig.~\ref{fig:fig_gates}(b)]. 
While each state has strong coupling to both $|g\rangle$ and $|e\rangle$ as desired, unfortunately the effective Rabi rates have opposite sign and therefore destructively interfere~\cite{}. 
Luckily, the computational states also have reasonable overlap with the excited states associated with the internal nodes of the circuit, which have energy just below $4$ GHz. 
We thus find that with a 4 GHz tone, we can drive $\pi/2$  pulses in approximately $150$ nanoseconds [Fig.~\ref{fig:fig_gates}(c)]. 

Although Fig.~\ref{fig:fig_gates}(c) demonstrates that single-qubit gates are in principle possible, there are two significant outstanding issues that we will not fully address here. 
The first is that, during the gate, there is significant population in the higher excited states of the circuit. 
Although this population coherently returns to the qubit manifold by the end of the gate (leakage $< 10^{-7}$), the lower coherence times of the higher excited states will result in incoherent error during the gate. 
Within our noise model, the lifetime of these states is approximately 20 $\mu$s, resulting in a decay probability during the gate of $5\times10^{-4}$. 
However, as the dominant decay channel of these higher excited states is not directly back to the computational states but rather to other excited states of the system, these incoherent errors could potentially be flagged as erasure errors; we leave a detailed exploration of this to future work. 

The second and perhaps larger issue is that, although the qubit sensitivity to gate charge dephasing is small, the matrix elements (and therefore the Rabi rate) are still gate-charge dependent. 
This can be understood as arising from a gate-charge-mediated parity symmetry of the Hamiltonian. 
Consider, as an example, the $|g\rangle, |e\rangle$ wavefunctions of the BO potential [Fig.~\ref{fig:fig_erasure}(a)]; these can be taken as the wavefunctions at $n_g = 0$. 
Because both wavefunctions have even parity (they are even under $\varphi \rightarrow - \varphi$), the charge matrix element is zero $\langle e|\hat{n}|g\rangle = 0$. 
For non-zero gate charge, this parity symmetry is modified. 
To see this, one can move the gate-charge dependence from the Hamiltonian to the wavefunctions via the gauge transformation $H \rightarrow e^{+i\hat{\varphi} n_g} H e^{-i\hat{\varphi} n_g}$, $\psi(\varphi) \rightarrow e^{+i\varphi n_g} \psi(\varphi)$; the wavefunctions are thus no longer invariant under $\varphi \rightarrow - \varphi$ and therefore $\langle e|\hat{n}|g\rangle \neq 0$. 
We note that a similar effect can be a found in the $0-\pi$ qubit~\cite{gyenis2021experimental}. 
As such, the simulations we present in Fig.~\ref{fig:fig_gates}(c) are with $n_g = 0$. 
But because in reality $n_g$ will drift on the timescale of seconds, our Raman-based approach to single-qubit gates will require frequent recalibration. 
We leave a full analysis of this issue and its significance for extensibility to future work. 

Finally, we note that non-adiabatic control could be achieved by using fast flux pulses to bring the qubit states into resonance, but an appreciable avoided crossing would likely necessitate reducing the qubit protection~\cite{campbell2020universal,zhang2021universal}.  
While here we have optimized the qubit circuit against decoherence only, a more complete treatment would optimize the circuit parameters against gate fidelity, and would preferably be informed by experimental measurements of harmonium qubits. 

\subsubsection{Two-qubit gates}

Two-qubit gates are less challenging than single-qubit gates, primarily because we are free to choose a controlled-Z gate as our entangling gate; this does not require population transfer between qubit states, but only relative phases. 
As a proof-of-principle example, we borrow a proposal for two-qubit gates between $0-\pi$ qubits [Fig.~\ref{fig:fig_gates}(d)]~\cite{brooks2013protected}. 
Due to the similarities in the qubit wavefunctions of the two qubits, namely that $|g\rangle$ is localized around $\varphi = 0$ and $|e\rangle$ around $\varphi = \pi$, the gate architecture also applies to harmonium.

This gate can be understood as follows. 
A symmetric squid with JJ energies $E_{Jc}$ and flux $\varphi_{\mathrm{e},c}$ couples the two qubits [Fig.~\ref{fig:fig_gates}(d)].
Labeling the phases of the two qubits as $\hat{\varphi}_1, \hat{\varphi}_2$ and using the fact that the qubit states are localized around $0,\pi$, we can write the squid potential as 

\begin{align*}
U_c & =   -E_{Jc} \cos (\varphi_{\mathrm{e},c} - (\hat{\varphi}_1 - \hat{\varphi}_2)) -E_{Jc} \cos (\hat{\varphi}_1 - \hat{\varphi}_2) \\
& \approx    -E_{Jc} (1 + \cos \varphi_{\mathrm{e},c}) \cos \hat{\varphi}_1 \cos \hat{\varphi}_2 \\
& \approx    -E_{Jc} (1 + \cos \varphi_{\mathrm{e},c}) \sigma_{z1} \sigma_{z2} \\
\end{align*}

\noindent where $\sigma_{z,i}$ is the $Z$ Pauli matrix. 
The last line shows that we can think of $U_c$ as a tunable $ZZ$ interaction between the qubits, exactly what we need to create a CZ gate.
When $\varphi_{\mathrm{e},c} = \pi$, $U_c = 0$ and the interaction is off. 
Tuning away from full frustration turns the interaction on, and causes the qubits to acquire conditional phases. 

We simulate the action of such a coupler in Fig.~\ref{fig:fig_gates}(e) under a cosine-shaped flux pulse keeping the full potential of the coupler (though we neglect any irrotational terms that may come from the coupler JJ capacitors~\cite{you2019circuit}). 
Averaging over the two-qubit Hilbert space, we compute a infidelity of $4 \times 10^{-7}$ without decoherence, primarily due to leakage. 
Since we expect the qubit to be limited by $1/f$ noise channels (charge and critical current), accurate simulation of infidelity due to decoherence would require a careful treatment beyond the standard Lindblad approach~\cite{groszkowski2023simple}. 
Still, we can estimate the incoherent infidelity as $\frac{4}{5}\frac{t_g}{T_\phi} = 8 \times 10^{-6}$ assuming $T_\phi = $ 10 milliseconds~\cite{pedersen2007fidelity}. 

While these estimated infidelities are several orders of magnitude below the state of the art, there are two potential problems with this gate architecture. 
The first is that, in practice, the squid coupler will have on average a roughly 1\% asymmetry between the two JJs, leading to an always-on entangling interaction between the two qubits even when $\Phi_c = \pi$. 
However, this could likely be dealt with by tailoring pulse sequences to absorb this interaction into desired entangling gates~\cite{goldschmidt2025crosstalk}. 
Perhaps the bigger issue is that, if tiled in a 2D array, this gate architecture would lead to large flux loops between qubits. 
As such, a coupler that does not introduce additional closed superconducting loops would be more desirable. 

\subsubsection{Readout}\label{readout}

\begin{figure}
\includegraphics[width=\columnwidth]{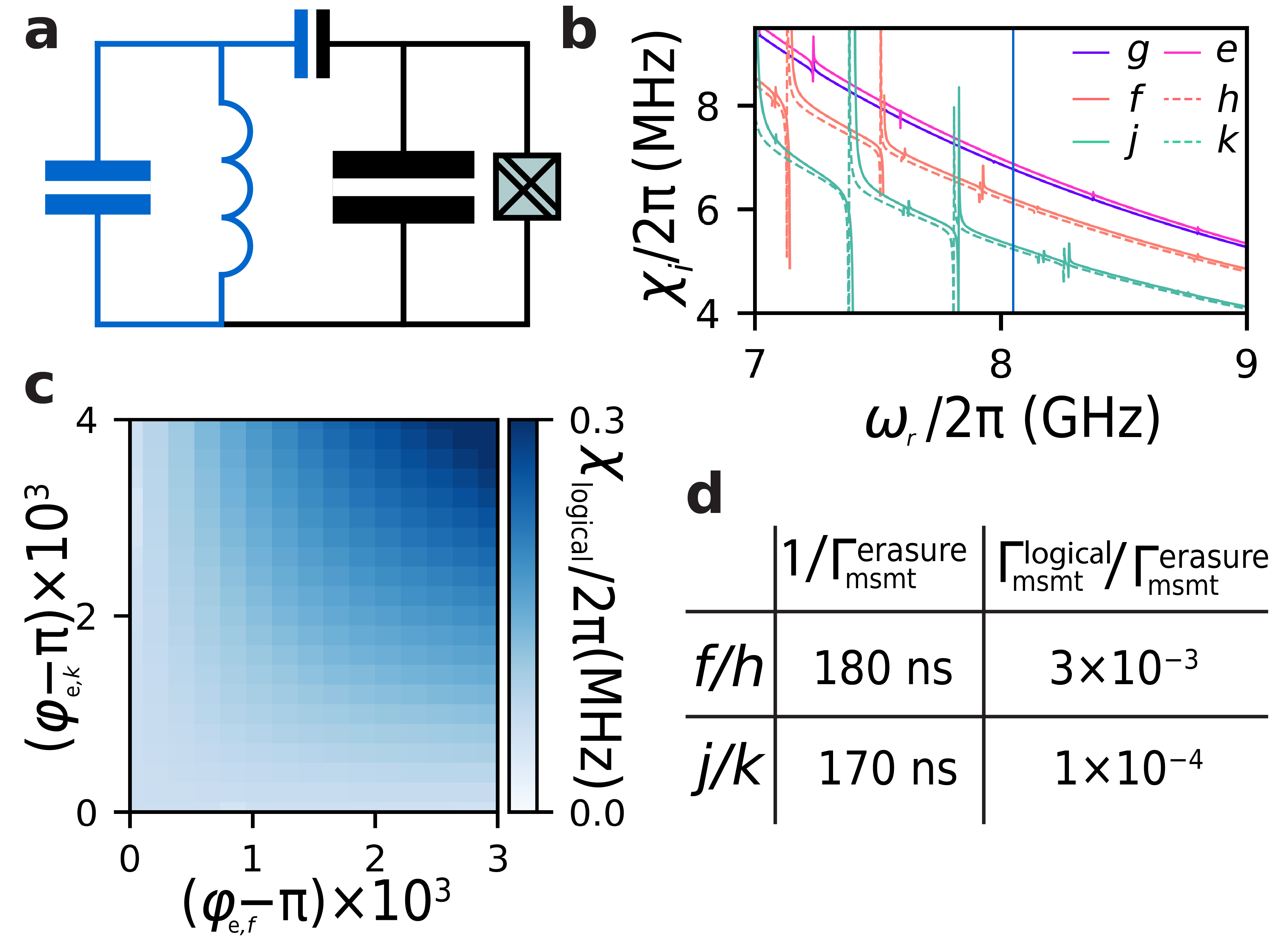}
\caption{\textbf{Logical and erasure readout.} (a) A resonator with frequency $\omega_r$ is capacitively coupled to a harmonium qubit. (b) The dispersive shift $\chi_i$ of the first six levels for $g_r = 2\pi \times 300$ MHz as a function of $\omega_r$. (c) The computational dispersive shift $\chi_{ge} = \chi_e - \chi_g$ as a function of the kite/fluxonium fluxes. (d) Estimated erasure measurement timescales for $\kappa = 2\pi \times 0.5$ MHz and $\bar{n}_\mathrm{erasure} = 5$. 
} 
\label{fig:readout}
\end{figure}

The most common form of qubit measurement for contemporary superconducting circuits is known as dispersive readout~\cite{blais2021circuit, gambetta2006qubit}. 
In dispersive readout, the qubit circuit is perturbatively coupled to a superconducting resonator such that the resonator frequency is shifted by the $\chi_i$ when the circuit is in state $|i\rangle$. 
The dispersive shift $\chi_i$ can then be detected by probing the resonator with a microwave tone. 
Here we will discuss how dispersive readout could be used in harmonium both for logical readout and for erasure detection. 

First, we review the physics of the dispersive shifts and apply them to harmonium. 
Here we will consider a readout resonator (bare frequency $\omega_r$) with a capacitive coupling to the qubit circuit such that the coupling is approximately given by $ g_r ( \hat{a}+ \hat{a}^\dag) \hat{n}$ where $\hat{a}$ is the annihilation operator of the resonator [Fig.~\ref{fig:readout}(a)]. 
At second order in $g_r$, we have~\cite{manucharyanThesis}
 
\begin{equation*}
\chi_i = g_r^2 \sum_{j \neq i} |n_{ij}|^2\frac{2\omega_{ij}}{\omega_{ij}^2 - \omega_r^2}
\end{equation*}

\noindent where $\omega_{ij} = \epsilon_i - \epsilon_j$ and $n_{ij} = \langle j| \hat{n} | i\rangle$. 
This formula can be understood as a level repulsion effect: the transition $|i \rangle \leftrightarrow |j\rangle$ ``pushes'' on the resonator frequency with a strength determined by $n_{ij}$ as well as the proximity of $\omega_{ij}$ to $\omega_r$. 
In harmonium, the approximate symmetry between the wells of the $\hat{\varphi}$ potential  at $0,\pi$ leads to a pairing up of the dispersive shifts $\chi_i \approx \chi_{i+1}$ similar to the pairing up energy levels seen in Fig.~\ref{fig:fig_erasure}(a). 
In what follows, the dispersive shifts of the first six levels of the circuit will be relevant; we plot these as a function of $\omega_r$ in Fig.~\ref{fig:readout}(b). 

We now turn to dispersive readout of the logical states $|g\rangle$, $|e\rangle$. 
Optimal readout requires $\chi_\mathrm{logical} = (\chi_e - \chi_g)/2 \sim \kappa$, where $\kappa$ is the single-photon loss rate of the resonator to the readout line. 
However, a fundamental problem with dispersive readout is that unwanted thermal photons $\bar{n}_\mathrm{th}$ in the readout resonator can also ``measure'' the qubit state. 
This dephasing channel often limits the pure dephasing time of modern charge and flux qubits. 
To mitigate thermal shot-noise dephasing, one strategy is to engineer a tunable $\chi_\mathrm{logical}$ such that it is only comparable to $\kappa$ during qubit readout. 
In harmonium, this can be achieved using flux-assisted dispersive readout~[Fig.~\ref{fig:readout}(c)]. 
At nominal bias $\varphi_{\mathrm{e},k} = \varphi_{\mathrm{e},f} = \pi$, we have $\chi_e \approx \chi_g \rightarrow \chi_\mathrm{logical} \approx 0$, resulting in natural protection from thermal shot-noise dephasing. 
However, $\chi_\mathrm{logical}$ can be increased by tuning $\varphi_{\mathrm{e},k}, \varphi_{\mathrm{e},f} \neq \pi$ to break the symmetry between the wells $\varphi = 0,\pi$. 
As such, we propose to use fast-flux pulses to bias away from the $\varphi_{\mathrm{e},k}, \varphi_{\mathrm{e},f} = \pi$ during readout. 
Note that this does not affect the bit-flip protection of the qubit.
Another strategy~\cite{zhang2021universal} besides fast-flux readout could be to apply a selective pulse from $|g\rangle \rightarrow |f\rangle$. 
The relevant $\chi$ would then be $\chi_f - \chi_e$, which is much larger than $\chi_e - \chi_g$. 

As previously discussed, our primary proposed strategy to combat heating events is to convert them to erasure errors. 
We define the erasure dispersive shift $\chi_\mathrm{erasure} = (\chi_f - \chi_g)/2 \approx (\chi_h - \chi_g)/2 \approx (\chi_f - \chi_e)/2 \approx (\chi_h - \chi_e)/2$. 
At nominal bias $\varphi_{\mathrm{e},k} =\varphi_{\mathrm{e},f} = \pi$ and perfect readout efficiency, a readout tone on resonance with the erasure states $\omega_r + \chi_f \approx \omega_r + \chi_h$ can detect errors at a rate $\Gamma_\mathrm{msmt}^{\mathrm{erasure}} \approx \frac{1}{2}\bar{n}_\mathrm{erasure} \kappa$. 
However, this readout tone will also result in unwanted dephasing of the logical information at a rate $\Gamma_\mathrm{msmt}^{\mathrm{logical}}$. 
For $\chi_\mathrm{erasure} \gg \kappa$, the probability of an erasure-check-induced error is given by $p_\mathrm{err} \approx \frac{\Gamma_\mathrm{msmt}^{\mathrm{logical}}}{\Gamma_\mathrm{msmt}^{\mathrm{erasure}}} \approx \frac{\kappa^2 \chi_\mathrm{logical}^2}{32 \chi_\mathrm{erasure}^4}$. 
The error rate can thus be reduced by reducing $\kappa$ compared to $\chi$, but this also reduces the overall speed of measurement and increases the thermal shot-noise dephasing rate on the logical manifold $\Gamma^\mathrm{th}_\phi \approx 4 \bar{n}_\mathrm{th} \chi_\mathrm{logical}^2/\kappa$. 

There is thus a balance between thermal shot-noise dephasing, erasure-induced error probability, and overall measurement rate. 
With this in mind, we present estimates of $\Gamma^\mathrm{erasure}_\mathrm{msmt}$ and $p_\mathrm{err}$ for experimentally accessible values of $\omega_r, g_r$ and $\kappa$ and with $\Delta E_{Jk}/E_{Jk}$ [Fig.~\ref{readout}(d)]. These trade-offs could be partially mitigated by applying a pre-readout pulse to move population from $f,h\rightarrow j,k$. 
This more than doubles the effective $\chi_\mathrm{erasure}$, reducing $p_\mathrm{err}$ by an order of magnitude. 
For smaller values of $\Delta E_{Jk}/E_{Jk}$, the value of $\chi_\mathrm{logical}$ drops linearly while $\chi_\mathrm{erasure}$ is relatively stable; this is the reason for the dependence of the shot-noise dephasing in Fig.~\ref{fig:fig_erasure}(b). 
Smaller values of asymmetry will also thus result in a strong suppression of $p_\mathrm{err} \propto \chi_\mathrm{logical}^2$. 

\section{C\MakeLowercase{onclusion}}

We have presented a design for a noise-resilient qubit based on engineering of the Josephson harmonics $E_{J1} \cos \varphi + E_{J2}\cos \varphi$. 
While the low-energy structure of the qubit is similar to the $0-\pi$, we have shown that, unlike the $0-\pi$, the noise-resilient regime of the qubit is within reach of modern fabrication techniques. 
We have introduced methods for qubit control with predicted gate times on the order of 100 nanoseconds.  
Dispersive readout could be implemented without introducing significant shot-noise dephasing, and could be used to convert out-of-manifold heating events into erasure errors. 

Although the harmonium circuit is certainly complex in terms of number of degrees of freedom, as long as these degrees of freedom are treated accurately as we have done here, this does not pose a problem. 
Perhaps a more practically-relevant metric of circuit complexity is the number of flux loops in the circuit; in general, each additional loop will require an additional on-chip bias line. 
While not strictly proven, it is likely that there does not exist a single-loop circuit that is superior to charge/flux qubits. 
Indeed, one recent work found that according to a certain metric, fluxonium is the optimum single-loop qubit~\cite{rajabzadeh2024general}. 
Compared to candidate two-loop next-generation qubits such as the dual-rail transmon~\cite{campbell2020universal, levine2024demonstrating} or fluxonium molecule~\cite{kou2017fluxonium,thibodeau2024floquet,kumar2024protomon}, harmonium theoretically possesses superior coherence times and erasure rates. 

Ultimately, the noise models we have used here must be validated experimentally; even among charge and flux qubits, the explanatory power of these models is far from perfect. 
For harmonium in particular, the limitation imposed by critical current noise is an open question. 
Informed by experiments, the optimization we have presented here could be refined to boost device coherence or a higher-level metric such as gate fidelity or even logical error rate within an erasure-based error correction code.

\textbf{Acknowledgments} We are grateful to Jamie Kerman, Jahn Belter, Terry Orlando, Aranya Goswami, Kyle Serniak, and Jeff Grover for helpful conversations, and Willem Wyndham for providing access to compute used for the numerical diagonalizations. 
This research was funded in part by the U.S. Army Research Office under Award No. W911NF-23-1-0045 and in part by the U.S. Department of Energy, Office of Science, National Quantum Information Science Research Centers, Co-design Center for Quantum Advantage (C2QA) under Contract No. DE-SC0012704. 
MH is supported by an appointment to the Intelligence Community Postdoctoral Research Fellowship Program at the Massachusetts Institute of Technology administered by Oak Ridge Institute for Science and Education (ORISE) through an interagency agreement between the U.S. Department of Energy and the Office of the Director of National Intelligence (ODNI).
J.K. gratefully acknowledges support from the Korea Foundation for Advanced Studies.

\appendix 

\section{Hamiltonian construction and diagonalization} \label{complete_hamiltonian}

\subsection{Hamiltonian construction}

\begin{figure}
\includegraphics[width=\columnwidth]{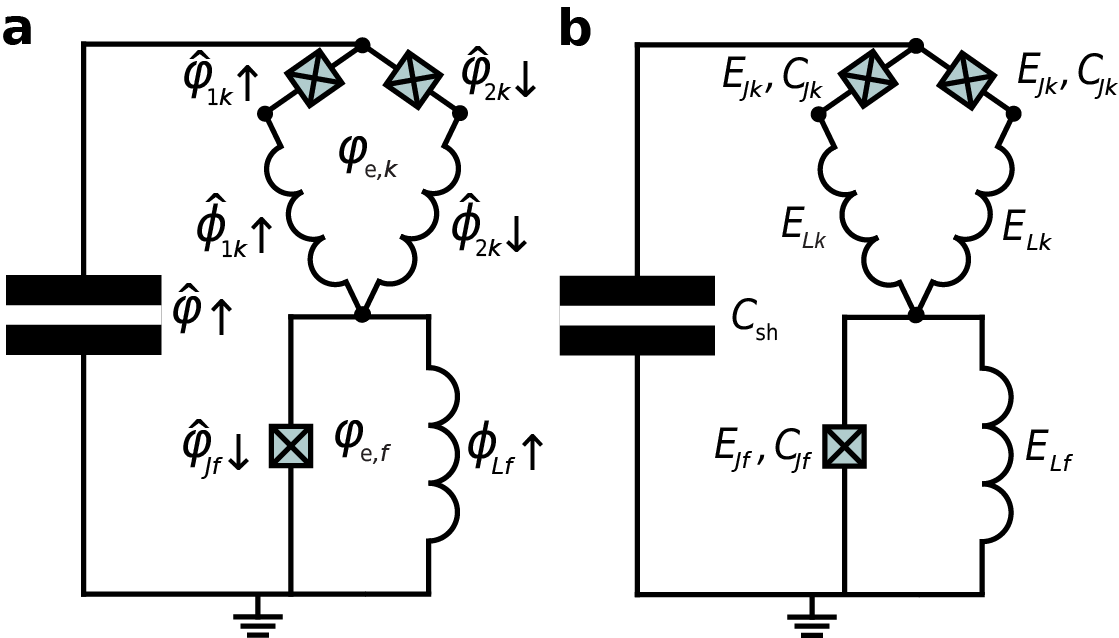}
\caption{\textbf{Complete circuit diagram.} (a) Definition of branch phases and external fluxes. Arrows indicate the direction of each branch phase. (b) Definition of inductive energies and capacitances. 
}
\label{fig:fig4}
\end{figure}

We begin by depicting the full circuit diagram in Fig.~\ref{fig:fig4}. 
There are four nodes besides the ground node; there are thus four degrees of freedom. 
We choose the node phase $\varphi$ at the top of the circuit as one of these phases, and label the remaining phases as $\varphi_f$, $\varphi_{\Sigma k}$, $\varphi_{\Delta k}$ (these are linear combinations of the remaining node phases and fluxes); we will refer to these four degrees of freedom as the ``dynamical'' phases. 
In terms of these dynamical phases and the external fluxes, the six branch phases across the inductive elements are

\begin{align*}
\phi_{Lf} & = +\varphi_f + \delta \varphi_{\mathrm{e},f} \\
\varphi_{Jf} & = -\hat{\varphi}_f + \varphi_{\mathrm{e},f} \\
\phi_{1k} & = +\varphi_{\Sigma k} +\varphi_{\Delta k}\\
\phi_{2k} & = +\varphi_{\Sigma k} -\varphi_{\Delta k} - \delta \varphi_{\mathrm{e},k}\\
\varphi_{1k} & = +\varphi -\hat{\varphi}_f-\varphi_{\Sigma k}-\varphi_{\Delta k} + \varphi_{\mathrm{e},f}\\
\varphi_{2k} & = -\varphi +\hat{\varphi}_f-\varphi_{\Sigma k}+\varphi_{\Delta k} - \varphi_{\mathrm{e},f}- \varphi_{\mathrm{e},k}\\
\end{align*}

\noindent where we have  split the two external fluxes into a static and dynamical part $\varphi_{\mathrm{e},i} \rightarrow \varphi_{\mathrm{e},i} + \delta \varphi_{\mathrm{e},i}$. For some further calculations, it is convenient to define the vector of branch phases $\overrightarrow{\varphi}_B^T = (\phi_{Lf}, \varphi_{Jf}, \phi_{1k}, \phi_{2k}, \varphi_{1k}, \varphi_{2k}, \varphi)$ and the vector of dynamical phases $\overrightarrow{\varphi}_D^T = (\varphi, \hat{\varphi}_f, \varphi_{\Sigma k}, \varphi_{\Delta k})$.  With $\varphi_{\mathrm{e},i}, \delta\varphi_{\mathrm{e},i} \rightarrow 0$, we also define a matrix $\overleftrightarrow{M}$ as $\overrightarrow{\varphi}_B = \overleftrightarrow{M} \overrightarrow{\varphi}_D$ for calculations below. 
Throughout this analysis, a symbol with an over-arrow in addition to a capital subscript $B$ or $D$ denotes a vector of branch or dynamical degrees of freedom, respectively. 
For symbols without an over-arrow, we use $B$ and $D$ to indicate an index over the branch and dynamical degrees of freedom, respectively. 

This choice of dynamical phases was made in order to satisfy several conditions. 
First, they ensure that the sum of the branch phases around a closed loop in the circuit is equal to the enclosed external flux (fluxoid quantization). 
Second, all dynamical fluxes $\delta \Phi_i$ appear on branch phases of inductors in the circuit, and not JJs. 
This ensures that the irrotional constraint associated with time-dependent fluxes is automatically satisfied~\cite{you2019circuit,riwar2022circuit,bryon2022experimental}. 
Finally, $\varphi$ only participates in branch phases of JJs, enforcing periodicity and therefore gate charge sensitivity. 
The other three dynamical phases are all non-periodic degrees of freedom. 

We next construct the Lagrangian as $L = T - U$. 
The ``potential'' part of the circuit is determined by the linear inductors and Josephson junctions: 

\begin{align*}
U = & +\frac{E_{Lf}}{2} \phi_{Lf}^2 + \frac{E_{Lk}}{2} \phi_{1k}^2 + \frac{E_{Lk}}{2} \phi_{2k}^2 \\ 
& - E_{Jf} \cos \varphi_{Jf} - E_{Jk} \cos \varphi_{1k}- E_{Jk} \cos \varphi_{2k} \\
\end{align*}

\noindent While this expression is reasonably simple as written, once we re-express it in terms of the four dynamical phases and two external fluxes, it becomes quite complex. 
As such, we compute it in Mathematica before feeding to our numerical diagonalization routine. 

The ``kinetic'' part of the Hamiltonian is determined by the four capacitors of the circuit: 

\begin{align*}
T = & +\frac{1}{2}\frac{\Phi_0^2}{(2\pi)^2}\bigg ( C_\mathrm{sh} \dot\varphi^2 + C_{Jf}\dot\varphi_{Jf}^2 + C_{Jk}\dot\varphi_{1k}^2+ C_{Jk}\dot\varphi_{2k}^2\bigg ) \\
& +\frac{\Phi_0}{2\pi}C_g V_g \dot\varphi\\
= &  +\frac{1}{2}\frac{\Phi_0^2}{(2\pi)^2} \overrightarrow{\dot\varphi}_D^T \overleftrightarrow{C} \overrightarrow{\dot\varphi}_D +\frac{\Phi_0}{2\pi}C_g V_g \dot\varphi\\
\end{align*}

\noindent where $\overleftrightarrow{C}$ is the capacitance matrix and the last term models the noisy gate charge $n_g = C_g V_g$ coupled to $\dot \varphi$ (not depicted in Fig. \ref{fig:fig4}). 
The conjugate charge operators can then be constructed as $\hbar n_i = \frac{\partial L}{\partial \dot \varphi_i}$ and the Hamiltonian as $H = \hbar \sum_i n_i \dot\varphi_i - L$. With $V_g \rightarrow 0$, we have $2e\overrightarrow{n}_D = \frac{\Phi_0}{2\pi} \overleftrightarrow{C}\overrightarrow{\dot \varphi}_D$. 
As with the potential energy, we compute $\overleftrightarrow{C}^{-1}$ in Mathematica before feeding it to our numerical diagonalization routine. 
Finally, we quantize the Hamiltonian by promoting all dynamical phases and charges to operators. 

With the Hamiltonian constructed, we can numerically calculate the qubit properties. 
We employ the standard technique of expressing the circuit Hamiltonian in a discrete basis so that the corresponding matrix can be numerically diagonalized. 
Here, we choose to describe the four dynamical phases using discrete phase bases. 
One reason for this choice is that, in the discrete phase basis, the Hamiltonian matrix can be expressed in sparse form; this significantly accelerates diagonalization. 
For a non-periodic dynamical phase $\hat{\varphi}_D$ (that is, $\hat{\varphi}_f, \hat{\varphi}_{\Sigma,k}$ or $\hat{\varphi}_{\Delta,k}$), the discrete phase basis vectors $|\varphi_{D}\rangle$ have eigenvalues evenly spaced from $-N_{\varphi_D} \Delta_{\varphi_D}$ to $N_{\varphi_D} \Delta_{\varphi_D}$ with spacing $\Delta_{\varphi_D}$ for a total of $2N_{\varphi_D}+1$ states. 
For the lone periodic dynamical phase $\varphi$, the basis vectors  $|\varphi\rangle$ have eigenvalues running from $-\pi$ to $+\pi - \Delta_{\varphi}$ with $\Delta_{\varphi} = 2\pi /(N_{\varphi} + 1)$; we enforce periodic boundary conditions on $\varphi$ and $n$. 
With the system discretized, the number and density of basis vectors is increased until all calculated quantities of interest have converged. 

\section{Decoherence channels}

\subsection{State transition channels} \label{state_transitions}

State transitions occur due to energy exchange between the circuit and its environment. 
Here we are interested in two types of state transitions. 
The first are those that leave the system within the computational subspace, i.e. transitions from $g$ to $e$ or vice versa. 
Using the noise models detailed below, we summarize these via the bit-flip rate $\Gamma_1 = \Gamma_{g \rightarrow e} + \Gamma_{e \rightarrow g}$. 
The second type of transitions are those that take us out of the computational subspace, i.e. heating events from either $g$ or $e$ to some final higher-energy state of the circuit $f$. 
These we summarize via the heating rates to all final states $\Gamma_{g,\uparrow} = \sum_{f > e} \Gamma_{g\rightarrow f}$ and $\Gamma_{e,\uparrow} = \sum_{f > e} \Gamma_{e\rightarrow f}$. 

We estimate these transition rates using the following framework. 
Suppose we have a circuit/environment Hamiltonian interaction term $\hat{\mathcal{O}} \mathcal{E}$, where $\hat{\mathcal{O}}$ is a circuit operator and $\mathcal{E}$ is an observable of the environment. 
We can estimate the transition rate between an initial state of the circuit $i$ and final state $f$ using Fermi's golden rule~\cite{smith2019design}: 

\begin{equation*}
 \Gamma_{i\rightarrow f} = \frac{1}{\hbar^2} |\langle f|\mathcal{O}|i\rangle |^2  S_{\mathcal{E}\mathcal{E}}[\omega_{if}]
\end{equation*}

\noindent where $S_{\mathcal{E}\mathcal{E}}[\omega]$ is the spectral density of $\mathcal{E}$ and $\omega_{if} = (\epsilon_i - \epsilon_f)/\hbar$ is the frequency difference between the initial and final states.
Here, we will consider the three intrinsic forms of loss that are believed to dominant state transitions in superconducting circuits: charge noise, flux noise, and non-equilibrium quasiparticles. 
Unless otherwise specified, we assume a noise temperature of $T = $ 20 mK. 

\textbf{Charge noise} induced transitions can be understood as arising from the interaction between the branch voltages $\hat{\mathcal{O}} = \hat{V}_b$ across capacitors in the circuit and noisy charges $\mathcal{E} = q$ in the circuit environment.
Note that this is formulation is distinct from a common approach~\cite{smith2019design} (charge operators coupled to noisy voltages), but our formulation ensures that the loss associated with a particular capacitor goes to zero as the capacitance (and therefore the stored energy) goes to zero. 
In harmonium, we consider the transition rates associated with each of the four capacitors separately, before summing them up to compute the total transition rate due to charge noise. 
To compute the matrix elements in Fermi's golden rule, we re-express the branch voltages in terms of the dynamical charges: 

\begin{equation*}
\overrightarrow{V}_B = 2e \overleftrightarrow{M} \overleftrightarrow{C}^{-1} \overrightarrow{n}_D
\end{equation*}

\noindent where we have dropped a term proportional to $n_g$ because it will evaluate to zero in Fermi's golden rule. 
To constrain the form of the spectral density, we note that it has been experimentally observed that charge-noise-induced loss in approximately linear circuits (resonators and transmons) 
varies linearly with the transition frequency $\Gamma_{1\rightarrow 0} \approx \omega_{10}/Q$, where $Q$ is the capacitor quality factor.
At zero temperature, this constrains the form of the charge spectral density to $S_{qq}[\omega] =\frac{2\hbar C}{Q} \Theta(\omega)$ where $\Theta(\omega)$ is the Heaviside theta function. 
In order to take heating effects into account, we model the temperature dependence as a bosonic bath, namely $\Theta(\omega) \rightarrow \Theta(-\omega)n_\mathrm{BE}(\omega) + \Theta(+\omega)n_\mathrm{BE}(\omega)e^{\hbar \omega/k_B T}$ where $n_\mathrm{BE}(\omega) = \frac{1}{e^{\hbar |\omega|/k_B T} - 1}$. 
Note that this is equivalent to the rate derived from the fluctuation-dissipation theorem, assuming a temperature-independent susceptibility. 
For the quality factor we assume $Q = 3 \times 10^6$, corresponding to a $T_1\approx 95 \; \mu \mathrm{s}$ for a 5~$\mathrm{GHz}$ transmon.
We note that while we also use this model for the sub-MHz qubit transition, we use a slightly different (more typical) $1/f$ model for low-frequency charge noise dephasing [section \ref{dephasing}]. 
While these models are not in principle inconsistent with each other since $\coth(x) \approx 1/x$ for small $x$, our high-frequency model predicts a higher noise amplitude at the qubit frequency and therefore gives a more conservative estimate of the charge-induced bit-flip rate. 

\textbf{Flux noise} can be understood as arising from the interaction between the noisy external fluxes $\mathcal{E} = \delta \Phi_i$ and the dynamical phases of the circuit $\hat{\mathcal{O}} = \frac{dH}{d\delta\Phi_i}$. 
Again, note that we have chosen our definitions of the dynamical phases such that the irrotational constraint is automatically satisfied~\cite{you2019circuit}. 
Experimentally, it has been observed that, with proper magnetic shielding, flux noise is dominated by microscopic defects living on the device itself, likely on the surface of any superconducting loop metallization~\cite{braumuller2020characterizing}. 
These defects give rise to a flux noise spectral density of the form $S_{\Phi \Phi}^+[\omega] = \frac{A_\Phi}{\omega}$ with $\sqrt{A_\Phi}$ as low as $1 \mu \Phi_0$. 
In this analysis we take this classic $1/f$ form for the symmetrized spectral density $S^+[\omega] = (S[+\omega] + S[-\omega])/2$ \cite{quintana2017observation}. 
Detailed balance and thermal equilibrium constrain $\frac{S[\omega]}{S[-\omega]} = e^{\hbar \omega /k_B T}$, so that the full form of the spectral density is $S_{\Phi \Phi}[\omega] = \frac{A_\Phi}{|\omega|} \frac{2}{1 + e^{- \hbar \omega /k_B T}}$. 

\textbf{Non-equilibrium quasiparticles} can cause state transitions when they tunnel across JJs in a qubit circuit~\cite{catelani2011relaxation}. 
The quasiparticle-induced transition rates can be written as

\begin{align*}\label{G_qp}
 \Gamma_{\mathrm{qp}, i\rightarrow f} = & + \frac{16 E_J}{h} |\langle f|\cos \frac{\hat{\varphi}_B}{2}|i\rangle |^2  S_{\mathrm{qp},c}[\omega_{fi}] \\ & + \frac{16 E_J}{h} |\langle f|\sin \frac{\hat{\varphi}_B}{2}|i\rangle |^2  S_{\mathrm{qp},s}[\omega_{fi}] \\
\end{align*}

\noindent where $S_{\mathrm{qp},c}[\omega]$ and $S_{\mathrm{qp},s}[\omega]$ are the quasiparticle spectral densities relevant for the $\cos \frac{\hat{\varphi}_B}{2}$ and $\sin \frac{\hat{\varphi}_B}{2}$ operators, respectively, and are proportional to the unitless quasiparticle density $x_\mathrm{qp}$. 
The spectral densities also depend on the effective quasiparticle temperature and the superconducting gap $\Delta$, as well as the gap difference $\delta \Delta$ on the two sides of a JJ. 
It was recently experimentally demonstrated that, although the number of quasiparticles is far above what is expected from thermal equilibrium at typical cryostat temperatures, the quasiparticle energy distribution is well-thermalized with the cryostat~\cite{connolly2024coexistence}. 
In the same work, the authors provide analytic forms for the spectral densities in the presence of a gap difference $\delta \Delta$, which we use for our calculations but do not reproduce here. 
Motivated by experiments~\cite{diamond2022distinguishing, marchegiani2022quasiparticles, mcewen2402resisting, connolly2024coexistence}, we take the gap to be that of bulk aluminum $\Delta = h \times 44 \; \mathrm{GHz}$, a gap asymmetry of $\delta \Delta = h \times 5 \; \mathrm{GHz}$, and take $x_\mathrm{qp} = 10^{-8}$. 

For quasiparticle tunneling across the JJs of the circuit, the above equations can be applied directly. 
However, we also take quasiparticle in the superinductors into account. 
In this case, the phase across each JJ internal to the superinductor can be linearized so that Fermi's golden rule can be written as $\Gamma_{\mathrm{qp}, i\rightarrow f} =  \frac{4E_L}{h}|\langle f|\hat{\varphi}_B|i\rangle |^2  S_{\mathrm{qp},+}[\omega_{fi}]$.

We note that here we do not quantitatively account for photon-assisted tunneling across the circuit JJs due to infrared photons, which has been shown to be an important source of non-equilibrium quasiparticles~\cite{houzet2019photon, diamond2022distinguishing,liu2024quasiparticle}. 
These processes can be described by an a Fermi's golden rule similar to that used for tunneling, but with different spectral densities that depend on the mode structure and population of the circuit's electromagnetic environment. 
However, we note that the maximum computational space quasiparticle tunneling matrix elements $\sin\frac{\hat{\varphi}_B}{2}, \cos\frac{\hat{\varphi}_B}{2}$ are four orders of magnitude smaller than typical transmon matrix elements, respectively. 
Filtering and device/package design have been shown to reduce these photon-assisted tunneling events in transmons to the sub-Hz regime~\cite{connolly2024coexistence}. 

\subsection{Dephasing channels} \label{dephasing}

Pure dephasing arises due to temporal fluctuation of the qubit frequency $\omega_q(\mathcal{E})$ as the noisy parameter $\mathcal{E}$ varies. 
In general, the rate of phase flips will depend on the particulars of the logic circuit being run on a quantum processor.
For instance, in some situations high-order dynamical decoupling can mitigate the effects of dephasing noise, while in other situations, it might be that no form of noise echo is possible. 
Here we will present single-echo dephasing rates; i.e. the dephasing rate that would be measured in Hahn-echo experiment with one $\pi$ pulse. 

The dephasing rate due to a noise channel $\mathcal{E}$ is proportional to the first-order derivative $|\frac{d\omega_q}{d\mathcal{E}}|$. 
If the first-order derivative is zero, then one can consider the second-order contribution $|\frac{d^2\omega_q}{d\mathcal{E}}|$. 
Several of the pure dephasing noise channels we consider here (charge, flux, and critical current) posses a $1/f$ spectral density $S_{\mathcal{E}\mathcal{E}}[\omega] = \frac{A_\mathcal{E}}{\omega}$. 
For $1/f$-type noise, the echo dephasing rate can be approximated as $\Gamma_{\phi,\mathcal{E}} \approx \sqrt{A_\mathcal{E}} |\frac{d \omega_{q}}{d \mathcal{E}}|$ for first-order-coupled noise (here, charge and critical current noise) and $\Gamma_{\phi,\mathcal{E}} \approx A_\mathcal{E} |\frac{d^2 \omega_{q}}{d \mathcal{E}^2}|$ for second-order-coupled noise (here, flux noise). 
As discussed in section \ref{readout}, we propose to readout the qubit using typical dispersive readout; as such, we also consider shot-noise dephasing. 

\textbf{Flux noise} only affects harmonium at 2nd order because the qubit is operated at the double sweet spot $\varphi_{\mathrm{e},k} = \varphi_{\mathrm{e},f} = \pi$. 
With proper shielding, global flux noise (i.e. noise common to multiple loops of a circuit) can be mitigated, leaving mainly local flux noise generated by defect spins. 
The amplitude of flux noise $\sqrt{A_\Phi}$ generated by these defect spins has been found to scale linearly with the flux loop perimeter and logarithmically with the loop width~\cite{braumuller2020characterizing}, and has been measured as low as $\sqrt{A_\Phi}\approx $ 1 $\mu \Phi_0$. 
Due to this geometry dependence, the amplitude will depend on the exact details of a fabricated device; as such, we compute estimates $\sqrt{A_\Phi} = $ 1 $\mu \Phi_0$ as a lower-bound on the flux noise dephasing rate. 
We have also computed the sensitivity of the qubit to global flux noise between the two loops, and find that as long as the global noise amplitude is an order of magnitude below the defect spin noise amplitude, global flux noise will not be limiting. 

\textbf{Charge noise} affects harmonium in a manner similar to the transmon; the dependence of $\omega_{q}$ in harmonium is a periodic function of $n_g$: 

\begin{equation*}
\omega_{q}(n_g) = \omega_{ge,0} + \delta \omega_{ge,0} \cos 2\pi n_g
\end{equation*}

\noindent where $\delta \omega_{ge,0}$ is commonly referred to as the ``charge dispersion''. 
Unlike flux noise, the gate charge $n_g$ cannot be relied upon to fluctuate around a bias value, and typically diffuses on the timescale of seconds; as such, the instantaneous dephasing rate, which is proportional $\frac{d\omega_{q}}{dn_g} = 2\pi \delta \omega_{ge, 0} \sin 2\pi n_g$, will vary sinusoidally with the gate charge. 
In Table~\ref{table:2}, we make the conservative assumption that $n_g = 0.25$, i.e. that the qubit is biased to the point of maximum dephasing; averaging this result over gate charge would result in a modest improvement of $\sim50\%$. 

\textbf{Critical current noise} is perhaps the most poorly understood of the dephasing channels we discuss here; this is in part due to the fact that it has not typically been a limiting decoherence channel in charge and flux qubits. 
Critical current noise is believed to be well-described by a $1/f$ spectral density with $\sqrt{A_{E_J}} \propto E_J$.
Twenty years ago, it was estimated that the noise amplitude could be as high as $\sqrt{A_{E_J}}/E_J = 5 \times 10^{-7}$; in the original transmon proposal this lead the authors to suggest that this noise channel might limit the dephasing time of the qubit to tens of microseconds\cite{van2004decoherence,koch2007charge}. 
A decade later, pure dephasing times of transmons have been measured at much larger values, as high as 10 milliseconds~\cite{wang2019cavity}. 
Here, we use these observations as a bound on the critical current noise amplitude. With a pure dephasing time of $\Gamma_\phi  = 1/5 \; \mathrm{ms}$, we estimate for a 5 GHz transmon $\Gamma_\phi = \frac{\sqrt{A_{E_J}}}{E_J}\omega_q/2 \quad \rightarrow \quad \sqrt{A_{E_J}}/E_J = 1.2\times 10^{-8}$. 

\textbf{Readout shot-noise dephasing} arises for qubits employing dispersive readout due to thermal population of the readout resonator. 
As discussed in section \ref{readout}, we propose to use dispersive readout for harmonium, and so must contend with this noise channel. 
The dephasing rate is given by~\cite{clerk2007using,wang2019cavity}

\begin{equation*}
\Gamma_\phi^\mathrm{sn} = \frac{\bar{n}_\mathrm{th}\kappa \chi^2}{\kappa^2/4 + \chi^2}
\end{equation*}

\noindent where $\chi$ is the dispersive shift between the resonator and qubit, $\kappa$ is the single-photon loss rate of the resonator, and $\bar{n}_\mathrm{th}$ is the thermal population of the resonator. 
Typically, resonator temperatures are much higher than 20 mK due to noise on the readout line; here we use 40 mK. 
While this is often the limiting dephasing channel for superconducting qubits measured with dispersively-coupled resonators, this is not the case for harmonium because $\chi \ll \kappa$ at the qubit bias point $\varphi_{\mathrm{e},k} = \varphi_{\mathrm{e},f} = \pi$. 
We discuss this in more detail in section \ref{readout}.

\textbf{Aharonov-Casher} dephasing is fundamentally due to low-frequency charge noise~\cite{manucharyan2012evidence, randeria2024dephasing}. 
But unlike the noisy gate charge $n_g$ that we have so far focused on, here the gate charges in question are those coupled to the internal degrees of freedom of the JJs that comprise the superinductors. 
For large $E_{J,\mathrm{array}}E_{C,\mathrm{array}}$ with $E_{J,\mathrm{array}}$ and $E_{C,\mathrm{array}}$ that of an individual JJ of a superinductor array, the amplitude of the periodic dispersion with the gate charge associated with one of JJs of the superinductor is $\epsilon_\mathrm{ps} = \frac{4\sqrt{2}}{\pi} \hbar \omega_\mathrm{p} \sqrt{\frac{1}{z}}e^{-4/\pi z}$ where $\hbar \omega_\mathrm{p} = \sqrt{8 E_{J,\mathrm{array}} E_{C,\mathrm{array}}}$ and $z = \sqrt{8E_{C,\mathrm{array}}/E_{J,\mathrm{array}}}/(2\pi)$. 
It can be shown that, when averaged over all $N_\mathrm{array}$ JJs of the array, the Ramsey dephasing time is given by~\cite{manucharyan2012evidence}

\begin{equation*}
\Gamma_{\phi, \mathrm{AC}} = \pi\sqrt{N_\mathrm{array}} \epsilon_\mathrm{ps} |\mathcal{F}_{ge}|
\end{equation*}

\noindent The structure factor $|\mathcal{F}_{ge}|$ characterizes the overlaps of the qubit states under a phase slip event in the superinductor. 
For phase slips in the superinductors of the fluxonium and the kite, it is respectively

\begin{equation*}
F_{ge}^a = \langle e| e, \hat{\varphi}_f \rightarrow \hat{\varphi}_f -2\pi \rangle - \langle g| g, \hat{\varphi}_f \rightarrow \hat{\varphi}_f -2\pi \rangle
\end{equation*}

\begin{align*}
F_{ge}^b =& +  \langle e| e, \hat{\varphi}_{\Sigma k}  \rightarrow \hat{\varphi}_{\Sigma k} -\pi, \hat{\varphi}_{\Delta k}  \rightarrow \hat{\varphi}_{\Delta k} + \pi \rangle \\ & - \langle g| g, \hat{\varphi}_{\Sigma k}  \rightarrow \hat{\varphi}_{\Sigma k} -\pi, \hat{\varphi}_{\Delta k}  \rightarrow \hat{\varphi}_{\Delta k} + \pi \rangle
\end{align*}

\noindent where $\hat{\varphi}_D \rightarrow \hat{\varphi}_D +\theta$ indicates a wavefunction with $\hat{\varphi}_D$ shifted by $\theta$. 
For the results presented in Table.~\ref{table:2}, we assume twenty array JJs in the fluxonium superinductor and thirty array JJs in the kite superinductors.

\bibliography{refs}

\begin{thebibliography}{78}%
\makeatletter
\providecommand \@ifxundefined [1]{%
 \@ifx{#1\undefined}
}%
\providecommand \@ifnum [1]{%
 \ifnum #1\expandafter \@firstoftwo
 \else \expandafter \@secondoftwo
 \fi
}%
\providecommand \@ifx [1]{%
 \ifx #1\expandafter \@firstoftwo
 \else \expandafter \@secondoftwo
 \fi
}%
\providecommand \natexlab [1]{#1}%
\providecommand \enquote  [1]{``#1''}%
\providecommand \bibnamefont  [1]{#1}%
\providecommand \bibfnamefont [1]{#1}%
\providecommand \citenamefont [1]{#1}%
\providecommand \href@noop [0]{\@secondoftwo}%
\providecommand \href [0]{\begingroup \@sanitize@url \@href}%
\providecommand \@href[1]{\@@startlink{#1}\@@href}%
\providecommand \@@href[1]{\endgroup#1\@@endlink}%
\providecommand \@sanitize@url [0]{\catcode `\\12\catcode `\$12\catcode
  `\&12\catcode `\#12\catcode `\^12\catcode `\_12\catcode `\%12\relax}%
\providecommand \@@startlink[1]{}%
\providecommand \@@endlink[0]{}%
\providecommand \url  [0]{\begingroup\@sanitize@url \@url }%
\providecommand \@url [1]{\endgroup\@href {#1}{\urlprefix }}%
\providecommand \urlprefix  [0]{URL }%
\providecommand \Eprint [0]{\href }%
\providecommand \doibase [0]{https://doi.org/}%
\providecommand \selectlanguage [0]{\@gobble}%
\providecommand \bibinfo  [0]{\@secondoftwo}%
\providecommand \bibfield  [0]{\@secondoftwo}%
\providecommand \translation [1]{[#1]}%
\providecommand \BibitemOpen [0]{}%
\providecommand \bibitemStop [0]{}%
\providecommand \bibitemNoStop [0]{.\EOS\space}%
\providecommand \EOS [0]{\spacefactor3000\relax}%
\providecommand \BibitemShut  [1]{\csname bibitem#1\endcsname}%
\let\auto@bib@innerbib\@empty
\bibitem [{\citenamefont {Krantz}\ \emph {et~al.}(2019)\citenamefont {Krantz},
  \citenamefont {Kjaergaard}, \citenamefont {Yan}, \citenamefont {Orlando},
  \citenamefont {Gustavsson},\ and\ \citenamefont
  {Oliver}}]{krantz2019quantum}%
  \BibitemOpen
  \bibfield  {author} {\bibinfo {author} {\bibfnamefont {P.}~\bibnamefont
  {Krantz}}, \bibinfo {author} {\bibfnamefont {M.}~\bibnamefont {Kjaergaard}},
  \bibinfo {author} {\bibfnamefont {F.}~\bibnamefont {Yan}}, \bibinfo {author}
  {\bibfnamefont {T.~P.}\ \bibnamefont {Orlando}}, \bibinfo {author}
  {\bibfnamefont {S.}~\bibnamefont {Gustavsson}},\ and\ \bibinfo {author}
  {\bibfnamefont {W.~D.}\ \bibnamefont {Oliver}},\ }\href@noop {} {\bibfield
  {journal} {\bibinfo  {journal} {Applied physics reviews}\ }\textbf {\bibinfo
  {volume} {6}} (\bibinfo {year} {2019})}\BibitemShut {NoStop}%
\bibitem [{\citenamefont {Devoret}\ and\ \citenamefont
  {Schoelkopf}(2013)}]{devoret2013superconducting}%
  \BibitemOpen
  \bibfield  {author} {\bibinfo {author} {\bibfnamefont {M.~H.}\ \bibnamefont
  {Devoret}}\ and\ \bibinfo {author} {\bibfnamefont {R.~J.}\ \bibnamefont
  {Schoelkopf}},\ }\href@noop {} {\bibfield  {journal} {\bibinfo  {journal}
  {Science}\ }\textbf {\bibinfo {volume} {339}},\ \bibinfo {pages} {1169}
  (\bibinfo {year} {2013})}\BibitemShut {NoStop}%
\bibitem [{\citenamefont {Blais}\ \emph {et~al.}(2021)\citenamefont {Blais},
  \citenamefont {Grimsmo}, \citenamefont {Girvin},\ and\ \citenamefont
  {Wallraff}}]{blais2021circuit}%
  \BibitemOpen
  \bibfield  {author} {\bibinfo {author} {\bibfnamefont {A.}~\bibnamefont
  {Blais}}, \bibinfo {author} {\bibfnamefont {A.~L.}\ \bibnamefont {Grimsmo}},
  \bibinfo {author} {\bibfnamefont {S.~M.}\ \bibnamefont {Girvin}},\ and\
  \bibinfo {author} {\bibfnamefont {A.}~\bibnamefont {Wallraff}},\ }\href@noop
  {} {\bibfield  {journal} {\bibinfo  {journal} {Reviews of Modern Physics}\
  }\textbf {\bibinfo {volume} {93}},\ \bibinfo {pages} {025005} (\bibinfo
  {year} {2021})}\BibitemShut {NoStop}%
\bibitem [{\citenamefont {Nakamura}\ \emph {et~al.}(1999)\citenamefont
  {Nakamura}, \citenamefont {Pashkin},\ and\ \citenamefont
  {Tsai}}]{nakamura1999coherent}%
  \BibitemOpen
  \bibfield  {author} {\bibinfo {author} {\bibfnamefont {Y.}~\bibnamefont
  {Nakamura}}, \bibinfo {author} {\bibfnamefont {Y.~A.}\ \bibnamefont
  {Pashkin}},\ and\ \bibinfo {author} {\bibfnamefont {J.}~\bibnamefont
  {Tsai}},\ }\href@noop {} {\bibfield  {journal} {\bibinfo  {journal} {nature}\
  }\textbf {\bibinfo {volume} {398}},\ \bibinfo {pages} {786} (\bibinfo {year}
  {1999})}\BibitemShut {NoStop}%
\bibitem [{\citenamefont {Mooij}\ \emph {et~al.}(1999)\citenamefont {Mooij},
  \citenamefont {Orlando}, \citenamefont {Levitov}, \citenamefont {Tian},
  \citenamefont {Van~der Wal},\ and\ \citenamefont
  {Lloyd}}]{mooij1999josephson}%
  \BibitemOpen
  \bibfield  {author} {\bibinfo {author} {\bibfnamefont {J.}~\bibnamefont
  {Mooij}}, \bibinfo {author} {\bibfnamefont {T.}~\bibnamefont {Orlando}},
  \bibinfo {author} {\bibfnamefont {L.}~\bibnamefont {Levitov}}, \bibinfo
  {author} {\bibfnamefont {L.}~\bibnamefont {Tian}}, \bibinfo {author}
  {\bibfnamefont {C.~H.}\ \bibnamefont {Van~der Wal}},\ and\ \bibinfo {author}
  {\bibfnamefont {S.}~\bibnamefont {Lloyd}},\ }\href@noop {} {\bibfield
  {journal} {\bibinfo  {journal} {Science}\ }\textbf {\bibinfo {volume}
  {285}},\ \bibinfo {pages} {1036} (\bibinfo {year} {1999})}\BibitemShut
  {NoStop}%
\bibitem [{\citenamefont {Koch}\ \emph {et~al.}(2007)\citenamefont {Koch},
  \citenamefont {Terri}, \citenamefont {Gambetta}, \citenamefont {Houck},
  \citenamefont {Schuster}, \citenamefont {Majer}, \citenamefont {Blais},
  \citenamefont {Devoret}, \citenamefont {Girvin},\ and\ \citenamefont
  {Schoelkopf}}]{koch2007charge}%
  \BibitemOpen
  \bibfield  {author} {\bibinfo {author} {\bibfnamefont {J.}~\bibnamefont
  {Koch}}, \bibinfo {author} {\bibfnamefont {M.~Y.}\ \bibnamefont {Terri}},
  \bibinfo {author} {\bibfnamefont {J.}~\bibnamefont {Gambetta}}, \bibinfo
  {author} {\bibfnamefont {A.~A.}\ \bibnamefont {Houck}}, \bibinfo {author}
  {\bibfnamefont {D.~I.}\ \bibnamefont {Schuster}}, \bibinfo {author}
  {\bibfnamefont {J.}~\bibnamefont {Majer}}, \bibinfo {author} {\bibfnamefont
  {A.}~\bibnamefont {Blais}}, \bibinfo {author} {\bibfnamefont {M.~H.}\
  \bibnamefont {Devoret}}, \bibinfo {author} {\bibfnamefont {S.~M.}\
  \bibnamefont {Girvin}},\ and\ \bibinfo {author} {\bibfnamefont {R.~J.}\
  \bibnamefont {Schoelkopf}},\ }\href@noop {} {\bibfield  {journal} {\bibinfo
  {journal} {Physical Review A}\ }\textbf {\bibinfo {volume} {76}},\ \bibinfo
  {pages} {042319} (\bibinfo {year} {2007})}\BibitemShut {NoStop}%
\bibitem [{\citenamefont {Arute}\ \emph {et~al.}(2019)\citenamefont {Arute},
  \citenamefont {Arya}, \citenamefont {Babbush}, \citenamefont {Bacon},
  \citenamefont {Bardin}, \citenamefont {Barends}, \citenamefont {Biswas},
  \citenamefont {Boixo}, \citenamefont {Brandao}, \citenamefont {Buell} \emph
  {et~al.}}]{arute2019quantum}%
  \BibitemOpen
  \bibfield  {author} {\bibinfo {author} {\bibfnamefont {F.}~\bibnamefont
  {Arute}}, \bibinfo {author} {\bibfnamefont {K.}~\bibnamefont {Arya}},
  \bibinfo {author} {\bibfnamefont {R.}~\bibnamefont {Babbush}}, \bibinfo
  {author} {\bibfnamefont {D.}~\bibnamefont {Bacon}}, \bibinfo {author}
  {\bibfnamefont {J.~C.}\ \bibnamefont {Bardin}}, \bibinfo {author}
  {\bibfnamefont {R.}~\bibnamefont {Barends}}, \bibinfo {author} {\bibfnamefont
  {R.}~\bibnamefont {Biswas}}, \bibinfo {author} {\bibfnamefont
  {S.}~\bibnamefont {Boixo}}, \bibinfo {author} {\bibfnamefont {F.~G.}\
  \bibnamefont {Brandao}}, \bibinfo {author} {\bibfnamefont {D.~A.}\
  \bibnamefont {Buell}}, \emph {et~al.},\ }\href@noop {} {\bibfield  {journal}
  {\bibinfo  {journal} {Nature}\ }\textbf {\bibinfo {volume} {574}},\ \bibinfo
  {pages} {505} (\bibinfo {year} {2019})}\BibitemShut {NoStop}%
\bibitem [{\citenamefont {Krinner}\ \emph {et~al.}(2022)\citenamefont
  {Krinner}, \citenamefont {Lacroix}, \citenamefont {Remm}, \citenamefont
  {Di~Paolo}, \citenamefont {Genois}, \citenamefont {Leroux}, \citenamefont
  {Hellings}, \citenamefont {Lazar}, \citenamefont {Swiadek}, \citenamefont
  {Herrmann} \emph {et~al.}}]{krinner2022realizing}%
  \BibitemOpen
  \bibfield  {author} {\bibinfo {author} {\bibfnamefont {S.}~\bibnamefont
  {Krinner}}, \bibinfo {author} {\bibfnamefont {N.}~\bibnamefont {Lacroix}},
  \bibinfo {author} {\bibfnamefont {A.}~\bibnamefont {Remm}}, \bibinfo {author}
  {\bibfnamefont {A.}~\bibnamefont {Di~Paolo}}, \bibinfo {author}
  {\bibfnamefont {E.}~\bibnamefont {Genois}}, \bibinfo {author} {\bibfnamefont
  {C.}~\bibnamefont {Leroux}}, \bibinfo {author} {\bibfnamefont
  {C.}~\bibnamefont {Hellings}}, \bibinfo {author} {\bibfnamefont
  {S.}~\bibnamefont {Lazar}}, \bibinfo {author} {\bibfnamefont
  {F.}~\bibnamefont {Swiadek}}, \bibinfo {author} {\bibfnamefont
  {J.}~\bibnamefont {Herrmann}}, \emph {et~al.},\ }\href@noop {} {\bibfield
  {journal} {\bibinfo  {journal} {Nature}\ }\textbf {\bibinfo {volume} {605}},\
  \bibinfo {pages} {669} (\bibinfo {year} {2022})}\BibitemShut {NoStop}%
\bibitem [{\citenamefont {Acharya}(2024)}]{googleEC}%
  \BibitemOpen
  \bibfield  {author} {\bibinfo {author} {\bibfnamefont {R.~e.~a.}\
  \bibnamefont {Acharya}},\ }\href@noop {} {\bibfield  {journal} {\bibinfo
  {journal} {Nature}\ } (\bibinfo {year} {2024})}\BibitemShut {NoStop}%
\bibitem [{\citenamefont {Manucharyan}\ \emph {et~al.}(2009)\citenamefont
  {Manucharyan}, \citenamefont {Koch}, \citenamefont {Glazman},\ and\
  \citenamefont {Devoret}}]{manucharyan2009fluxonium}%
  \BibitemOpen
  \bibfield  {author} {\bibinfo {author} {\bibfnamefont {V.~E.}\ \bibnamefont
  {Manucharyan}}, \bibinfo {author} {\bibfnamefont {J.}~\bibnamefont {Koch}},
  \bibinfo {author} {\bibfnamefont {L.~I.}\ \bibnamefont {Glazman}},\ and\
  \bibinfo {author} {\bibfnamefont {M.~H.}\ \bibnamefont {Devoret}},\
  }\href@noop {} {\bibfield  {journal} {\bibinfo  {journal} {Science}\ }\textbf
  {\bibinfo {volume} {326}},\ \bibinfo {pages} {113} (\bibinfo {year}
  {2009})}\BibitemShut {NoStop}%
\bibitem [{\citenamefont {Somoroff}\ \emph {et~al.}(2023)\citenamefont
  {Somoroff}, \citenamefont {Ficheux}, \citenamefont {Mencia}, \citenamefont
  {Xiong}, \citenamefont {Kuzmin},\ and\ \citenamefont
  {Manucharyan}}]{somoroff2023millisecond}%
  \BibitemOpen
  \bibfield  {author} {\bibinfo {author} {\bibfnamefont {A.}~\bibnamefont
  {Somoroff}}, \bibinfo {author} {\bibfnamefont {Q.}~\bibnamefont {Ficheux}},
  \bibinfo {author} {\bibfnamefont {R.~A.}\ \bibnamefont {Mencia}}, \bibinfo
  {author} {\bibfnamefont {H.}~\bibnamefont {Xiong}}, \bibinfo {author}
  {\bibfnamefont {R.}~\bibnamefont {Kuzmin}},\ and\ \bibinfo {author}
  {\bibfnamefont {V.~E.}\ \bibnamefont {Manucharyan}},\ }\href@noop {}
  {\bibfield  {journal} {\bibinfo  {journal} {Physical Review Letters}\
  }\textbf {\bibinfo {volume} {130}},\ \bibinfo {pages} {267001} (\bibinfo
  {year} {2023})}\BibitemShut {NoStop}%
\bibitem [{\citenamefont {Ding}\ \emph {et~al.}(2023)\citenamefont {Ding},
  \citenamefont {Hays}, \citenamefont {Sung}, \citenamefont {Kannan},
  \citenamefont {An}, \citenamefont {Di~Paolo}, \citenamefont {Karamlou},
  \citenamefont {Hazard}, \citenamefont {Azar}, \citenamefont {Kim} \emph
  {et~al.}}]{ding2023high}%
  \BibitemOpen
  \bibfield  {author} {\bibinfo {author} {\bibfnamefont {L.}~\bibnamefont
  {Ding}}, \bibinfo {author} {\bibfnamefont {M.}~\bibnamefont {Hays}}, \bibinfo
  {author} {\bibfnamefont {Y.}~\bibnamefont {Sung}}, \bibinfo {author}
  {\bibfnamefont {B.}~\bibnamefont {Kannan}}, \bibinfo {author} {\bibfnamefont
  {J.}~\bibnamefont {An}}, \bibinfo {author} {\bibfnamefont {A.}~\bibnamefont
  {Di~Paolo}}, \bibinfo {author} {\bibfnamefont {A.~H.}\ \bibnamefont
  {Karamlou}}, \bibinfo {author} {\bibfnamefont {T.~M.}\ \bibnamefont
  {Hazard}}, \bibinfo {author} {\bibfnamefont {K.}~\bibnamefont {Azar}},
  \bibinfo {author} {\bibfnamefont {D.~K.}\ \bibnamefont {Kim}}, \emph
  {et~al.},\ }\href@noop {} {\bibfield  {journal} {\bibinfo  {journal}
  {Physical Review X}\ }\textbf {\bibinfo {volume} {13}},\ \bibinfo {pages}
  {031035} (\bibinfo {year} {2023})}\BibitemShut {NoStop}%
\bibitem [{\citenamefont {Zhang}\ \emph {et~al.}(2024)\citenamefont {Zhang},
  \citenamefont {Ding}, \citenamefont {Weiss}, \citenamefont {Huang},
  \citenamefont {Ma}, \citenamefont {Guinn}, \citenamefont {Sussman},
  \citenamefont {Chitta}, \citenamefont {Chen}, \citenamefont {Houck} \emph
  {et~al.}}]{zhang2024tunable}%
  \BibitemOpen
  \bibfield  {author} {\bibinfo {author} {\bibfnamefont {H.}~\bibnamefont
  {Zhang}}, \bibinfo {author} {\bibfnamefont {C.}~\bibnamefont {Ding}},
  \bibinfo {author} {\bibfnamefont {D.}~\bibnamefont {Weiss}}, \bibinfo
  {author} {\bibfnamefont {Z.}~\bibnamefont {Huang}}, \bibinfo {author}
  {\bibfnamefont {Y.}~\bibnamefont {Ma}}, \bibinfo {author} {\bibfnamefont
  {C.}~\bibnamefont {Guinn}}, \bibinfo {author} {\bibfnamefont
  {S.}~\bibnamefont {Sussman}}, \bibinfo {author} {\bibfnamefont {S.~P.}\
  \bibnamefont {Chitta}}, \bibinfo {author} {\bibfnamefont {D.}~\bibnamefont
  {Chen}}, \bibinfo {author} {\bibfnamefont {A.~A.}\ \bibnamefont {Houck}},
  \emph {et~al.},\ }\href@noop {} {\bibfield  {journal} {\bibinfo  {journal}
  {PRX Quantum}\ }\textbf {\bibinfo {volume} {5}},\ \bibinfo {pages} {020326}
  (\bibinfo {year} {2024})}\BibitemShut {NoStop}%
\bibitem [{\citenamefont {Lin}\ \emph {et~al.}(2024)\citenamefont {Lin},
  \citenamefont {Cho}, \citenamefont {Chen}, \citenamefont {Vavilov},
  \citenamefont {Wang},\ and\ \citenamefont {Manucharyan}}]{lin202424}%
  \BibitemOpen
  \bibfield  {author} {\bibinfo {author} {\bibfnamefont {W.-J.}\ \bibnamefont
  {Lin}}, \bibinfo {author} {\bibfnamefont {H.}~\bibnamefont {Cho}}, \bibinfo
  {author} {\bibfnamefont {Y.}~\bibnamefont {Chen}}, \bibinfo {author}
  {\bibfnamefont {M.~G.}\ \bibnamefont {Vavilov}}, \bibinfo {author}
  {\bibfnamefont {C.}~\bibnamefont {Wang}},\ and\ \bibinfo {author}
  {\bibfnamefont {V.~E.}\ \bibnamefont {Manucharyan}},\ }\href@noop {}
  {\bibfield  {journal} {\bibinfo  {journal} {arXiv preprint arXiv:2407.15783}\
  } (\bibinfo {year} {2024})}\BibitemShut {NoStop}%
\bibitem [{\citenamefont {Gyenis}\ \emph
  {et~al.}(2021{\natexlab{a}})\citenamefont {Gyenis}, \citenamefont {Di~Paolo},
  \citenamefont {Koch}, \citenamefont {Blais}, \citenamefont {Houck},\ and\
  \citenamefont {Schuster}}]{gyenis2021moving}%
  \BibitemOpen
  \bibfield  {author} {\bibinfo {author} {\bibfnamefont {A.}~\bibnamefont
  {Gyenis}}, \bibinfo {author} {\bibfnamefont {A.}~\bibnamefont {Di~Paolo}},
  \bibinfo {author} {\bibfnamefont {J.}~\bibnamefont {Koch}}, \bibinfo {author}
  {\bibfnamefont {A.}~\bibnamefont {Blais}}, \bibinfo {author} {\bibfnamefont
  {A.~A.}\ \bibnamefont {Houck}},\ and\ \bibinfo {author} {\bibfnamefont
  {D.~I.}\ \bibnamefont {Schuster}},\ }\href@noop {} {\bibfield  {journal}
  {\bibinfo  {journal} {PRX Quantum}\ }\textbf {\bibinfo {volume} {2}},\
  \bibinfo {pages} {030101} (\bibinfo {year} {2021}{\natexlab{a}})}\BibitemShut
  {NoStop}%
\bibitem [{\citenamefont {Kitaev}(2006)}]{kitaevCurrent}%
  \BibitemOpen
  \bibfield  {author} {\bibinfo {author} {\bibfnamefont {A.}~\bibnamefont
  {Kitaev}},\ }\href@noop {} {\bibfield  {journal} {\bibinfo  {journal}
  {arXiv:0609441}\ } (\bibinfo {year} {2006})}\BibitemShut {NoStop}%
\bibitem [{\citenamefont {Brooks}\ \emph {et~al.}(2013)\citenamefont {Brooks},
  \citenamefont {Kitaev},\ and\ \citenamefont
  {Preskill}}]{brooks2013protected}%
  \BibitemOpen
  \bibfield  {author} {\bibinfo {author} {\bibfnamefont {P.}~\bibnamefont
  {Brooks}}, \bibinfo {author} {\bibfnamefont {A.}~\bibnamefont {Kitaev}},\
  and\ \bibinfo {author} {\bibfnamefont {J.}~\bibnamefont {Preskill}},\
  }\href@noop {} {\bibfield  {journal} {\bibinfo  {journal} {Physical Review
  A—Atomic, Molecular, and Optical Physics}\ }\textbf {\bibinfo {volume}
  {87}},\ \bibinfo {pages} {052306} (\bibinfo {year} {2013})}\BibitemShut
  {NoStop}%
\bibitem [{\citenamefont {Kalashnikov}\ \emph {et~al.}(2020)\citenamefont
  {Kalashnikov}, \citenamefont {Hsieh}, \citenamefont {Zhang}, \citenamefont
  {Lu}, \citenamefont {Kamenov}, \citenamefont {Di~Paolo}, \citenamefont
  {Blais}, \citenamefont {Gershenson},\ and\ \citenamefont
  {Bell}}]{kalashnikov2020bifluxon}%
  \BibitemOpen
  \bibfield  {author} {\bibinfo {author} {\bibfnamefont {K.}~\bibnamefont
  {Kalashnikov}}, \bibinfo {author} {\bibfnamefont {W.~T.}\ \bibnamefont
  {Hsieh}}, \bibinfo {author} {\bibfnamefont {W.}~\bibnamefont {Zhang}},
  \bibinfo {author} {\bibfnamefont {W.-S.}\ \bibnamefont {Lu}}, \bibinfo
  {author} {\bibfnamefont {P.}~\bibnamefont {Kamenov}}, \bibinfo {author}
  {\bibfnamefont {A.}~\bibnamefont {Di~Paolo}}, \bibinfo {author}
  {\bibfnamefont {A.}~\bibnamefont {Blais}}, \bibinfo {author} {\bibfnamefont
  {M.~E.}\ \bibnamefont {Gershenson}},\ and\ \bibinfo {author} {\bibfnamefont
  {M.}~\bibnamefont {Bell}},\ }\href@noop {} {\bibfield  {journal} {\bibinfo
  {journal} {PRX Quantum}\ }\textbf {\bibinfo {volume} {1}},\ \bibinfo {pages}
  {010307} (\bibinfo {year} {2020})}\BibitemShut {NoStop}%
\bibitem [{\citenamefont {Gyenis}\ \emph
  {et~al.}(2021{\natexlab{b}})\citenamefont {Gyenis}, \citenamefont {Mundada},
  \citenamefont {Di~Paolo}, \citenamefont {Hazard}, \citenamefont {You},
  \citenamefont {Schuster}, \citenamefont {Koch}, \citenamefont {Blais},\ and\
  \citenamefont {Houck}}]{gyenis2021experimental}%
  \BibitemOpen
  \bibfield  {author} {\bibinfo {author} {\bibfnamefont {A.}~\bibnamefont
  {Gyenis}}, \bibinfo {author} {\bibfnamefont {P.~S.}\ \bibnamefont {Mundada}},
  \bibinfo {author} {\bibfnamefont {A.}~\bibnamefont {Di~Paolo}}, \bibinfo
  {author} {\bibfnamefont {T.~M.}\ \bibnamefont {Hazard}}, \bibinfo {author}
  {\bibfnamefont {X.}~\bibnamefont {You}}, \bibinfo {author} {\bibfnamefont
  {D.~I.}\ \bibnamefont {Schuster}}, \bibinfo {author} {\bibfnamefont
  {J.}~\bibnamefont {Koch}}, \bibinfo {author} {\bibfnamefont {A.}~\bibnamefont
  {Blais}},\ and\ \bibinfo {author} {\bibfnamefont {A.~A.}\ \bibnamefont
  {Houck}},\ }\href@noop {} {\bibfield  {journal} {\bibinfo  {journal} {PRX
  Quantum}\ }\textbf {\bibinfo {volume} {2}},\ \bibinfo {pages} {010339}
  (\bibinfo {year} {2021}{\natexlab{b}})}\BibitemShut {NoStop}%
\bibitem [{\citenamefont {Ioffe}\ and\ \citenamefont
  {Feigel’man}(2002)}]{ioffe2002possible}%
  \BibitemOpen
  \bibfield  {author} {\bibinfo {author} {\bibfnamefont {L.}~\bibnamefont
  {Ioffe}}\ and\ \bibinfo {author} {\bibfnamefont {M.}~\bibnamefont
  {Feigel’man}},\ }\href@noop {} {\bibfield  {journal} {\bibinfo  {journal}
  {Physical Review B}\ }\textbf {\bibinfo {volume} {66}},\ \bibinfo {pages}
  {224503} (\bibinfo {year} {2002})}\BibitemShut {NoStop}%
\bibitem [{\citenamefont {Dou{\c{c}}ot}\ \emph {et~al.}(2005)\citenamefont
  {Dou{\c{c}}ot}, \citenamefont {Feigel’Man}, \citenamefont {Ioffe},\ and\
  \citenamefont {Ioselevich}}]{douccot2005protected}%
  \BibitemOpen
  \bibfield  {author} {\bibinfo {author} {\bibfnamefont {B.}~\bibnamefont
  {Dou{\c{c}}ot}}, \bibinfo {author} {\bibfnamefont {M.}~\bibnamefont
  {Feigel’Man}}, \bibinfo {author} {\bibfnamefont {L.}~\bibnamefont
  {Ioffe}},\ and\ \bibinfo {author} {\bibfnamefont {A.}~\bibnamefont
  {Ioselevich}},\ }\href@noop {} {\bibfield  {journal} {\bibinfo  {journal}
  {Physical Review B—Condensed Matter and Materials Physics}\ }\textbf
  {\bibinfo {volume} {71}},\ \bibinfo {pages} {024505} (\bibinfo {year}
  {2005})}\BibitemShut {NoStop}%
\bibitem [{\citenamefont {Gladchenko}\ \emph {et~al.}(2009)\citenamefont
  {Gladchenko}, \citenamefont {Olaya}, \citenamefont {Dupont-Ferrier},
  \citenamefont {Dou{\c{c}}ot}, \citenamefont {Ioffe},\ and\ \citenamefont
  {Gershenson}}]{gladchenko2009superconducting}%
  \BibitemOpen
  \bibfield  {author} {\bibinfo {author} {\bibfnamefont {S.}~\bibnamefont
  {Gladchenko}}, \bibinfo {author} {\bibfnamefont {D.}~\bibnamefont {Olaya}},
  \bibinfo {author} {\bibfnamefont {E.}~\bibnamefont {Dupont-Ferrier}},
  \bibinfo {author} {\bibfnamefont {B.}~\bibnamefont {Dou{\c{c}}ot}}, \bibinfo
  {author} {\bibfnamefont {L.~B.}\ \bibnamefont {Ioffe}},\ and\ \bibinfo
  {author} {\bibfnamefont {M.~E.}\ \bibnamefont {Gershenson}},\ }\href@noop {}
  {\bibfield  {journal} {\bibinfo  {journal} {Nature Physics}\ }\textbf
  {\bibinfo {volume} {5}},\ \bibinfo {pages} {48} (\bibinfo {year}
  {2009})}\BibitemShut {NoStop}%
\bibitem [{\citenamefont {Bell}\ \emph {et~al.}(2014)\citenamefont {Bell},
  \citenamefont {Paramanandam}, \citenamefont {Ioffe},\ and\ \citenamefont
  {Gershenson}}]{bell2014protected}%
  \BibitemOpen
  \bibfield  {author} {\bibinfo {author} {\bibfnamefont {M.~T.}\ \bibnamefont
  {Bell}}, \bibinfo {author} {\bibfnamefont {J.}~\bibnamefont {Paramanandam}},
  \bibinfo {author} {\bibfnamefont {L.~B.}\ \bibnamefont {Ioffe}},\ and\
  \bibinfo {author} {\bibfnamefont {M.~E.}\ \bibnamefont {Gershenson}},\
  }\href@noop {} {\bibfield  {journal} {\bibinfo  {journal} {Physical review
  letters}\ }\textbf {\bibinfo {volume} {112}},\ \bibinfo {pages} {167001}
  (\bibinfo {year} {2014})}\BibitemShut {NoStop}%
\bibitem [{\citenamefont {Smith}\ \emph {et~al.}(2020)\citenamefont {Smith},
  \citenamefont {Kou}, \citenamefont {Xiao}, \citenamefont {Vool},\ and\
  \citenamefont {Devoret}}]{smith2020superconducting}%
  \BibitemOpen
  \bibfield  {author} {\bibinfo {author} {\bibfnamefont {W.}~\bibnamefont
  {Smith}}, \bibinfo {author} {\bibfnamefont {A.}~\bibnamefont {Kou}}, \bibinfo
  {author} {\bibfnamefont {X.}~\bibnamefont {Xiao}}, \bibinfo {author}
  {\bibfnamefont {U.}~\bibnamefont {Vool}},\ and\ \bibinfo {author}
  {\bibfnamefont {M.}~\bibnamefont {Devoret}},\ }\href@noop {} {\bibfield
  {journal} {\bibinfo  {journal} {npj Quantum Information}\ }\textbf {\bibinfo
  {volume} {6}},\ \bibinfo {pages} {8} (\bibinfo {year} {2020})}\BibitemShut
  {NoStop}%
\bibitem [{\citenamefont {Larsen}\ \emph {et~al.}(2020)\citenamefont {Larsen},
  \citenamefont {Gershenson}, \citenamefont {Casparis}, \citenamefont
  {Kringh{\o}j}, \citenamefont {Pearson}, \citenamefont {McNeil}, \citenamefont
  {Kuemmeth}, \citenamefont {Krogstrup}, \citenamefont {Petersson},\ and\
  \citenamefont {Marcus}}]{larsen2020parity}%
  \BibitemOpen
  \bibfield  {author} {\bibinfo {author} {\bibfnamefont {T.~W.}\ \bibnamefont
  {Larsen}}, \bibinfo {author} {\bibfnamefont {M.~E.}\ \bibnamefont
  {Gershenson}}, \bibinfo {author} {\bibfnamefont {L.}~\bibnamefont
  {Casparis}}, \bibinfo {author} {\bibfnamefont {A.}~\bibnamefont
  {Kringh{\o}j}}, \bibinfo {author} {\bibfnamefont {N.~J.}\ \bibnamefont
  {Pearson}}, \bibinfo {author} {\bibfnamefont {R.~P.}\ \bibnamefont {McNeil}},
  \bibinfo {author} {\bibfnamefont {F.}~\bibnamefont {Kuemmeth}}, \bibinfo
  {author} {\bibfnamefont {P.}~\bibnamefont {Krogstrup}}, \bibinfo {author}
  {\bibfnamefont {K.~D.}\ \bibnamefont {Petersson}},\ and\ \bibinfo {author}
  {\bibfnamefont {C.~M.}\ \bibnamefont {Marcus}},\ }\href@noop {} {\bibfield
  {journal} {\bibinfo  {journal} {Physical review letters}\ }\textbf {\bibinfo
  {volume} {125}},\ \bibinfo {pages} {056801} (\bibinfo {year}
  {2020})}\BibitemShut {NoStop}%
\bibitem [{\citenamefont {Schrade}\ \emph {et~al.}(2022)\citenamefont
  {Schrade}, \citenamefont {Marcus},\ and\ \citenamefont
  {Gyenis}}]{schrade2022protected}%
  \BibitemOpen
  \bibfield  {author} {\bibinfo {author} {\bibfnamefont {C.}~\bibnamefont
  {Schrade}}, \bibinfo {author} {\bibfnamefont {C.~M.}\ \bibnamefont
  {Marcus}},\ and\ \bibinfo {author} {\bibfnamefont {A.}~\bibnamefont
  {Gyenis}},\ }\href@noop {} {\bibfield  {journal} {\bibinfo  {journal} {PRX
  Quantum}\ }\textbf {\bibinfo {volume} {3}},\ \bibinfo {pages} {030303}
  (\bibinfo {year} {2022})}\BibitemShut {NoStop}%
\bibitem [{\citenamefont {Stace}\ \emph {et~al.}(2009)\citenamefont {Stace},
  \citenamefont {Barrett},\ and\ \citenamefont
  {Doherty}}]{stace2009thresholds}%
  \BibitemOpen
  \bibfield  {author} {\bibinfo {author} {\bibfnamefont {T.~M.}\ \bibnamefont
  {Stace}}, \bibinfo {author} {\bibfnamefont {S.~D.}\ \bibnamefont {Barrett}},\
  and\ \bibinfo {author} {\bibfnamefont {A.~C.}\ \bibnamefont {Doherty}},\
  }\href@noop {} {\bibfield  {journal} {\bibinfo  {journal} {Physical review
  letters}\ }\textbf {\bibinfo {volume} {102}},\ \bibinfo {pages} {200501}
  (\bibinfo {year} {2009})}\BibitemShut {NoStop}%
\bibitem [{\citenamefont {Shim}\ and\ \citenamefont
  {Tahan}(2016)}]{shim2016semiconductor}%
  \BibitemOpen
  \bibfield  {author} {\bibinfo {author} {\bibfnamefont {Y.-P.}\ \bibnamefont
  {Shim}}\ and\ \bibinfo {author} {\bibfnamefont {C.}~\bibnamefont {Tahan}},\
  }\href@noop {} {\bibfield  {journal} {\bibinfo  {journal} {Nature
  communications}\ }\textbf {\bibinfo {volume} {7}},\ \bibinfo {pages} {11059}
  (\bibinfo {year} {2016})}\BibitemShut {NoStop}%
\bibitem [{\citenamefont {Campbell}\ \emph {et~al.}(2020)\citenamefont
  {Campbell}, \citenamefont {Shim}, \citenamefont {Kannan}, \citenamefont
  {Winik}, \citenamefont {Kim}, \citenamefont {Melville}, \citenamefont
  {Niedzielski}, \citenamefont {Yoder}, \citenamefont {Tahan}, \citenamefont
  {Gustavsson} \emph {et~al.}}]{campbell2020universal}%
  \BibitemOpen
  \bibfield  {author} {\bibinfo {author} {\bibfnamefont {D.~L.}\ \bibnamefont
  {Campbell}}, \bibinfo {author} {\bibfnamefont {Y.-P.}\ \bibnamefont {Shim}},
  \bibinfo {author} {\bibfnamefont {B.}~\bibnamefont {Kannan}}, \bibinfo
  {author} {\bibfnamefont {R.}~\bibnamefont {Winik}}, \bibinfo {author}
  {\bibfnamefont {D.~K.}\ \bibnamefont {Kim}}, \bibinfo {author} {\bibfnamefont
  {A.}~\bibnamefont {Melville}}, \bibinfo {author} {\bibfnamefont {B.~M.}\
  \bibnamefont {Niedzielski}}, \bibinfo {author} {\bibfnamefont {J.~L.}\
  \bibnamefont {Yoder}}, \bibinfo {author} {\bibfnamefont {C.}~\bibnamefont
  {Tahan}}, \bibinfo {author} {\bibfnamefont {S.}~\bibnamefont {Gustavsson}},
  \emph {et~al.},\ }\href@noop {} {\bibfield  {journal} {\bibinfo  {journal}
  {Physical Review X}\ }\textbf {\bibinfo {volume} {10}},\ \bibinfo {pages}
  {041051} (\bibinfo {year} {2020})}\BibitemShut {NoStop}%
\bibitem [{\citenamefont {Chou}\ \emph {et~al.}(2023)\citenamefont {Chou},
  \citenamefont {Shemma}, \citenamefont {McCarrick}, \citenamefont {Chien},
  \citenamefont {Teoh}, \citenamefont {Winkel}, \citenamefont {Anderson},
  \citenamefont {Chen}, \citenamefont {Curtis}, \citenamefont {de~Graaf} \emph
  {et~al.}}]{chou2023demonstrating}%
  \BibitemOpen
  \bibfield  {author} {\bibinfo {author} {\bibfnamefont {K.~S.}\ \bibnamefont
  {Chou}}, \bibinfo {author} {\bibfnamefont {T.}~\bibnamefont {Shemma}},
  \bibinfo {author} {\bibfnamefont {H.}~\bibnamefont {McCarrick}}, \bibinfo
  {author} {\bibfnamefont {T.-C.}\ \bibnamefont {Chien}}, \bibinfo {author}
  {\bibfnamefont {J.~D.}\ \bibnamefont {Teoh}}, \bibinfo {author}
  {\bibfnamefont {P.}~\bibnamefont {Winkel}}, \bibinfo {author} {\bibfnamefont
  {A.}~\bibnamefont {Anderson}}, \bibinfo {author} {\bibfnamefont
  {J.}~\bibnamefont {Chen}}, \bibinfo {author} {\bibfnamefont {J.}~\bibnamefont
  {Curtis}}, \bibinfo {author} {\bibfnamefont {S.~J.}\ \bibnamefont
  {de~Graaf}}, \emph {et~al.},\ }\href@noop {} {\bibfield  {journal} {\bibinfo
  {journal} {arXiv preprint arXiv:2307.03169}\ } (\bibinfo {year}
  {2023})}\BibitemShut {NoStop}%
\bibitem [{\citenamefont {Ma}\ \emph {et~al.}(2023)\citenamefont {Ma},
  \citenamefont {Liu}, \citenamefont {Peng}, \citenamefont {Zhang},
  \citenamefont {Jandura}, \citenamefont {Claes}, \citenamefont {Burgers},
  \citenamefont {Pupillo}, \citenamefont {Puri},\ and\ \citenamefont
  {Thompson}}]{ma2023high}%
  \BibitemOpen
  \bibfield  {author} {\bibinfo {author} {\bibfnamefont {S.}~\bibnamefont
  {Ma}}, \bibinfo {author} {\bibfnamefont {G.}~\bibnamefont {Liu}}, \bibinfo
  {author} {\bibfnamefont {P.}~\bibnamefont {Peng}}, \bibinfo {author}
  {\bibfnamefont {B.}~\bibnamefont {Zhang}}, \bibinfo {author} {\bibfnamefont
  {S.}~\bibnamefont {Jandura}}, \bibinfo {author} {\bibfnamefont
  {J.}~\bibnamefont {Claes}}, \bibinfo {author} {\bibfnamefont {A.~P.}\
  \bibnamefont {Burgers}}, \bibinfo {author} {\bibfnamefont {G.}~\bibnamefont
  {Pupillo}}, \bibinfo {author} {\bibfnamefont {S.}~\bibnamefont {Puri}},\ and\
  \bibinfo {author} {\bibfnamefont {J.~D.}\ \bibnamefont {Thompson}},\
  }\href@noop {} {\bibfield  {journal} {\bibinfo  {journal} {Nature}\ }\textbf
  {\bibinfo {volume} {622}},\ \bibinfo {pages} {279} (\bibinfo {year}
  {2023})}\BibitemShut {NoStop}%
\bibitem [{\citenamefont {Levine}\ \emph {et~al.}(2024)\citenamefont {Levine},
  \citenamefont {Haim}, \citenamefont {Hung}, \citenamefont {Alidoust},
  \citenamefont {Kalaee}, \citenamefont {DeLorenzo}, \citenamefont {Wollack},
  \citenamefont {Arrangoiz-Arriola}, \citenamefont {Khalajhedayati},
  \citenamefont {Sanil} \emph {et~al.}}]{levine2024demonstrating}%
  \BibitemOpen
  \bibfield  {author} {\bibinfo {author} {\bibfnamefont {H.}~\bibnamefont
  {Levine}}, \bibinfo {author} {\bibfnamefont {A.}~\bibnamefont {Haim}},
  \bibinfo {author} {\bibfnamefont {J.~S.}\ \bibnamefont {Hung}}, \bibinfo
  {author} {\bibfnamefont {N.}~\bibnamefont {Alidoust}}, \bibinfo {author}
  {\bibfnamefont {M.}~\bibnamefont {Kalaee}}, \bibinfo {author} {\bibfnamefont
  {L.}~\bibnamefont {DeLorenzo}}, \bibinfo {author} {\bibfnamefont {E.~A.}\
  \bibnamefont {Wollack}}, \bibinfo {author} {\bibfnamefont {P.}~\bibnamefont
  {Arrangoiz-Arriola}}, \bibinfo {author} {\bibfnamefont {A.}~\bibnamefont
  {Khalajhedayati}}, \bibinfo {author} {\bibfnamefont {R.}~\bibnamefont
  {Sanil}}, \emph {et~al.},\ }\href@noop {} {\bibfield  {journal} {\bibinfo
  {journal} {Physical Review X}\ }\textbf {\bibinfo {volume} {14}},\ \bibinfo
  {pages} {011051} (\bibinfo {year} {2024})}\BibitemShut {NoStop}%
\bibitem [{\citenamefont {Koottandavida}\ \emph {et~al.}(2024)\citenamefont
  {Koottandavida}, \citenamefont {Tsioutsios}, \citenamefont {Kargioti},
  \citenamefont {Smith}, \citenamefont {Joshi}, \citenamefont {Dai},
  \citenamefont {Teoh}, \citenamefont {Curtis}, \citenamefont {Frunzio},
  \citenamefont {Schoelkopf} \emph {et~al.}}]{koottandavida2024erasure}%
  \BibitemOpen
  \bibfield  {author} {\bibinfo {author} {\bibfnamefont {A.}~\bibnamefont
  {Koottandavida}}, \bibinfo {author} {\bibfnamefont {I.}~\bibnamefont
  {Tsioutsios}}, \bibinfo {author} {\bibfnamefont {A.}~\bibnamefont
  {Kargioti}}, \bibinfo {author} {\bibfnamefont {C.~R.}\ \bibnamefont {Smith}},
  \bibinfo {author} {\bibfnamefont {V.~R.}\ \bibnamefont {Joshi}}, \bibinfo
  {author} {\bibfnamefont {W.}~\bibnamefont {Dai}}, \bibinfo {author}
  {\bibfnamefont {J.~D.}\ \bibnamefont {Teoh}}, \bibinfo {author}
  {\bibfnamefont {J.~C.}\ \bibnamefont {Curtis}}, \bibinfo {author}
  {\bibfnamefont {L.}~\bibnamefont {Frunzio}}, \bibinfo {author} {\bibfnamefont
  {R.~J.}\ \bibnamefont {Schoelkopf}}, \emph {et~al.},\ }\href@noop {}
  {\bibfield  {journal} {\bibinfo  {journal} {Physical Review Letters}\
  }\textbf {\bibinfo {volume} {132}},\ \bibinfo {pages} {180601} (\bibinfo
  {year} {2024})}\BibitemShut {NoStop}%
\bibitem [{\citenamefont {DiVincenzo}\ \emph {et~al.}(2006)\citenamefont
  {DiVincenzo}, \citenamefont {Brito},\ and\ \citenamefont
  {Koch}}]{divincenzo2006decoherence}%
  \BibitemOpen
  \bibfield  {author} {\bibinfo {author} {\bibfnamefont {D.~P.}\ \bibnamefont
  {DiVincenzo}}, \bibinfo {author} {\bibfnamefont {F.}~\bibnamefont {Brito}},\
  and\ \bibinfo {author} {\bibfnamefont {R.~H.}\ \bibnamefont {Koch}},\
  }\href@noop {} {\bibfield  {journal} {\bibinfo  {journal} {Physical Review
  B—Condensed Matter and Materials Physics}\ }\textbf {\bibinfo {volume}
  {74}},\ \bibinfo {pages} {014514} (\bibinfo {year} {2006})}\BibitemShut
  {NoStop}%
\bibitem [{\citenamefont {Rymarz}\ and\ \citenamefont
  {DiVincenzo}(2023)}]{rymarz2023consistent}%
  \BibitemOpen
  \bibfield  {author} {\bibinfo {author} {\bibfnamefont {M.}~\bibnamefont
  {Rymarz}}\ and\ \bibinfo {author} {\bibfnamefont {D.~P.}\ \bibnamefont
  {DiVincenzo}},\ }\href@noop {} {\bibfield  {journal} {\bibinfo  {journal}
  {Physical Review X}\ }\textbf {\bibinfo {volume} {13}},\ \bibinfo {pages}
  {021017} (\bibinfo {year} {2023})}\BibitemShut {NoStop}%
\bibitem [{\citenamefont {Egusquiza}\ and\ \citenamefont
  {Parra-Rodriguez}(2024)}]{egusquiza2024consistent}%
  \BibitemOpen
  \bibfield  {author} {\bibinfo {author} {\bibfnamefont {I.}~\bibnamefont
  {Egusquiza}}\ and\ \bibinfo {author} {\bibfnamefont {A.}~\bibnamefont
  {Parra-Rodriguez}},\ }\href@noop {} {\bibfield  {journal} {\bibinfo
  {journal} {arXiv preprint arXiv:2408.05174}\ } (\bibinfo {year}
  {2024})}\BibitemShut {NoStop}%
\bibitem [{\citenamefont {Gustavsson}\ \emph {et~al.}(2011)\citenamefont
  {Gustavsson}, \citenamefont {Bylander}, \citenamefont {Yan}, \citenamefont
  {Oliver}, \citenamefont {Yoshihara},\ and\ \citenamefont
  {Nakamura}}]{gustavsson2011noise}%
  \BibitemOpen
  \bibfield  {author} {\bibinfo {author} {\bibfnamefont {S.}~\bibnamefont
  {Gustavsson}}, \bibinfo {author} {\bibfnamefont {J.}~\bibnamefont
  {Bylander}}, \bibinfo {author} {\bibfnamefont {F.}~\bibnamefont {Yan}},
  \bibinfo {author} {\bibfnamefont {W.~D.}\ \bibnamefont {Oliver}}, \bibinfo
  {author} {\bibfnamefont {F.}~\bibnamefont {Yoshihara}},\ and\ \bibinfo
  {author} {\bibfnamefont {Y.}~\bibnamefont {Nakamura}},\ }\href@noop {}
  {\bibfield  {journal} {\bibinfo  {journal} {Physical Review B—Condensed
  Matter and Materials Physics}\ }\textbf {\bibinfo {volume} {84}},\ \bibinfo
  {pages} {014525} (\bibinfo {year} {2011})}\BibitemShut {NoStop}%
\bibitem [{\citenamefont {Kou}\ \emph {et~al.}(2017)\citenamefont {Kou},
  \citenamefont {Smith}, \citenamefont {Vool}, \citenamefont {Brierley},
  \citenamefont {Meier}, \citenamefont {Frunzio}, \citenamefont {Girvin},
  \citenamefont {Glazman},\ and\ \citenamefont {Devoret}}]{kou2017fluxonium}%
  \BibitemOpen
  \bibfield  {author} {\bibinfo {author} {\bibfnamefont {A.}~\bibnamefont
  {Kou}}, \bibinfo {author} {\bibfnamefont {W.}~\bibnamefont {Smith}}, \bibinfo
  {author} {\bibfnamefont {U.}~\bibnamefont {Vool}}, \bibinfo {author}
  {\bibfnamefont {R.}~\bibnamefont {Brierley}}, \bibinfo {author}
  {\bibfnamefont {H.}~\bibnamefont {Meier}}, \bibinfo {author} {\bibfnamefont
  {L.}~\bibnamefont {Frunzio}}, \bibinfo {author} {\bibfnamefont
  {S.}~\bibnamefont {Girvin}}, \bibinfo {author} {\bibfnamefont
  {L.}~\bibnamefont {Glazman}},\ and\ \bibinfo {author} {\bibfnamefont
  {M.}~\bibnamefont {Devoret}},\ }\href@noop {} {\bibfield  {journal} {\bibinfo
   {journal} {Physical Review X}\ }\textbf {\bibinfo {volume} {7}},\ \bibinfo
  {pages} {031037} (\bibinfo {year} {2017})}\BibitemShut {NoStop}%
\bibitem [{\citenamefont {Braum{\"u}ller}\ \emph {et~al.}(2020)\citenamefont
  {Braum{\"u}ller}, \citenamefont {Ding}, \citenamefont {Veps{\"a}l{\"a}inen},
  \citenamefont {Sung}, \citenamefont {Kjaergaard}, \citenamefont {Menke},
  \citenamefont {Winik}, \citenamefont {Kim}, \citenamefont {Niedzielski},
  \citenamefont {Melville} \emph {et~al.}}]{braumuller2020characterizing}%
  \BibitemOpen
  \bibfield  {author} {\bibinfo {author} {\bibfnamefont {J.}~\bibnamefont
  {Braum{\"u}ller}}, \bibinfo {author} {\bibfnamefont {L.}~\bibnamefont
  {Ding}}, \bibinfo {author} {\bibfnamefont {A.~P.}\ \bibnamefont
  {Veps{\"a}l{\"a}inen}}, \bibinfo {author} {\bibfnamefont {Y.}~\bibnamefont
  {Sung}}, \bibinfo {author} {\bibfnamefont {M.}~\bibnamefont {Kjaergaard}},
  \bibinfo {author} {\bibfnamefont {T.}~\bibnamefont {Menke}}, \bibinfo
  {author} {\bibfnamefont {R.}~\bibnamefont {Winik}}, \bibinfo {author}
  {\bibfnamefont {D.}~\bibnamefont {Kim}}, \bibinfo {author} {\bibfnamefont
  {B.~M.}\ \bibnamefont {Niedzielski}}, \bibinfo {author} {\bibfnamefont
  {A.}~\bibnamefont {Melville}}, \emph {et~al.},\ }\href@noop {} {\bibfield
  {journal} {\bibinfo  {journal} {Physical Review Applied}\ }\textbf {\bibinfo
  {volume} {13}},\ \bibinfo {pages} {054079} (\bibinfo {year}
  {2020})}\BibitemShut {NoStop}%
\bibitem [{\citenamefont {Van Der~Zant}\ \emph {et~al.}(1994)\citenamefont {Van
  Der~Zant}, \citenamefont {Receveur}, \citenamefont {Orlando},\ and\
  \citenamefont {Kleinsasser}}]{van1994one}%
  \BibitemOpen
  \bibfield  {author} {\bibinfo {author} {\bibfnamefont {H.}~\bibnamefont {Van
  Der~Zant}}, \bibinfo {author} {\bibfnamefont {R.}~\bibnamefont {Receveur}},
  \bibinfo {author} {\bibfnamefont {T.}~\bibnamefont {Orlando}},\ and\ \bibinfo
  {author} {\bibfnamefont {A.}~\bibnamefont {Kleinsasser}},\ }\href@noop {}
  {\bibfield  {journal} {\bibinfo  {journal} {Applied physics letters}\
  }\textbf {\bibinfo {volume} {65}},\ \bibinfo {pages} {2102} (\bibinfo {year}
  {1994})}\BibitemShut {NoStop}%
\bibitem [{\citenamefont {Randeria}\ \emph {et~al.}(2024)\citenamefont
  {Randeria}, \citenamefont {Hazard}, \citenamefont {Di~Paolo}, \citenamefont
  {Azar}, \citenamefont {Hays}, \citenamefont {Ding}, \citenamefont {An},
  \citenamefont {Gingras}, \citenamefont {Niedzielski}, \citenamefont
  {Stickler} \emph {et~al.}}]{randeria2024dephasing}%
  \BibitemOpen
  \bibfield  {author} {\bibinfo {author} {\bibfnamefont {M.~T.}\ \bibnamefont
  {Randeria}}, \bibinfo {author} {\bibfnamefont {T.~M.}\ \bibnamefont
  {Hazard}}, \bibinfo {author} {\bibfnamefont {A.}~\bibnamefont {Di~Paolo}},
  \bibinfo {author} {\bibfnamefont {K.}~\bibnamefont {Azar}}, \bibinfo {author}
  {\bibfnamefont {M.}~\bibnamefont {Hays}}, \bibinfo {author} {\bibfnamefont
  {L.}~\bibnamefont {Ding}}, \bibinfo {author} {\bibfnamefont {J.}~\bibnamefont
  {An}}, \bibinfo {author} {\bibfnamefont {M.}~\bibnamefont {Gingras}},
  \bibinfo {author} {\bibfnamefont {B.~M.}\ \bibnamefont {Niedzielski}},
  \bibinfo {author} {\bibfnamefont {H.}~\bibnamefont {Stickler}}, \emph
  {et~al.},\ }\href@noop {} {\bibfield  {journal} {\bibinfo  {journal} {PRX
  Quantum}\ }\textbf {\bibinfo {volume} {5}},\ \bibinfo {pages} {030341}
  (\bibinfo {year} {2024})}\BibitemShut {NoStop}%
\bibitem [{\citenamefont {Moskalev}\ \emph {et~al.}(2023)\citenamefont
  {Moskalev}, \citenamefont {Zikiy}, \citenamefont {Pishchimova}, \citenamefont
  {Ezenkova}, \citenamefont {Smirnov}, \citenamefont {Ivanov}, \citenamefont
  {Korshakov},\ and\ \citenamefont {Rodionov}}]{moskalev2023optimization}%
  \BibitemOpen
  \bibfield  {author} {\bibinfo {author} {\bibfnamefont {D.~O.}\ \bibnamefont
  {Moskalev}}, \bibinfo {author} {\bibfnamefont {E.~V.}\ \bibnamefont {Zikiy}},
  \bibinfo {author} {\bibfnamefont {A.~A.}\ \bibnamefont {Pishchimova}},
  \bibinfo {author} {\bibfnamefont {D.~A.}\ \bibnamefont {Ezenkova}}, \bibinfo
  {author} {\bibfnamefont {N.~S.}\ \bibnamefont {Smirnov}}, \bibinfo {author}
  {\bibfnamefont {A.~I.}\ \bibnamefont {Ivanov}}, \bibinfo {author}
  {\bibfnamefont {N.~D.}\ \bibnamefont {Korshakov}},\ and\ \bibinfo {author}
  {\bibfnamefont {I.~A.}\ \bibnamefont {Rodionov}},\ }\href@noop {} {\bibfield
  {journal} {\bibinfo  {journal} {Scientific Reports}\ }\textbf {\bibinfo
  {volume} {13}},\ \bibinfo {pages} {4174} (\bibinfo {year}
  {2023})}\BibitemShut {NoStop}%
\bibitem [{\citenamefont {Anferov}\ \emph {et~al.}(2024)\citenamefont
  {Anferov}, \citenamefont {Harvey}, \citenamefont {Wan}, \citenamefont
  {Simon},\ and\ \citenamefont {Schuster}}]{anferov2024superconducting}%
  \BibitemOpen
  \bibfield  {author} {\bibinfo {author} {\bibfnamefont {A.}~\bibnamefont
  {Anferov}}, \bibinfo {author} {\bibfnamefont {S.~P.}\ \bibnamefont {Harvey}},
  \bibinfo {author} {\bibfnamefont {F.}~\bibnamefont {Wan}}, \bibinfo {author}
  {\bibfnamefont {J.}~\bibnamefont {Simon}},\ and\ \bibinfo {author}
  {\bibfnamefont {D.~I.}\ \bibnamefont {Schuster}},\ }\href@noop {} {\bibfield
  {journal} {\bibinfo  {journal} {PRX Quantum}\ }\textbf {\bibinfo {volume}
  {5}},\ \bibinfo {pages} {030347} (\bibinfo {year} {2024})}\BibitemShut
  {NoStop}%
\bibitem [{\citenamefont {De~Lange}\ \emph {et~al.}(2015)\citenamefont
  {De~Lange}, \citenamefont {Van~Heck}, \citenamefont {Bruno}, \citenamefont
  {Van~Woerkom}, \citenamefont {Geresdi}, \citenamefont {Plissard},
  \citenamefont {Bakkers}, \citenamefont {Akhmerov},\ and\ \citenamefont
  {DiCarlo}}]{de2015realization}%
  \BibitemOpen
  \bibfield  {author} {\bibinfo {author} {\bibfnamefont {G.}~\bibnamefont
  {De~Lange}}, \bibinfo {author} {\bibfnamefont {B.}~\bibnamefont {Van~Heck}},
  \bibinfo {author} {\bibfnamefont {A.}~\bibnamefont {Bruno}}, \bibinfo
  {author} {\bibfnamefont {D.}~\bibnamefont {Van~Woerkom}}, \bibinfo {author}
  {\bibfnamefont {A.}~\bibnamefont {Geresdi}}, \bibinfo {author} {\bibfnamefont
  {S.}~\bibnamefont {Plissard}}, \bibinfo {author} {\bibfnamefont
  {E.}~\bibnamefont {Bakkers}}, \bibinfo {author} {\bibfnamefont
  {A.}~\bibnamefont {Akhmerov}},\ and\ \bibinfo {author} {\bibfnamefont
  {L.}~\bibnamefont {DiCarlo}},\ }\href@noop {} {\bibfield  {journal} {\bibinfo
   {journal} {Physical review letters}\ }\textbf {\bibinfo {volume} {115}},\
  \bibinfo {pages} {127002} (\bibinfo {year} {2015})}\BibitemShut {NoStop}%
\bibitem [{\citenamefont {Larsen}\ \emph {et~al.}(2015)\citenamefont {Larsen},
  \citenamefont {Petersson}, \citenamefont {Kuemmeth}, \citenamefont
  {Jespersen}, \citenamefont {Krogstrup}, \citenamefont {Nyg{\aa}rd},\ and\
  \citenamefont {Marcus}}]{larsen2015semiconductor}%
  \BibitemOpen
  \bibfield  {author} {\bibinfo {author} {\bibfnamefont {T.~W.}\ \bibnamefont
  {Larsen}}, \bibinfo {author} {\bibfnamefont {K.~D.}\ \bibnamefont
  {Petersson}}, \bibinfo {author} {\bibfnamefont {F.}~\bibnamefont {Kuemmeth}},
  \bibinfo {author} {\bibfnamefont {T.~S.}\ \bibnamefont {Jespersen}}, \bibinfo
  {author} {\bibfnamefont {P.}~\bibnamefont {Krogstrup}}, \bibinfo {author}
  {\bibfnamefont {J.}~\bibnamefont {Nyg{\aa}rd}},\ and\ \bibinfo {author}
  {\bibfnamefont {C.~M.}\ \bibnamefont {Marcus}},\ }\href@noop {} {\bibfield
  {journal} {\bibinfo  {journal} {Physical review letters}\ }\textbf {\bibinfo
  {volume} {115}},\ \bibinfo {pages} {127001} (\bibinfo {year}
  {2015})}\BibitemShut {NoStop}%
\bibitem [{\citenamefont {Casparis}\ \emph {et~al.}(2018)\citenamefont
  {Casparis}, \citenamefont {Connolly}, \citenamefont {Kjaergaard},
  \citenamefont {Pearson}, \citenamefont {Kringh{\o}j}, \citenamefont {Larsen},
  \citenamefont {Kuemmeth}, \citenamefont {Wang}, \citenamefont {Thomas},
  \citenamefont {Gronin} \emph {et~al.}}]{casparis2018superconducting}%
  \BibitemOpen
  \bibfield  {author} {\bibinfo {author} {\bibfnamefont {L.}~\bibnamefont
  {Casparis}}, \bibinfo {author} {\bibfnamefont {M.~R.}\ \bibnamefont
  {Connolly}}, \bibinfo {author} {\bibfnamefont {M.}~\bibnamefont
  {Kjaergaard}}, \bibinfo {author} {\bibfnamefont {N.~J.}\ \bibnamefont
  {Pearson}}, \bibinfo {author} {\bibfnamefont {A.}~\bibnamefont
  {Kringh{\o}j}}, \bibinfo {author} {\bibfnamefont {T.~W.}\ \bibnamefont
  {Larsen}}, \bibinfo {author} {\bibfnamefont {F.}~\bibnamefont {Kuemmeth}},
  \bibinfo {author} {\bibfnamefont {T.}~\bibnamefont {Wang}}, \bibinfo {author}
  {\bibfnamefont {C.}~\bibnamefont {Thomas}}, \bibinfo {author} {\bibfnamefont
  {S.}~\bibnamefont {Gronin}}, \emph {et~al.},\ }\href@noop {} {\bibfield
  {journal} {\bibinfo  {journal} {Nature nanotechnology}\ }\textbf {\bibinfo
  {volume} {13}},\ \bibinfo {pages} {915} (\bibinfo {year} {2018})}\BibitemShut
  {NoStop}%
\bibitem [{\citenamefont {Paolo}\ \emph {et~al.}(2019)\citenamefont {Paolo},
  \citenamefont {Grimsmo}, \citenamefont {Groszkowski}, \citenamefont {Koch},\
  and\ \citenamefont {Blais}}]{paolo2019control}%
  \BibitemOpen
  \bibfield  {author} {\bibinfo {author} {\bibfnamefont {A.~D.}\ \bibnamefont
  {Paolo}}, \bibinfo {author} {\bibfnamefont {A.~L.}\ \bibnamefont {Grimsmo}},
  \bibinfo {author} {\bibfnamefont {P.}~\bibnamefont {Groszkowski}}, \bibinfo
  {author} {\bibfnamefont {J.}~\bibnamefont {Koch}},\ and\ \bibinfo {author}
  {\bibfnamefont {A.}~\bibnamefont {Blais}},\ }\href@noop {} {\bibfield
  {journal} {\bibinfo  {journal} {New Journal of Physics}\ }\textbf {\bibinfo
  {volume} {21}},\ \bibinfo {pages} {043002} (\bibinfo {year}
  {2019})}\BibitemShut {NoStop}%
\bibitem [{\citenamefont {Harrington}\ \emph {et~al.}(2022)\citenamefont
  {Harrington}, \citenamefont {Mueller},\ and\ \citenamefont
  {Murch}}]{harrington2022engineered}%
  \BibitemOpen
  \bibfield  {author} {\bibinfo {author} {\bibfnamefont {P.~M.}\ \bibnamefont
  {Harrington}}, \bibinfo {author} {\bibfnamefont {E.~J.}\ \bibnamefont
  {Mueller}},\ and\ \bibinfo {author} {\bibfnamefont {K.~W.}\ \bibnamefont
  {Murch}},\ }\href@noop {} {\bibfield  {journal} {\bibinfo  {journal} {Nature
  Reviews Physics}\ }\textbf {\bibinfo {volume} {4}},\ \bibinfo {pages} {660}
  (\bibinfo {year} {2022})}\BibitemShut {NoStop}%
\bibitem [{\citenamefont {Premkumar}(2023)}]{premkumar2023hamiltonian}%
  \BibitemOpen
  \bibfield  {author} {\bibinfo {author} {\bibfnamefont {A.}~\bibnamefont
  {Premkumar}},\ }\emph {\bibinfo {title} {Hamiltonian and Materials
  Engineering for Superconducting Qubit Lifetime Enhancement}},\ \href@noop {}
  {Ph.D. thesis},\ \bibinfo  {school} {Princeton University} (\bibinfo {year}
  {2023})\BibitemShut {NoStop}%
\bibitem [{\citenamefont {Malherbe}\ and\ \citenamefont
  {Vayatis}(2017)}]{malherbe2017global}%
  \BibitemOpen
  \bibfield  {author} {\bibinfo {author} {\bibfnamefont {C.}~\bibnamefont
  {Malherbe}}\ and\ \bibinfo {author} {\bibfnamefont {N.}~\bibnamefont
  {Vayatis}},\ }in\ \href@noop {} {\emph {\bibinfo {booktitle} {International
  Conference on Machine Learning}}}\ (\bibinfo {organization} {PMLR},\ \bibinfo
  {year} {2017})\ pp.\ \bibinfo {pages} {2314--2323}\BibitemShut {NoStop}%
\bibitem [{\citenamefont {Beitner}()}]{pyLIPO}%
  \BibitemOpen
  \bibfield  {author} {\bibinfo {author} {\bibfnamefont {J.}~\bibnamefont
  {Beitner}},\ }\href@noop {} {\bibinfo {title} {Lipo python implementation}},\
  \bibinfo {howpublished} {\url{https://github.com/jdb78/lipo}}\BibitemShut
  {NoStop}%
\bibitem [{\citenamefont {Van~Harlingen}\ \emph {et~al.}(2004)\citenamefont
  {Van~Harlingen}, \citenamefont {Robertson}, \citenamefont {Plourde},
  \citenamefont {Reichardt}, \citenamefont {Crane},\ and\ \citenamefont
  {Clarke}}]{van2004decoherence}%
  \BibitemOpen
  \bibfield  {author} {\bibinfo {author} {\bibfnamefont {D.}~\bibnamefont
  {Van~Harlingen}}, \bibinfo {author} {\bibfnamefont {T.}~\bibnamefont
  {Robertson}}, \bibinfo {author} {\bibfnamefont {B.}~\bibnamefont {Plourde}},
  \bibinfo {author} {\bibfnamefont {P.}~\bibnamefont {Reichardt}}, \bibinfo
  {author} {\bibfnamefont {T.}~\bibnamefont {Crane}},\ and\ \bibinfo {author}
  {\bibfnamefont {J.}~\bibnamefont {Clarke}},\ }\href@noop {} {\bibfield
  {journal} {\bibinfo  {journal} {Physical Review B—Condensed Matter and
  Materials Physics}\ }\textbf {\bibinfo {volume} {70}},\ \bibinfo {pages}
  {064517} (\bibinfo {year} {2004})}\BibitemShut {NoStop}%
\bibitem [{\citenamefont {Wang}\ \emph {et~al.}(2019)\citenamefont {Wang},
  \citenamefont {Shankar}, \citenamefont {Minev}, \citenamefont
  {Campagne-Ibarcq}, \citenamefont {Narla},\ and\ \citenamefont
  {Devoret}}]{wang2019cavity}%
  \BibitemOpen
  \bibfield  {author} {\bibinfo {author} {\bibfnamefont {Z.}~\bibnamefont
  {Wang}}, \bibinfo {author} {\bibfnamefont {S.}~\bibnamefont {Shankar}},
  \bibinfo {author} {\bibfnamefont {Z.}~\bibnamefont {Minev}}, \bibinfo
  {author} {\bibfnamefont {P.}~\bibnamefont {Campagne-Ibarcq}}, \bibinfo
  {author} {\bibfnamefont {A.}~\bibnamefont {Narla}},\ and\ \bibinfo {author}
  {\bibfnamefont {M.~H.}\ \bibnamefont {Devoret}},\ }\href@noop {} {\bibfield
  {journal} {\bibinfo  {journal} {Physical Review Applied}\ }\textbf {\bibinfo
  {volume} {11}},\ \bibinfo {pages} {014031} (\bibinfo {year}
  {2019})}\BibitemShut {NoStop}%
\bibitem [{\citenamefont {Bylander}\ \emph {et~al.}(2011)\citenamefont
  {Bylander}, \citenamefont {Gustavsson}, \citenamefont {Yan}, \citenamefont
  {Yoshihara}, \citenamefont {Harrabi}, \citenamefont {Fitch}, \citenamefont
  {Cory}, \citenamefont {Nakamura}, \citenamefont {Tsai},\ and\ \citenamefont
  {Oliver}}]{bylander2011noise}%
  \BibitemOpen
  \bibfield  {author} {\bibinfo {author} {\bibfnamefont {J.}~\bibnamefont
  {Bylander}}, \bibinfo {author} {\bibfnamefont {S.}~\bibnamefont
  {Gustavsson}}, \bibinfo {author} {\bibfnamefont {F.}~\bibnamefont {Yan}},
  \bibinfo {author} {\bibfnamefont {F.}~\bibnamefont {Yoshihara}}, \bibinfo
  {author} {\bibfnamefont {K.}~\bibnamefont {Harrabi}}, \bibinfo {author}
  {\bibfnamefont {G.}~\bibnamefont {Fitch}}, \bibinfo {author} {\bibfnamefont
  {D.~G.}\ \bibnamefont {Cory}}, \bibinfo {author} {\bibfnamefont
  {Y.}~\bibnamefont {Nakamura}}, \bibinfo {author} {\bibfnamefont {J.-S.}\
  \bibnamefont {Tsai}},\ and\ \bibinfo {author} {\bibfnamefont {W.~D.}\
  \bibnamefont {Oliver}},\ }\href@noop {} {\bibfield  {journal} {\bibinfo
  {journal} {Nature Physics}\ }\textbf {\bibinfo {volume} {7}},\ \bibinfo
  {pages} {565} (\bibinfo {year} {2011})}\BibitemShut {NoStop}%
\bibitem [{\citenamefont {Rower}\ \emph {et~al.}(2024)\citenamefont {Rower},
  \citenamefont {Ding}, \citenamefont {Zhang}, \citenamefont {Hays},
  \citenamefont {An}, \citenamefont {Harrington}, \citenamefont {Rosen},
  \citenamefont {Gertler}, \citenamefont {Hazard}, \citenamefont {Niedzielski}
  \emph {et~al.}}]{rower2024suppressing}%
  \BibitemOpen
  \bibfield  {author} {\bibinfo {author} {\bibfnamefont {D.~A.}\ \bibnamefont
  {Rower}}, \bibinfo {author} {\bibfnamefont {L.}~\bibnamefont {Ding}},
  \bibinfo {author} {\bibfnamefont {H.}~\bibnamefont {Zhang}}, \bibinfo
  {author} {\bibfnamefont {M.}~\bibnamefont {Hays}}, \bibinfo {author}
  {\bibfnamefont {J.}~\bibnamefont {An}}, \bibinfo {author} {\bibfnamefont
  {P.~M.}\ \bibnamefont {Harrington}}, \bibinfo {author} {\bibfnamefont
  {I.~T.}\ \bibnamefont {Rosen}}, \bibinfo {author} {\bibfnamefont {J.~M.}\
  \bibnamefont {Gertler}}, \bibinfo {author} {\bibfnamefont {T.~M.}\
  \bibnamefont {Hazard}}, \bibinfo {author} {\bibfnamefont {B.~M.}\
  \bibnamefont {Niedzielski}}, \emph {et~al.},\ }\href@noop {} {\bibfield
  {journal} {\bibinfo  {journal} {arXiv preprint arXiv:2406.08295}\ } (\bibinfo
  {year} {2024})}\BibitemShut {NoStop}%
\bibitem [{\citenamefont {Zhang}\ \emph {et~al.}(2021)\citenamefont {Zhang},
  \citenamefont {Chakram}, \citenamefont {Roy}, \citenamefont {Earnest},
  \citenamefont {Lu}, \citenamefont {Huang}, \citenamefont {Weiss},
  \citenamefont {Koch},\ and\ \citenamefont {Schuster}}]{zhang2021universal}%
  \BibitemOpen
  \bibfield  {author} {\bibinfo {author} {\bibfnamefont {H.}~\bibnamefont
  {Zhang}}, \bibinfo {author} {\bibfnamefont {S.}~\bibnamefont {Chakram}},
  \bibinfo {author} {\bibfnamefont {T.}~\bibnamefont {Roy}}, \bibinfo {author}
  {\bibfnamefont {N.}~\bibnamefont {Earnest}}, \bibinfo {author} {\bibfnamefont
  {Y.}~\bibnamefont {Lu}}, \bibinfo {author} {\bibfnamefont {Z.}~\bibnamefont
  {Huang}}, \bibinfo {author} {\bibfnamefont {D.}~\bibnamefont {Weiss}},
  \bibinfo {author} {\bibfnamefont {J.}~\bibnamefont {Koch}},\ and\ \bibinfo
  {author} {\bibfnamefont {D.~I.}\ \bibnamefont {Schuster}},\ }\href@noop {}
  {\bibfield  {journal} {\bibinfo  {journal} {Physical Review X}\ }\textbf
  {\bibinfo {volume} {11}},\ \bibinfo {pages} {011010} (\bibinfo {year}
  {2021})}\BibitemShut {NoStop}%
\bibitem [{\citenamefont {You}\ \emph {et~al.}(2019)\citenamefont {You},
  \citenamefont {Sauls},\ and\ \citenamefont {Koch}}]{you2019circuit}%
  \BibitemOpen
  \bibfield  {author} {\bibinfo {author} {\bibfnamefont {X.}~\bibnamefont
  {You}}, \bibinfo {author} {\bibfnamefont {J.~A.}\ \bibnamefont {Sauls}},\
  and\ \bibinfo {author} {\bibfnamefont {J.}~\bibnamefont {Koch}},\ }\href@noop
  {} {\bibfield  {journal} {\bibinfo  {journal} {Physical Review B}\ }\textbf
  {\bibinfo {volume} {99}},\ \bibinfo {pages} {174512} (\bibinfo {year}
  {2019})}\BibitemShut {NoStop}%
\bibitem [{\citenamefont {Groszkowski}\ \emph {et~al.}(2023)\citenamefont
  {Groszkowski}, \citenamefont {Seif}, \citenamefont {Koch},\ and\
  \citenamefont {Clerk}}]{groszkowski2023simple}%
  \BibitemOpen
  \bibfield  {author} {\bibinfo {author} {\bibfnamefont {P.}~\bibnamefont
  {Groszkowski}}, \bibinfo {author} {\bibfnamefont {A.}~\bibnamefont {Seif}},
  \bibinfo {author} {\bibfnamefont {J.}~\bibnamefont {Koch}},\ and\ \bibinfo
  {author} {\bibfnamefont {A.}~\bibnamefont {Clerk}},\ }\href@noop {}
  {\bibfield  {journal} {\bibinfo  {journal} {Quantum}\ }\textbf {\bibinfo
  {volume} {7}},\ \bibinfo {pages} {972} (\bibinfo {year} {2023})}\BibitemShut
  {NoStop}%
\bibitem [{\citenamefont {Pedersen}\ \emph {et~al.}(2007)\citenamefont
  {Pedersen}, \citenamefont {M{\o}ller},\ and\ \citenamefont
  {M{\o}lmer}}]{pedersen2007fidelity}%
  \BibitemOpen
  \bibfield  {author} {\bibinfo {author} {\bibfnamefont {L.~H.}\ \bibnamefont
  {Pedersen}}, \bibinfo {author} {\bibfnamefont {N.~M.}\ \bibnamefont
  {M{\o}ller}},\ and\ \bibinfo {author} {\bibfnamefont {K.}~\bibnamefont
  {M{\o}lmer}},\ }\href@noop {} {\bibfield  {journal} {\bibinfo  {journal}
  {Physics Letters A}\ }\textbf {\bibinfo {volume} {367}},\ \bibinfo {pages}
  {47} (\bibinfo {year} {2007})}\BibitemShut {NoStop}%
\bibitem [{\citenamefont {Goldschmidt}\ \emph {et~al.}(2025)\citenamefont
  {Goldschmidt}, \citenamefont {Pel\'{a}ez~Cisneros}, \citenamefont {Sitler},
  \citenamefont {Olsson}, \citenamefont {Smith},\ and\ \citenamefont
  {Quiroz}}]{goldschmidt2025crosstalk}%
  \BibitemOpen
  \bibfield  {author} {\bibinfo {author} {\bibfnamefont {A.~J.}\ \bibnamefont
  {Goldschmidt}}, \bibinfo {author} {\bibfnamefont {E.}~\bibnamefont
  {Pel\'{a}ez~Cisneros}}, \bibinfo {author} {\bibfnamefont {R.}~\bibnamefont
  {Sitler}}, \bibinfo {author} {\bibfnamefont {K.}~\bibnamefont {Olsson}},
  \bibinfo {author} {\bibfnamefont {K.~N.}\ \bibnamefont {Smith}},\ and\
  \bibinfo {author} {\bibfnamefont {G.}~\bibnamefont {Quiroz}}} (\bibinfo
  {year} {2025})\BibitemShut {NoStop}%
\bibitem [{\citenamefont {Gambetta}\ \emph {et~al.}(2006)\citenamefont
  {Gambetta}, \citenamefont {Blais}, \citenamefont {Schuster}, \citenamefont
  {Wallraff}, \citenamefont {Frunzio}, \citenamefont {Majer}, \citenamefont
  {Devoret}, \citenamefont {Girvin},\ and\ \citenamefont
  {Schoelkopf}}]{gambetta2006qubit}%
  \BibitemOpen
  \bibfield  {author} {\bibinfo {author} {\bibfnamefont {J.}~\bibnamefont
  {Gambetta}}, \bibinfo {author} {\bibfnamefont {A.}~\bibnamefont {Blais}},
  \bibinfo {author} {\bibfnamefont {D.~I.}\ \bibnamefont {Schuster}}, \bibinfo
  {author} {\bibfnamefont {A.}~\bibnamefont {Wallraff}}, \bibinfo {author}
  {\bibfnamefont {L.}~\bibnamefont {Frunzio}}, \bibinfo {author} {\bibfnamefont
  {J.}~\bibnamefont {Majer}}, \bibinfo {author} {\bibfnamefont {M.~H.}\
  \bibnamefont {Devoret}}, \bibinfo {author} {\bibfnamefont {S.~M.}\
  \bibnamefont {Girvin}},\ and\ \bibinfo {author} {\bibfnamefont {R.~J.}\
  \bibnamefont {Schoelkopf}},\ }\href@noop {} {\bibfield  {journal} {\bibinfo
  {journal} {Physical Review A—Atomic, Molecular, and Optical Physics}\
  }\textbf {\bibinfo {volume} {74}},\ \bibinfo {pages} {042318} (\bibinfo
  {year} {2006})}\BibitemShut {NoStop}%
\bibitem [{\citenamefont {Manucharyan}(2012)}]{manucharyanThesis}%
  \BibitemOpen
  \bibfield  {author} {\bibinfo {author} {\bibfnamefont {V.~E.}\ \bibnamefont
  {Manucharyan}},\ }\emph {\bibinfo {title} {Superinductance}},\ \href@noop {}
  {Ph.D. thesis},\ \bibinfo  {school} {Yale University} (\bibinfo {year}
  {2012})\BibitemShut {NoStop}%
\bibitem [{\citenamefont {Rajabzadeh}\ \emph {et~al.}(2024)\citenamefont
  {Rajabzadeh}, \citenamefont {Boulton-McKeehan}, \citenamefont {Bonkowsky},
  \citenamefont {Schuster},\ and\ \citenamefont
  {Safavi-Naeini}}]{rajabzadeh2024general}%
  \BibitemOpen
  \bibfield  {author} {\bibinfo {author} {\bibfnamefont {T.}~\bibnamefont
  {Rajabzadeh}}, \bibinfo {author} {\bibfnamefont {A.}~\bibnamefont
  {Boulton-McKeehan}}, \bibinfo {author} {\bibfnamefont {S.}~\bibnamefont
  {Bonkowsky}}, \bibinfo {author} {\bibfnamefont {D.~I.}\ \bibnamefont
  {Schuster}},\ and\ \bibinfo {author} {\bibfnamefont {A.~H.}\ \bibnamefont
  {Safavi-Naeini}},\ }\href@noop {} {\bibfield  {journal} {\bibinfo  {journal}
  {arXiv preprint arXiv:2408.12704}\ } (\bibinfo {year} {2024})}\BibitemShut
  {NoStop}%
\bibitem [{\citenamefont {Thibodeau}\ \emph {et~al.}(2024)\citenamefont
  {Thibodeau}, \citenamefont {Kou},\ and\ \citenamefont
  {Clark}}]{thibodeau2024floquet}%
  \BibitemOpen
  \bibfield  {author} {\bibinfo {author} {\bibfnamefont {M.}~\bibnamefont
  {Thibodeau}}, \bibinfo {author} {\bibfnamefont {A.}~\bibnamefont {Kou}},\
  and\ \bibinfo {author} {\bibfnamefont {B.~K.}\ \bibnamefont {Clark}},\
  }\href@noop {} {\bibfield  {journal} {\bibinfo  {journal} {PRX Quantum}\
  }\textbf {\bibinfo {volume} {5}},\ \bibinfo {pages} {040314} (\bibinfo {year}
  {2024})}\BibitemShut {NoStop}%
\bibitem [{\citenamefont {Kumar}\ \emph {et~al.}(2024)\citenamefont {Kumar},
  \citenamefont {You}, \citenamefont {Croot}, \citenamefont {Zhao},
  \citenamefont {Chen}, \citenamefont {Sussman}, \citenamefont {Premkumar},
  \citenamefont {Bryon}, \citenamefont {Koch},\ and\ \citenamefont
  {Houck}}]{kumar2024protomon}%
  \BibitemOpen
  \bibfield  {author} {\bibinfo {author} {\bibfnamefont {S.}~\bibnamefont
  {Kumar}}, \bibinfo {author} {\bibfnamefont {X.}~\bibnamefont {You}}, \bibinfo
  {author} {\bibfnamefont {X.}~\bibnamefont {Croot}}, \bibinfo {author}
  {\bibfnamefont {T.}~\bibnamefont {Zhao}}, \bibinfo {author} {\bibfnamefont
  {D.}~\bibnamefont {Chen}}, \bibinfo {author} {\bibfnamefont {S.}~\bibnamefont
  {Sussman}}, \bibinfo {author} {\bibfnamefont {A.}~\bibnamefont {Premkumar}},
  \bibinfo {author} {\bibfnamefont {J.}~\bibnamefont {Bryon}}, \bibinfo
  {author} {\bibfnamefont {J.}~\bibnamefont {Koch}},\ and\ \bibinfo {author}
  {\bibfnamefont {A.~A.}\ \bibnamefont {Houck}},\ }\href@noop {} {\bibfield
  {journal} {\bibinfo  {journal} {arXiv preprint arXiv:2411.16648}\ } (\bibinfo
  {year} {2024})}\BibitemShut {NoStop}%
\bibitem [{\citenamefont {Riwar}\ and\ \citenamefont
  {DiVincenzo}(2022)}]{riwar2022circuit}%
  \BibitemOpen
  \bibfield  {author} {\bibinfo {author} {\bibfnamefont {R.-P.}\ \bibnamefont
  {Riwar}}\ and\ \bibinfo {author} {\bibfnamefont {D.~P.}\ \bibnamefont
  {DiVincenzo}},\ }\href@noop {} {\bibfield  {journal} {\bibinfo  {journal}
  {npj Quantum Information}\ }\textbf {\bibinfo {volume} {8}},\ \bibinfo
  {pages} {36} (\bibinfo {year} {2022})}\BibitemShut {NoStop}%
\bibitem [{\citenamefont {Bryon}\ \emph {et~al.}(2022)\citenamefont {Bryon},
  \citenamefont {Weiss}, \citenamefont {You}, \citenamefont {Sussman},
  \citenamefont {Croot}, \citenamefont {Huang}, \citenamefont {Koch},\ and\
  \citenamefont {Houck}}]{bryon2022experimental}%
  \BibitemOpen
  \bibfield  {author} {\bibinfo {author} {\bibfnamefont {J.}~\bibnamefont
  {Bryon}}, \bibinfo {author} {\bibfnamefont {D.}~\bibnamefont {Weiss}},
  \bibinfo {author} {\bibfnamefont {X.}~\bibnamefont {You}}, \bibinfo {author}
  {\bibfnamefont {S.}~\bibnamefont {Sussman}}, \bibinfo {author} {\bibfnamefont
  {X.}~\bibnamefont {Croot}}, \bibinfo {author} {\bibfnamefont
  {Z.}~\bibnamefont {Huang}}, \bibinfo {author} {\bibfnamefont
  {J.}~\bibnamefont {Koch}},\ and\ \bibinfo {author} {\bibfnamefont
  {A.}~\bibnamefont {Houck}},\ }\href@noop {} {\bibfield  {journal} {\bibinfo
  {journal} {arXiv preprint arXiv:2208.03738}\ } (\bibinfo {year}
  {2022})}\BibitemShut {NoStop}%
\bibitem [{\citenamefont {Smith}(2019)}]{smith2019design}%
  \BibitemOpen
  \bibfield  {author} {\bibinfo {author} {\bibfnamefont {W.~C.}\ \bibnamefont
  {Smith}},\ }\emph {\bibinfo {title} {Design of protected superconducting
  qubits}},\ \href@noop {} {Ph.D. thesis},\ \bibinfo  {school} {Yale
  University} (\bibinfo {year} {2019})\BibitemShut {NoStop}%
\bibitem [{\citenamefont {Quintana}\ \emph {et~al.}(2017)\citenamefont
  {Quintana}, \citenamefont {Chen}, \citenamefont {Sank}, \citenamefont
  {Petukhov}, \citenamefont {White}, \citenamefont {Kafri}, \citenamefont
  {Chiaro}, \citenamefont {Megrant}, \citenamefont {Barends}, \citenamefont
  {Campbell} \emph {et~al.}}]{quintana2017observation}%
  \BibitemOpen
  \bibfield  {author} {\bibinfo {author} {\bibfnamefont {C.}~\bibnamefont
  {Quintana}}, \bibinfo {author} {\bibfnamefont {Y.}~\bibnamefont {Chen}},
  \bibinfo {author} {\bibfnamefont {D.}~\bibnamefont {Sank}}, \bibinfo {author}
  {\bibfnamefont {A.}~\bibnamefont {Petukhov}}, \bibinfo {author}
  {\bibfnamefont {T.}~\bibnamefont {White}}, \bibinfo {author} {\bibfnamefont
  {D.}~\bibnamefont {Kafri}}, \bibinfo {author} {\bibfnamefont
  {B.}~\bibnamefont {Chiaro}}, \bibinfo {author} {\bibfnamefont
  {A.}~\bibnamefont {Megrant}}, \bibinfo {author} {\bibfnamefont
  {R.}~\bibnamefont {Barends}}, \bibinfo {author} {\bibfnamefont
  {B.}~\bibnamefont {Campbell}}, \emph {et~al.},\ }\href@noop {} {\bibfield
  {journal} {\bibinfo  {journal} {Physical review letters}\ }\textbf {\bibinfo
  {volume} {118}},\ \bibinfo {pages} {057702} (\bibinfo {year}
  {2017})}\BibitemShut {NoStop}%
\bibitem [{\citenamefont {Catelani}\ \emph {et~al.}(2011)\citenamefont
  {Catelani}, \citenamefont {Schoelkopf}, \citenamefont {Devoret},\ and\
  \citenamefont {Glazman}}]{catelani2011relaxation}%
  \BibitemOpen
  \bibfield  {author} {\bibinfo {author} {\bibfnamefont {G.}~\bibnamefont
  {Catelani}}, \bibinfo {author} {\bibfnamefont {R.~J.}\ \bibnamefont
  {Schoelkopf}}, \bibinfo {author} {\bibfnamefont {M.~H.}\ \bibnamefont
  {Devoret}},\ and\ \bibinfo {author} {\bibfnamefont {L.~I.}\ \bibnamefont
  {Glazman}},\ }\href@noop {} {\bibfield  {journal} {\bibinfo  {journal}
  {Physical Review B—Condensed Matter and Materials Physics}\ }\textbf
  {\bibinfo {volume} {84}},\ \bibinfo {pages} {064517} (\bibinfo {year}
  {2011})}\BibitemShut {NoStop}%
\bibitem [{\citenamefont {Connolly}\ \emph {et~al.}(2024)\citenamefont
  {Connolly}, \citenamefont {Kurilovich}, \citenamefont {Diamond},
  \citenamefont {Nho}, \citenamefont {B{\o}ttcher}, \citenamefont {Glazman},
  \citenamefont {Fatemi},\ and\ \citenamefont
  {Devoret}}]{connolly2024coexistence}%
  \BibitemOpen
  \bibfield  {author} {\bibinfo {author} {\bibfnamefont {T.}~\bibnamefont
  {Connolly}}, \bibinfo {author} {\bibfnamefont {P.~D.}\ \bibnamefont
  {Kurilovich}}, \bibinfo {author} {\bibfnamefont {S.}~\bibnamefont {Diamond}},
  \bibinfo {author} {\bibfnamefont {H.}~\bibnamefont {Nho}}, \bibinfo {author}
  {\bibfnamefont {C.~G.}\ \bibnamefont {B{\o}ttcher}}, \bibinfo {author}
  {\bibfnamefont {L.~I.}\ \bibnamefont {Glazman}}, \bibinfo {author}
  {\bibfnamefont {V.}~\bibnamefont {Fatemi}},\ and\ \bibinfo {author}
  {\bibfnamefont {M.~H.}\ \bibnamefont {Devoret}},\ }\href@noop {} {\bibfield
  {journal} {\bibinfo  {journal} {Physical Review Letters}\ }\textbf {\bibinfo
  {volume} {132}},\ \bibinfo {pages} {217001} (\bibinfo {year}
  {2024})}\BibitemShut {NoStop}%
\bibitem [{\citenamefont {Diamond}\ \emph {et~al.}(2022)\citenamefont
  {Diamond}, \citenamefont {Fatemi}, \citenamefont {Hays}, \citenamefont {Nho},
  \citenamefont {Kurilovich}, \citenamefont {Connolly}, \citenamefont {Joshi},
  \citenamefont {Serniak}, \citenamefont {Frunzio}, \citenamefont {Glazman}
  \emph {et~al.}}]{diamond2022distinguishing}%
  \BibitemOpen
  \bibfield  {author} {\bibinfo {author} {\bibfnamefont {S.}~\bibnamefont
  {Diamond}}, \bibinfo {author} {\bibfnamefont {V.}~\bibnamefont {Fatemi}},
  \bibinfo {author} {\bibfnamefont {M.}~\bibnamefont {Hays}}, \bibinfo {author}
  {\bibfnamefont {H.}~\bibnamefont {Nho}}, \bibinfo {author} {\bibfnamefont
  {P.~D.}\ \bibnamefont {Kurilovich}}, \bibinfo {author} {\bibfnamefont
  {T.}~\bibnamefont {Connolly}}, \bibinfo {author} {\bibfnamefont {V.~R.}\
  \bibnamefont {Joshi}}, \bibinfo {author} {\bibfnamefont {K.}~\bibnamefont
  {Serniak}}, \bibinfo {author} {\bibfnamefont {L.}~\bibnamefont {Frunzio}},
  \bibinfo {author} {\bibfnamefont {L.~I.}\ \bibnamefont {Glazman}}, \emph
  {et~al.},\ }\href@noop {} {\bibfield  {journal} {\bibinfo  {journal} {PRX
  Quantum}\ }\textbf {\bibinfo {volume} {3}},\ \bibinfo {pages} {040304}
  (\bibinfo {year} {2022})}\BibitemShut {NoStop}%
\bibitem [{\citenamefont {Marchegiani}\ \emph {et~al.}(2022)\citenamefont
  {Marchegiani}, \citenamefont {Amico},\ and\ \citenamefont
  {Catelani}}]{marchegiani2022quasiparticles}%
  \BibitemOpen
  \bibfield  {author} {\bibinfo {author} {\bibfnamefont {G.}~\bibnamefont
  {Marchegiani}}, \bibinfo {author} {\bibfnamefont {L.}~\bibnamefont {Amico}},\
  and\ \bibinfo {author} {\bibfnamefont {G.}~\bibnamefont {Catelani}},\
  }\href@noop {} {\bibfield  {journal} {\bibinfo  {journal} {PRX Quantum}\
  }\textbf {\bibinfo {volume} {3}},\ \bibinfo {pages} {040338} (\bibinfo {year}
  {2022})}\BibitemShut {NoStop}%
\bibitem [{\citenamefont {McEwen}\ \emph {et~al.}()\citenamefont {McEwen},
  \citenamefont {Miao}, \citenamefont {Atalaya}, \citenamefont {Bilmes},
  \citenamefont {Crook}, \citenamefont {Bovaird}, \citenamefont {Kreikebaum},
  \citenamefont {Zobrist}, \citenamefont {Jeffrey}, \citenamefont {Ying} \emph
  {et~al.}}]{mcewen2402resisting}%
  \BibitemOpen
  \bibfield  {author} {\bibinfo {author} {\bibfnamefont {M.}~\bibnamefont
  {McEwen}}, \bibinfo {author} {\bibfnamefont {K.}~\bibnamefont {Miao}},
  \bibinfo {author} {\bibfnamefont {J.}~\bibnamefont {Atalaya}}, \bibinfo
  {author} {\bibfnamefont {A.}~\bibnamefont {Bilmes}}, \bibinfo {author}
  {\bibfnamefont {A.}~\bibnamefont {Crook}}, \bibinfo {author} {\bibfnamefont
  {J.}~\bibnamefont {Bovaird}}, \bibinfo {author} {\bibfnamefont
  {J.}~\bibnamefont {Kreikebaum}}, \bibinfo {author} {\bibfnamefont
  {N.}~\bibnamefont {Zobrist}}, \bibinfo {author} {\bibfnamefont
  {E.}~\bibnamefont {Jeffrey}}, \bibinfo {author} {\bibfnamefont
  {B.}~\bibnamefont {Ying}}, \emph {et~al.},\ }\href@noop {} {\bibinfo
  {journal} {arXiv preprint arXiv:2402.15644}\ }\BibitemShut {NoStop}%
\bibitem [{\citenamefont {Houzet}\ \emph {et~al.}(2019)\citenamefont {Houzet},
  \citenamefont {Serniak}, \citenamefont {Catelani}, \citenamefont {Devoret},\
  and\ \citenamefont {Glazman}}]{houzet2019photon}%
  \BibitemOpen
\bibfield  {journal} {  }\bibfield  {author} {\bibinfo {author} {\bibfnamefont
  {M.}~\bibnamefont {Houzet}}, \bibinfo {author} {\bibfnamefont
  {K.}~\bibnamefont {Serniak}}, \bibinfo {author} {\bibfnamefont
  {G.}~\bibnamefont {Catelani}}, \bibinfo {author} {\bibfnamefont
  {M.}~\bibnamefont {Devoret}},\ and\ \bibinfo {author} {\bibfnamefont
  {L.}~\bibnamefont {Glazman}},\ }\href@noop {} {\bibfield  {journal} {\bibinfo
   {journal} {Physical review letters}\ }\textbf {\bibinfo {volume} {123}},\
  \bibinfo {pages} {107704} (\bibinfo {year} {2019})}\BibitemShut {NoStop}%
\bibitem [{\citenamefont {Liu}\ \emph {et~al.}(2024)\citenamefont {Liu},
  \citenamefont {Harrison}, \citenamefont {Patel}, \citenamefont {Wilen},
  \citenamefont {Rafferty}, \citenamefont {Shearrow}, \citenamefont {Ballard},
  \citenamefont {Iaia}, \citenamefont {Ku}, \citenamefont {Plourde} \emph
  {et~al.}}]{liu2024quasiparticle}%
  \BibitemOpen
  \bibfield  {author} {\bibinfo {author} {\bibfnamefont {C.-H.}\ \bibnamefont
  {Liu}}, \bibinfo {author} {\bibfnamefont {D.~C.}\ \bibnamefont {Harrison}},
  \bibinfo {author} {\bibfnamefont {S.}~\bibnamefont {Patel}}, \bibinfo
  {author} {\bibfnamefont {C.~D.}\ \bibnamefont {Wilen}}, \bibinfo {author}
  {\bibfnamefont {O.}~\bibnamefont {Rafferty}}, \bibinfo {author}
  {\bibfnamefont {A.}~\bibnamefont {Shearrow}}, \bibinfo {author}
  {\bibfnamefont {A.}~\bibnamefont {Ballard}}, \bibinfo {author} {\bibfnamefont
  {V.}~\bibnamefont {Iaia}}, \bibinfo {author} {\bibfnamefont {J.}~\bibnamefont
  {Ku}}, \bibinfo {author} {\bibfnamefont {B.~L.}\ \bibnamefont {Plourde}},
  \emph {et~al.},\ }\href@noop {} {\bibfield  {journal} {\bibinfo  {journal}
  {Physical Review Letters}\ }\textbf {\bibinfo {volume} {132}},\ \bibinfo
  {pages} {017001} (\bibinfo {year} {2024})}\BibitemShut {NoStop}%
\bibitem [{\citenamefont {Clerk}\ and\ \citenamefont
  {Utami}(2007)}]{clerk2007using}%
  \BibitemOpen
  \bibfield  {author} {\bibinfo {author} {\bibfnamefont {A.}~\bibnamefont
  {Clerk}}\ and\ \bibinfo {author} {\bibfnamefont {D.~W.}\ \bibnamefont
  {Utami}},\ }\href@noop {} {\bibfield  {journal} {\bibinfo  {journal}
  {Physical Review A—Atomic, Molecular, and Optical Physics}\ }\textbf
  {\bibinfo {volume} {75}},\ \bibinfo {pages} {042302} (\bibinfo {year}
  {2007})}\BibitemShut {NoStop}%
\bibitem [{\citenamefont {Manucharyan}\ \emph {et~al.}(2012)\citenamefont
  {Manucharyan}, \citenamefont {Masluk}, \citenamefont {Kamal}, \citenamefont
  {Koch}, \citenamefont {Glazman},\ and\ \citenamefont
  {Devoret}}]{manucharyan2012evidence}%
  \BibitemOpen
  \bibfield  {author} {\bibinfo {author} {\bibfnamefont {V.~E.}\ \bibnamefont
  {Manucharyan}}, \bibinfo {author} {\bibfnamefont {N.~A.}\ \bibnamefont
  {Masluk}}, \bibinfo {author} {\bibfnamefont {A.}~\bibnamefont {Kamal}},
  \bibinfo {author} {\bibfnamefont {J.}~\bibnamefont {Koch}}, \bibinfo {author}
  {\bibfnamefont {L.~I.}\ \bibnamefont {Glazman}},\ and\ \bibinfo {author}
  {\bibfnamefont {M.~H.}\ \bibnamefont {Devoret}},\ }\href@noop {} {\bibfield
  {journal} {\bibinfo  {journal} {Physical Review B—Condensed Matter and
  Materials Physics}\ }\textbf {\bibinfo {volume} {85}},\ \bibinfo {pages}
  {024521} (\bibinfo {year} {2012})}\BibitemShut {NoStop}%
\end{thebibliography}%

\end{document}